\documentclass[12pt]{article}
\usepackage{url, natbib}
\usepackage{graphicx}
\usepackage{amsmath, color, adjustbox, bm}
\usepackage{rotating}
\usepackage{caption}

\author{ \hspace{-1cm}\small
Jingxiong Xu$^{1,2}$, Wei Xu$^{1,3}$, Laurent Briollais$^{1,2}$ \\ \\
\small \hspace{-1cm}
$^1$ Dalla Lana School of Public Health, University of Toronto, Toronto, Canada \\
\small \hspace{-2cm}
$^2$ Lunenfeld-Tanenbaum Research Institute, Mount Sinai Hospital, Toronto, Canada \\
\small\hspace{-2cm}
$^3$ Princess Margaret Cancer Center, Toronto, Canada \\
\\
}
\title {A Bayes Factor Approach with Informative Prior for Rare Genetic Variant Analysis from Next Generation Sequencing Data}
\date{}
\begin{document}
\maketitle

\begin{abstract}
{The discovery of rare genetic variants through Next Generation Sequencing is a very challenging issue in the field of human genetics. We propose a novel region-based statistical approach based on a Bayes Factor (BF) to assess evidence of association between a set of rare variants (RVs) located on the same genomic region and a disease outcome {in the context of case-control design}. Marginal likelihoods are computed under the null and alternative hypotheses assuming a binomial distribution for the RV count in the region and a beta or mixture of Dirac and beta prior distribution for the probability of RV. We derive the theoretical null distribution of the BF under our prior setting and show that a Bayesian control of the False Discovery Rate (BFDR) can be obtained for genome-wide inference. Informative priors are introduced using prior evidence of association from a Kolmogorov-Smirnov test statistic. We use our simulation program, sim1000G, to generate RV data similar to the 1,000 genomes sequencing project. Our simulation studies showed that the new BF statistic outperforms standard methods (SKAT, SKAT-O, Burden test) in case-control studies with moderate sample sizes and is equivalent to them under large sample size scenarios. Our real data application to a lung cancer case-control study found enrichment for RVs in known and novel cancer genes. It also suggests that using the BF with informative prior improves the overall gene discovery compared to the BF with non-informative prior. \\ \\
}
{Bayes Factor; Bayesian FDR; Gene-based analysis; Rare variant; Whole Exome Sequencing study.
}
\end{abstract}

\section{Introduction}
\label{s:intro}

The emergence of new high-throughput genotyping technologies, such as Next Generation Sequencing (NGS), allows the study of the human genome at an unprecedented depth and scale \citep{Lee2014}. They provide invaluable opportunities to decipher the biological processes involved in complex human diseases, such as cancers, in particular to understand the complex molecular mechanisms leading to cancer development and progression \citep{Mardis2009}.  With the advances of NGS technology and its deep coverage capacity, there have been increased interests in the analysis of rare {single-nucleotide variants (SNVs)},  e.g., minor allele frequency (MAF) less than 1\%, as reviewed in \cite{Gibson2012}. Besides, the genetic variability explained by common SNPs identified through genome-wide association studies (GWAS) has been generally low and it is hypothesized that some of the unexplained variability might be due to rare variants (RVs) \citep{Manolio2009}.

For example, a recent application of NGS to a lung cancer study has led to the successful identification of 48 germline RVs with deleterious effects on lung cancer in known candidate genes such as $BRCA2$ in a sample of 260 case patients with the disease and 318 controls \citep{Liu2018}. This study underlines the significant contribution of germline RVs to the genetic susceptibility of lung cancer but it also raises some concerns. The identification of RVs was based on the application of the Burden test statistic \citep{Lee2014}, which is powerful only when the RVs have the same direction of effects in the same gene (e.g. all RVs are deleterious), and it only focused on candidate genes. The motivation for our paper is to propose a more general statistical framework able to identify genes across the entire genome exhibiting RV count differences between cases and controls, where the RVs can be protective or deleterious within the same gene. It could therefore help disentangle the contribution of RVs to the genetic susceptibility of lung cancer and of other complex human diseases. 

The discovery of RVs through NGS raises many statistical challenges. Since RVs have extremely low frequencies, traditional strategies that analyze one variant at a time are underpowered for detecting associations with RVs. Gene-level statistics can provide a first step in the analysis of RVs. Once a gene has been implied as associated with a disease outcome, based on statistical evidence, a variant-level assessment could follow that relies on in-depth genetic, experimental and informatic analyses \citep{MacArthur2014}. Besides, gene-based groupings can lead to biological interpretation and be further validated by functional experiments, including studies on model organisms and cell lines \citep{Cirulli2016,Sung2014}. 

The general principle for RV analysis of these methods is usually to collapse information across multiple RVs from the same gene or genomic region and to test an association between the cumulative effects of multiple RVs with the response of interest. These methods can be broadly categorized into five main classes: burden tests, adaptive burden tests, variance-component tests, combined burden and variance component tests, and the exponential-combination (EC) test (See Table 2 in \cite{Lee2014} for a review). {Bayesian analysis is not well developed for RV analysis but has had relevant applications in the context of genetic association studies.  For instance, the Bayes factor (BF) has been attractive because it provides a natural framework for including prior information \citep{Wakefield2009, Spencer2015}. Of note, two region-based RV tests introduced hierarchical Bayesian models to integrate some prior information on genetic variants \citep{Yi2011, He2015} but they have not proposed any genome-wide inference control procedure specific to the BF itself and lack theoretical justifications for region-based inference.}

We propose here a novel region-based statistic based on a BF approach to assess evidence of association between a set of RVs located on same chromosomal region and a disease outcome {in the context of case-control design}. Section 2 describes the data generated by NGS technology. Section 3 defines our model setting, the BF derivation for gene-based analysis and some asymptotic results about the BF distribution. Section 4 introduces a Bayesian False Discovery Rate control procedure for genome-wide inference. Section 5 presents simulation results comparing the BF to popular competing approaches. Section 6 concludes with a real data application on a case-control lung cancer study.

\section{The NGS Data}

The analysis of NGS data involves a number of pre-processing steps. Once the data have been pre-processed, a sequenced genomic region can be displayed as in Figure 1 (left panel) using for example the sequence viewer software IGV \citep{Robinson2011}. The top bar represents the sites with genetic variations in a specific region of the genome. The genotypes of individuals are represented below, where the grey colour is the reference genotype, dark blue the heterozygote individuals and light blue the homozygotes for the rare allele, often coded as 0,1 or 2. The analysis of RVs generally focuses on bi-allelic sites (genetic locus with two possible alleles) with MAF less than $1\%$.  The genotypes can then be recoded in 0 or 1 since the homozygote category is less common, where 1 or 0 indicates whether an individual carries the RV or not. In association studies, the goal is often to compare the distribution of RV counts in cases and controls (Figure 1, right panel). An example of {distribution of RV proportions} for the gene CHEK2 in a sample of 260 cases and 259 controls is given in Web Appendix A. We observe that cases have a higher density corresponding to higher proportion of RVs, which could reflect an enrichment of RVs in this gene. 

\begin{figure}
	\centering
	\begin{minipage}[b]{0.45\textwidth}
		\includegraphics[width=\textwidth]{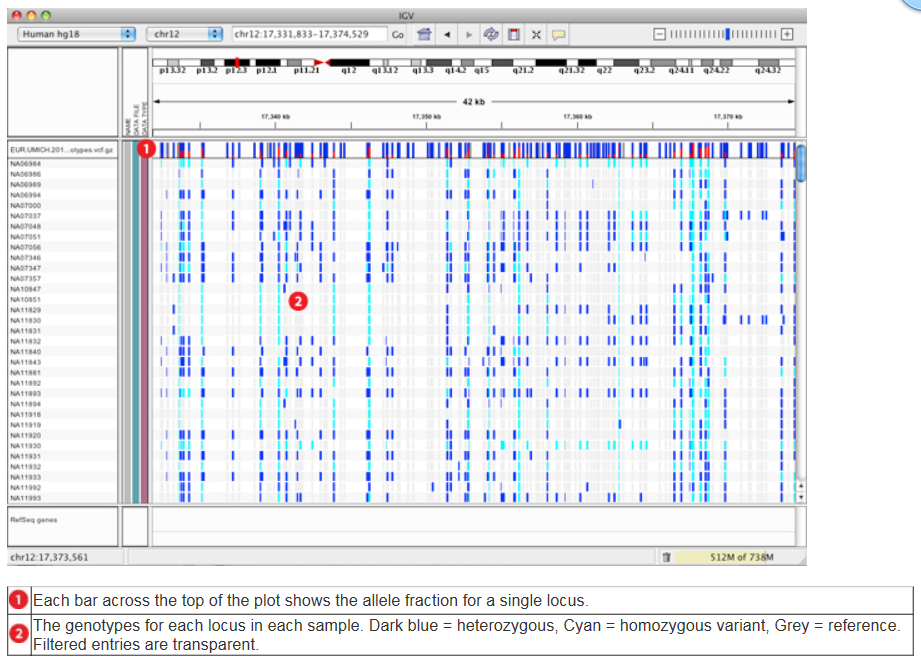}
	\end{minipage}
	\begin{minipage}[b]{0.45\textwidth}
		\includegraphics[width=\textwidth]{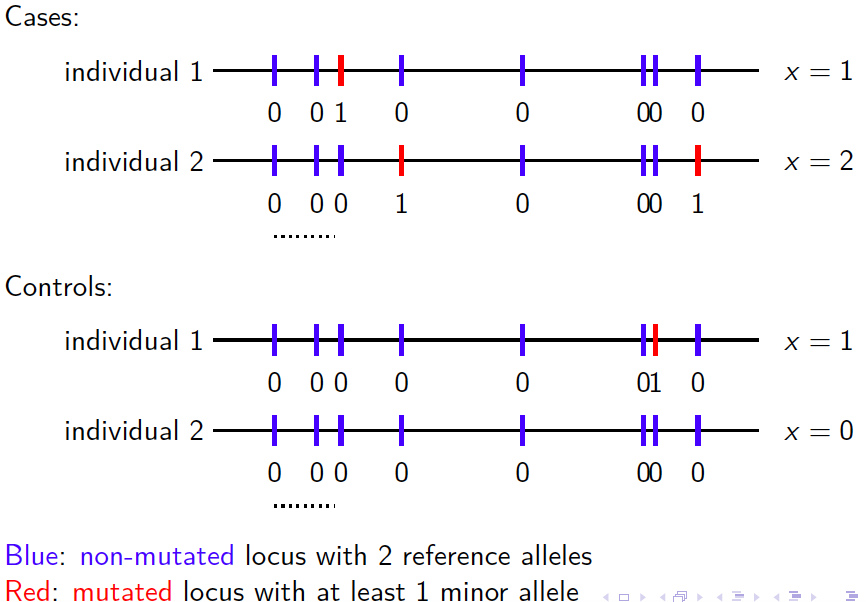}
		\vspace*{-0cm}
	\end{minipage}
	\caption{ Example of NGS data. The left panel displays the NGS data, presented by IGV software, a visualization tool for large scale genomic data. The right panel demonstrates an example region with 10 loci. To detect the genetic association, we derive the test statistic to compare the RV rate between cases and controls by assuming that the RV count of each individual $x$ follows Binomial distribution (see Section 3.1).}
\end{figure}
\section{Model}\label{s:model}
\subsection{Model Setting}
We propose a region-based statistic by modelling the count of RVs in a specific chromosomal region, e.g. a gene, as shown in Web Appendix A. Let $X_{ijk}$ be the count of RVs in the region $i$, for group $j$ and individual $k$, with $i \in \{1,...,m\}$, $j \in \{1,2\}$ (1 for the control group, 2 for the case group) and $k \in \{1, ...,N_j\}$. We assume that the occurrence of a RV at any given site of the region follows an independent Bernoulli process. The distribution of $X_{ijk}$ is therefore binomial
$$ X_{ijk} \sim binomial(n_{ijk},p_{ijk})$$
where  $p_{ijk}$ is the true, unobserved rate of RV at a single locus of the region and $n_{ijk}$ is total number of sites in the region $i$ for group $j$ and individual $k$.

We suppose that $p_{ijk}$ varies across genetic regions and individuals, according to a prior density function $g(p_{ijk} | {\boldsymbol \theta}_{ij} )$, with  ${\boldsymbol \theta}_{ij} \equiv {\boldsymbol \theta}_{i1}$ if $j$ is in the control group and  ${\boldsymbol \theta}_{ij} \equiv {\boldsymbol \theta}_{i2}$ if $j$ is in the case group. Our goal is to assess whether there is a difference in RV counts between cases and controls in a particular region $i$ by comparing : $H_{i0}: {\boldsymbol \theta}_{i1}= {\boldsymbol \theta}_{i2}= {\boldsymbol \theta}_{i}$ vs. $H_{i1}: {\boldsymbol \theta}_{i1} \neq {\boldsymbol \theta}_{i2}$ using the BF statistic. 
\subsection{BF derivation under case-control design}
By definition, the BF is the ratio of the marginal likelihoods of the observed data under $H_1$($m_1({\boldsymbol X})$) and $H_0$($m_0({\boldsymbol X})$). We derive the BF assuming a case-control sampling design.

Let ${\boldsymbol X} \equiv {\boldsymbol X}_N=(X_{1},...,X_{N})$ be the vector of RV counts and ${\boldsymbol P} \equiv {\boldsymbol P}_N=(p_{1},...,p_{N})$ the vector of RV proportions over $N (=N_1+N_2)$ individuals.  Under $H_0$, the marginal likelihood is
\begin{equation}
\begin{split}
m_0({\boldsymbol X})  
= &  \int_P f({\boldsymbol X}|{\boldsymbol P})g({\boldsymbol P})d{\boldsymbol P}, \\
= &  \int_P f( {\boldsymbol X}| {\boldsymbol P})\int_{\theta} g({\boldsymbol P}| {\boldsymbol \theta})\pi({\boldsymbol \theta})d {\boldsymbol \theta} d{\boldsymbol P}. \\
\end{split}
\end{equation}	
where $f$ denotes the binomial distribution probability mass function, $g$ the prior density function for ${\boldsymbol P}$, and $\pi$ the density function for the parameter ${\boldsymbol \theta}$ that we are interested to compare between cases and controls. 

Under $H_1$, the marginal likelihood is written as a product of two marginal likelihood functions over  cases and controls
\begin{equation}
\begin{split}
m_1({\boldsymbol X}) 
= &  \int_{P_1} f({\boldsymbol X}_1| {\boldsymbol P}_1)g( {\boldsymbol P}_1)d{\boldsymbol P}_1 \int_{P_2} f({\boldsymbol X}_2|{\boldsymbol P}_2)g({\boldsymbol P}_2)d{\boldsymbol P}_2, \\
= &  \int_{P_1} f({\boldsymbol X}_1| {\boldsymbol P}_1)\int_{\theta_1} g({\boldsymbol P}_1|{\boldsymbol \theta}_1)\pi({\boldsymbol \theta}_1)d{\boldsymbol \theta}_1 d{\boldsymbol P}_1 \int_{P_2} f({\boldsymbol X}_2|{\boldsymbol P}_2)\int_{\theta_2} g({\boldsymbol P}_2|{\boldsymbol \theta}_2)\pi({\boldsymbol \theta}_2)d{\boldsymbol \theta}_2 d{\boldsymbol P}_2.\\
\end{split}
\end{equation}
where ${\boldsymbol X}_1$ and ${\boldsymbol P}_1$ are the vector of RV counts and rates in controls and ${\boldsymbol X}_2$, ${\boldsymbol P}_2$ in cases. Thus, we have ${\boldsymbol X} \equiv ({\boldsymbol X}_1,{\boldsymbol X}_2)$ and ${\boldsymbol P} \equiv ({\boldsymbol P}_1,{\boldsymbol P}_2)$. The marginal likelihoods are calculated using the Laplace approximation.

\subsection{Prior specification}
Since the proportions of RVs across individuals  can be anywhere between 0 and 1, we assume that these proportions for each genomic region within each group of
individuals follow a beta distribution or a mixture of Dirac and beta distribution. {The mixture prior distribution is proposed to capture the excess of zero counts in the distribution of RV counts as illustrated in Web Appendix A}. The beta distribution has long been a natural choice to model binomial proportions
as it is a conjugate prior distribution of the binomial distribution with a support interval of $[0,1]$. 

\noindent For the beta prior, we assume 
$$ p_{ijk} | {\boldsymbol \theta}_{ij} \sim beta(\eta_{ij}, K_{ij}), \;\;\;\;\; \textrm{with    } {\boldsymbol \theta}_{ij} \equiv  (\eta_{ij}, K_{ij}).$$

Here the beta distribution is parametrized in terms of mean (denoted by $\eta_{ij}$) and precision (denoted by $K_{ij}$). Compared with the traditional parameterization of the $beta(\alpha, \beta)$ distribution, the parameters have the following relationship: 
$$\eta=\frac{\alpha}{(\alpha + \beta)}, \;\; K=\alpha + \beta.$$ 
With the beta prior, the marginal distribution of RV count in the region is beta-binomial. It assumes a similar pairwise correlation between the RV within the region measured by the intraclass correlation $\rho=1/(1+K)$. Our simulation studies and real data analysis showed that the beta-binomial distribution provided a very good fit of the RV rate (Web Appendix B) as well as the pairwise correlation coefficient between RVs (Web Appendix C) within a gene. 

For the mixture prior, we assume that $p_{ijk}$ follows a mixture distribution of a point mass at zero and a beta distribution with parameters $\eta_{ij}$ and $K_{ij}$, with probability $w_{0ij}$ and $w_{1ij}=1-w_{0ij}$, respectively. The distribution of $X_{ijk}$ becomes
\begin{eqnarray*}
	X_{ijk}=\left\{
	\begin{array}{@{}ll@{}}
		0, & \textrm{if}\ p_{ijk}=0 \textrm{ with } P(p_{ijk}=0)=w_{0ij}, \\
		X_{ijk} \sim Bin(n_{ijk},p_{ijk}), & \textrm{if}\ p_{ijk}>0 \textrm{ with } P(p_{ijk}>0)=1-w_{0ij}. 
	\end{array}\right.
\end{eqnarray*} 
Also when $p_{ijk}>0$, the prior density for $p_{ijk}$ is $beta(\eta_{ij},K_{ij})$.

For both the beta prior and mixture prior, we assume $(\eta_{ij}, K_{ij}) \equiv \theta_{i1} \equiv  (\eta_{i1}, K_{i1})$ if $j$ is in the control group or $(\eta_{ij}, K_{ij}) \equiv \theta_{i2} \equiv (\eta_{i2}, K_{i2})$ if $j$ is in the case group. In addition, for the mixture prior, we also have $w_{0ij} \equiv w_{0i1}$ or  $w_{0ij} \equiv w_{0i2}$ if $j$ is in the control or case group, respectively. Our BF statistic aims to compare $\eta_{i1}$ and ${\eta_{i2}}$ or both  ($\eta_{i1}$, $w_{0i1}$) and ( ${\eta_{i2}}$, $w_{0i2}$). We further assumed the same $K_{ij}\equiv K_{i}$ in cases and controls, which fitted well our data, according to the marginal likelihood criteria.

The precision parameter $K_{i}$ captures the variation of the proportion of RVs relative to the group mean. Under both beta and mixture prior, $K_{i}$ can be estimated for each region using the MLE. 
\subsection{BF Calculation}\label{BF_calculate}
The BF is calculated for each genomic region (or gene) $i$. For the sake of presentation, we omit the index $i$ in this section. 
\subsubsection{BF with beta prior}\label{BF_betaprior}
To compare the distribution of RVs in cases and controls, our hyperparameters of interest are $\eta$, $\eta_1$ and $\eta_2$. The null and alternative hypotheses can be formulated  as
$$H_0: \eta_{1}  = \eta_{2} = \eta$$
$$H_1: \eta_{1}  \neq \eta_{2}$$
We assume a hyper prior structure where each hyperparameter is assumed to follow a beta distribution with new mean parameters ($\eta^*$, $\eta_1^*$, $\eta_2^*$ for $\eta$, $\eta_1$, $\eta_2$, respectively) and new precision parameters ($K^*$, $K_1^*$, $K_2^*$ for $\eta$, $\eta_1$, $\eta_2$, respectively).

The marginal likelihood of the data $\boldsymbol X$ under $H_0$ is
\begin{equation}\label{m0_beta}
\begin{split}
m_0({\bf X}|K,\eta^*,K^*) 
= &\displaystyle\int_{\eta}  \displaystyle\int_{{ P}} f({\bf X}|{\bf P}) g({\bf P}|\eta,K) d{\bf P}\;\; \pi(\eta|\eta^*,K^*)d\eta,  \\
= & \prod_{j=1}^{2}\prod_{k=1}^{N_j} \binom{n_{jk}}{x_{jk}} \displaystyle\int_{\eta}\frac{\prod_{j=1}^{2}\prod_{k=1}^{N_j}B\{x_{jk}+K\eta,n_{jk}-x_{jk}+K(1-\eta)\}}{B\{K\eta,K(1-\eta)\}^N}  \\
&\displaystyle\int_{P} \prod_{j=1}^{2}\prod_{k=1}^{N_j}\frac{p^{x_{jk}+K\eta-1}_{jk}(1-p_{jk})^{n_{jk}-x_{jk}+K(1-\eta)-1}}{B\{x_{jk}+K\eta,n_{jk}-x_{jk}+K(1-\eta)\}} d{\bf P} \frac{\eta^{K^*\eta^*-1}(1-\eta)^{K^*(1-\eta^*)-1}}{B(K^*\eta^*,K^*(1-\eta^*))} d\eta,\\
= & \prod_{j=1}^{2}\prod_{k=1}^{N_j} \binom{n_{jk}}{x_{jk}} \displaystyle\int_{0}^{1}\frac{\prod_{j=1}^{2}\prod_{k=1}^{N_j}B\{x_{jk}+K\eta,n_{jk}-x_{jk}+K(1-\eta)\}}{B\{K\eta,K(1-\eta)\}^N}\\ 
&\frac{\eta^{K^*\eta^*-1}(1-\eta)^{K^*(1-\eta^*)-1}}{B\{K^*\eta^*,K^*(1-\eta^*)\}} d\eta.\\
\end{split}
\end{equation}
The marginal likelihood of the data $\boldsymbol X$ under $H_1$ is
\begin{equation}\label{m1_beta}
\begin{split}
m_1({\bf X}|K,\eta_1^*,K_1^*,\eta_2^*,K_2^*) = & \prod_{j=1}^{2}\prod_{k=1}^{N_j} \binom{n_{jk}}{x_{jk}} \times \prod_{j=1}^{2}\displaystyle\int_{0}^{1}\frac{\prod_{k=1}^{N_j}B\{x_{jk}+K\eta,n_{jk}-x_{jk}+K(1-\eta)\}}{B\{K\eta,K(1-\eta)\}^{N_j}}\\ &\frac{\eta^{K_j^*\eta_j^*-1}(1-\eta)^{K_j^*(1-\eta_j^*)-1}}{B\{K_j^*\eta_j^*,K_j^*(1-\eta_j^*)\}} d\eta .
\end{split}
\end{equation}
The BF is the ratio of marginal likelihood under $H_1$ and under $H_0$:
$BF = \frac{m_1({\bf X}|K,\eta_1^*,K_1^*,\eta_2^*,K_2^*)}{m_0({\bf X}|K,\eta^*,K^*)}.$ In this expression, we consider a fixed $K$ and estimate its MLE, $\tilde{K}$, from the whole sample {under the null hypothesis}. Therefore,
\begin{equation}\label{BF_last}
BF = \frac{m_1({\bf X}|\tilde{K},\eta_1^*,K_1^*,\eta_2^*,K_2^*)}{m_0({\bf X}|\tilde{K},\eta^*,K^*)}.
\end{equation}


The integral part in the above BF formula is calculated using Laplace approximation (see equations (1) and (2) of Web Appendix D). The accuracy of this approximation is evaluated in Web Appendix E. 

\subsubsection{hyperparameter specification}\label{Hyper_param}

We use empirical Bayes estimates for the parameters $\eta^*$,  $\eta_1^*$ and $\eta_2^*$ in the hyperprior distribution, i.e.,   $\eta^* = \hat{\eta}$, $\eta_1^* = \hat{\eta}_1$, $\eta_2^* = \hat{\eta}_2$, where $\hat{\eta}$ is the MLE of the parameter $\eta$ in the likelihood function $f({\bf X}|\eta,\tilde{K}) = \displaystyle\int_{P} f({\bf X}|{\bf P}) g({\bf P}|\eta,\tilde{K}) d{\bf P}$ obtained by 
$\hat{\eta}=arg\underset{\eta}{\operatorname{max}} \log f({\bf X}|\eta, \tilde{K}) $. 
The estimates of $\eta_1$ and $\eta_2$ are obtained in a similar way by the MLE computed separately in controls and cases but assuming the same $\tilde{K}$, i.e.  $\hat{\eta_1}=arg\underset{\eta}{\operatorname{max}} \log f({\bf X_1}|\eta, \tilde{K})  $ and $\hat{\eta_2}=arg\underset{\eta}{\operatorname{max}} \log f({\bf X_2}|\eta, \tilde{K})  $.  The other hyperparameters $K^{*}_1$, $K^{*}_2$ and $K^{*}$ are interpreted as precision parameters of the mean RV counts in controls, cases and whole sample, respectively. Our rationale for defining these hyperparameters is to get convenient asymptotic results for the BF distribution and to be able to introduce informative prior at the region level. Therefore, we set $K^* = p^2\hat{\eta}\hat{\Sigma}^{-1}$, $K_1^* = \hat{\eta}_1\hat{\Sigma}_1^{-1}$ and $K_2^* = \hat{\eta}_2\hat{\Sigma}_2^{-1}$, where $\hat\Sigma$, $\hat\Sigma_1$ and $\hat\Sigma_2$ are { the estimated variances of $\hat\eta$, $\hat\eta_1$ and $\hat\eta_2$, respectively}. In the expression of $K^*$, $p$ is a term that allows to introduce informative prior at the region (or gene) level. {The role of p is to influence the precision parameter K* under the null hypothesis (denominator of equation (\ref{BF_last})), where $p<1$ leads to a smaller K* in equation (\ref{BF_last}) and thus more uncertainty on the prior of η. This yields less evidence for the null hypothesis in comparison with the situation $p=1$ and thus stronger evidence for association. Because the effect of this prior is to force the evidence towards association (alternative hypothesis) when $p \rightarrow 0$, we refer to this prior as "informative" and the situation where p=1 as "non-informative", since in this latter case, no attempt to influence either hypothesis is made.} Indeed, in section 3.5 below, we explain that $p$ can be obtained from the p-value of a Kolmogorov Smirnov (KS) test, comparing the distribution of RV counts within the region of interest between cases and controls. We can prove that under $H_0$, when $p=1$, $2\log BF \xrightarrow{d} \chi^2(1)$. When $p<1$ and assuming independence between the BF with non-informative prior and the KS component, we show that with our definition of $K^*$ above (in particular the use of $p^2$ in this expression), we obtain $2\log BF \xrightarrow{d} \chi^2(3)$ (see section 3.6 and Web Appendix D). Thus, more information is added to the BF and it will give more evidence for association when it can overcome the additional 2 degrees of freedom provided by the KS test p-value. 


\subsubsection{BF with mixture prior}
We now assume that the RV counts among individuals are ordered with $x_{(1)}=x_{(2)}=...=x_{(M)}=0$ and $x_{(M+1)},...,x_{(N)}>0$. Let $0<M<N$ and {the variance} $Var(x_{(M+1)},...,x_{(N)})>0$, for the mixture prior to be defined.
As noted previously, the prior distribution for $p_{jk}$ can be written as

\begin{equation*}
	g(p_{jk}|w_{0j},\eta_{j})=\left\{
	\begin{array}{@{}ll@{}}
		w_{0j}, & \text{if}\ p_{jk}=0, \\
		(1-w_{0j})g(p_{jk}|\eta_j,K)=(1-w_{0j})\frac{p_{jk}^{\eta_j K-1}(1-p_{jk})^{K(1-\eta_j)-1}}{B(\eta_j K,K(1-\eta_j))}, & \text{if}\ p_{jk}>0. 
	\end{array}\right.
\end{equation*} 

Under the mixture prior, the hyperparameters of interest are $w_0$, $w_{01}$, $w_{02}$, $\eta$, $\eta_1$, and $\eta_2$. Similar to \ref{BF_betaprior}, we assume a hyper prior structure where each hyperparameter is assumed to follow a beta distribution
$$ \eta \sim beta(\eta^*,K^*), \eta_1 \sim beta(\eta_1^*,K_1^*), \;\text{and}\; \eta_2 \sim beta(\eta_2^*,K_2^*),$$
$$ w_0 \sim beta(\eta^{**},K^{**}), w_{01} \sim beta(\eta_1^{**},K_1^{**}), \;\text{and}\; w_{02} \sim beta(\eta_2^{**},K_2^{**}).$$

We propose to perform two comparisons, first compare $\eta_1$ and $\eta_2$ with fixed $w_{0j}$ and second, compare both ($\eta_1$,  $w_{01}$) to ($\eta_2$, $w_{02}$). {The derivation of this BF follows the same principle as the BF with beta prior shown before and the details are given in Web Appendix F.  }

\subsection{Informative prior}\label{s:inf_prior}
As explained in section 3.4.2, we choose $p$ as the probability of association from a Kolmogorov Smirnov (KS) test derived as follows. First, we conduct a single RV test between cases and controls for all variants in region $i$ as explained below; Second, we compare the set of region-specific $p$-values to a null $p$-value distribution using a one-sided one-sample KS test (ie. we expect a higher proportion of low $p$-values for associated genes). The null distribution is empirically estimated using all the genes simulated under $H_0$ in the simulation study or all the genes across the genome in the real data analysis. {The advantage of this prior is that it can capture RV allelic differences between cases and controls through the KS test and those RVs can have opposite effects within the same gene. The other advantage is that it is almost uncorrelated with the BF assuming non-informative prior and thus provides convenient asymptotic results for the BF with informative prior. {The rationale is that the BF with non-informative prior  compares the overall RV count between cases and controls but does not depend on the differences in RV allelic frequencies at individual sites between them, while the KS component is based on allelic distribution differences between cases and controls at individual sites within the region}. Evidence for the very low correlation is given in Web Appendix G. }

{For the single RV test, for each single variant $v$ in region $i$, let $Y_{i1v}$ and $Y_{i2v}$ be the RV counts within cases and controls. We transform the observed RV count $Y_{ijv}$ to obtain the count $\tilde{Y}_{ijv} = Y_{ijv}*C_v$, where $C_v$ is the largest integer $\leq \frac{0.01}{P_v}$ and $P_v$ is the MAF of variant $v$. In this way, all the single RV p-values can be comparable regardless of their MAF. We assume $\tilde{Y}_{ijv}$ has a Poisson distribution with
	$\tilde{Y}_{i1v} \sim Poisson(\lambda_{1v} N_1) \;\; \text{ and }\;\;  \tilde{Y}_{i2v} \sim Poisson(\lambda_{2v} N_2) . $
	The parameters $\lambda_{1v}$ and $\lambda_{2v}$ correspond to the probabilities of a control or a case to have a RV at site $v$. Our interest is to test $\lambda_{1v}=\lambda_{2v}$.  When both $\lambda_{1v} N_1$ and $\lambda_{2v} N_2$  are large, the normal distribution is a good approximation of the Poisson distribution. Define $r=\frac{N_2}{N_1}$ and $W_{iv} = r\tilde{Y}_{i1v}-\tilde{Y}_{i2v}$, under $H_0$, we have
	\begin{equation}\label{RV_normal}
	\frac{W_{iv}-E(W_{iv})}{\sqrt{Var(W_{iv})}} \approx \frac{r\tilde{Y}_{i1v}-\tilde{Y}_{i2v}}{\sqrt{r^2\tilde{Y}_{i1v}+\tilde{Y}_{i2v}}}\sim N(0,1).
	\end{equation}

	Since $\lambda_{1v}$ and $\lambda_{2v}$ are usually small, the condition that both $\lambda_1 N_1$ and $\lambda_2 N_2$  are large is met only when the sample sizes $N_1$ and $N_2$ are large.  Besides, for the validity of these variant level comparison tests, we recommend excluding the sites with $\tilde{y}_{1k} + \tilde{y}_{2k} < 5$ and those with extremely low MAF, e.g. $MAF<0.1\%$.  We conduct two-sided test against $H_0$: $\lambda_{1v}=\lambda_{2v}$ for each RV.}

\subsection{Asymptotic property of the BF statistic under the null hypothesis}
For region-based inference (i.e. testing one region at the time) or genome-wide inference (testing all regions simultaneously), it is critical to know the asymptotic distribution of the BF under the null. Given a known null distribution of BF, the {\it p}-value corresponding to each region-based BF  can be derived. The problem of genome-wide inference is addressed in the next section.

In this section, $\eta_0$ denotes the true value of $\eta$. Based on the likelihood functions $Pr({\bf X}|{\eta})$, $Pr({\bf X_1}|{\eta_1})$ and $Pr({\bf X_2}|{\eta_2})$, by definition of the MLE, we have under $H_0$, $\hat{\eta} \sim N(\eta_0,\Sigma)$, $\hat\eta_1 \sim N(\eta_0,\Sigma_1)$, and $\hat\eta_2 \sim N(\eta_0,\Sigma_2)$. In addition, we define
$\Sigma^{-1} \equiv  -\frac{d^2\log Pr({\bf X}|{\eta})}{d \eta^2}|_{\eta=\eta_0}$,
$\Sigma_1^{-1} \equiv  -\frac{d^2\log Pr({\bf X_1}|{\eta})}{d \eta^2}|_{\eta=\eta_0}$ and  $\Sigma_2^{-1} \equiv  -\frac{d^2\log Pr({\bf X_2}|{\eta})}{d \eta^2}|_{\eta=\eta_0}$.  \\
The asymptotic distribution depends on the prior specification. Under a particular  prior specification, we show the following result: \\
{\it Proposition 1:} 
Under $H_0$, 
$\hat\eta \approx \frac{\hat\eta_1\Sigma_2+\hat\eta_2\Sigma_1}{\Sigma_1+\Sigma_2}\;\;\text {and} \;\; \Sigma = \frac{\Sigma_1 \Sigma_2}{\Sigma_1+\Sigma_2}.$ (See Web Appendix H for a proof). \\ \ \\
{\it Theorem 1:}	For the BF with beta prior, assuming that the parameters in the hyperprior distribution are defined by $\eta^* = \hat{\eta}$, $\eta_1^* = \hat{\eta}_1$, $\eta_2^* = \hat{\eta}_2$, $K^* =p^2 \hat{\eta}\hat{\Sigma}^{-1}$, $K_1^* = \hat{\eta}_1\hat{\Sigma}_1^{-1}$ and $K_2^* = \hat{\eta}_2\hat{\Sigma}_2^{-1}$, where $p$ is a random variable with uniform distribution in (0,1). When $K^*\eta^* \rightarrow \infty$, $K^*_1\eta^*_1 \rightarrow \infty$ and $K^*_2\eta^*_2 \rightarrow \infty$, sample size of cases $N_1 \rightarrow \infty$ and of controls $N_2 \rightarrow \infty$, and also $\eta_0 \rightarrow 0$, we have $2\log BF \xrightarrow{d} \chi^2(3)$ , under $H_0$. Assuming $p=1$ leads to non-informative priors with $2\log BF \xrightarrow{d} \chi^2(1)$. The full proof of the Theorem 1 is given in the Web Appendix D. 

When using the BF with informative prior ($p\neq 1$), $p$ in the Theorem 1 corresponds to the $p$ value of the KS test (see section 3.5), assumed uniform distributed under the null hypothesis.  We also assumed the independence between the BF with non-informative prior and the KS test component ($-2\log p$). Although a formal proof is not available, we showed in the simulated data and real data application that these two components are almost uncorrelated (Web Appendix G). 

{The derivation of the asymptotic results under $H_0$ for the BF with mixture prior follows the same principles as for the BF with beta prior (see Web Appendix F).}
\section{Bayesian False Discovery Rate (FDR) for genome wide inference}
The goal of  genome-wide inference is to perform a simultaneous testing of multiple hypotheses (i.e. all the genes or all the genomic regions across the genome).
Following \citep{Wen2017}, we propose to use the Bayes factor as the test statistic in the Bayesian decision rule to reject the null hypothesis. 

Suppose, we want to test  $m$ null hypotheses {\bf $H_{0i}, i=1,\cdots,m$}, using data denoted as $Y$.  Let $Z_i=1$ if {\bf $H_{0i}$} is false and $Z_i=0$ if $H_{0i}$ is true,  $i=1,\cdots,m$. If we denote $\pi_0$ the proportion of data generated under the null hypothesis, we assume that $Z_i$ is unobserved indicator with $Z_i|\pi_0 \sim Bernoulli(1-\pi_0)$.

Let $\delta_i$ denote a decision rule in $(0,1)$ on $Z_i$ based on all the observed data and $D=\sum_{i=1}^{m}\delta_i$. Following \citep{Muller2006}, the False Discovery Proportion (FDP) is defined as

$$\text{FDP} \equiv \frac{\sum_{i=1}^{m}\delta_i(1-Z_i)}{\max(1, D)},$$ 
and the Bayesian FDR as:
$$\overline{FDR} \equiv E(FDP|Y)=\frac{\sum_{i=1}^{m}\delta_i(1-v_i)}{\max(1,D)}, $$ 

where $v_i \equiv Pr(Z_i=1|Y)$ is the posterior probability that the observed data is generated from $H_{i1}$ for each test $i$. The interest in the Bayesian control of the FDR is $v_i$,
\begin{equation}\label{vi_def}
\begin{split}
{v_i}&=\int Pr(Z_i=1|Y,\pi_0)p(\pi_0|Y)d\pi_0\\
\end{split}
\end{equation}
where 
\begin{equation}\label{vi_BF}
\begin{split}
Pr(Z_i=1|Y,\pi_0) &= \frac{(1-{\pi}_0)BF_i}{{\pi}_0+(1-{\pi}_0)BF_i}\\
\end{split}
\end{equation}
where $\text{BF}{_i}$ is the BF for $i$th gene, based on either the beta or mixture prior.\\
The Bayesian control of FDR is then based on the decision rule: $$\delta_i(t) = \mathnormal{I}(v_i>t)$$
It indicates that $H_{i0}$ is rejected if the posterior probability that the observed data is generated from $H_{i1}$ is high.
For a pre-defined FDR level $\alpha_0$, the threshold $t_{\alpha_0}$ that controls the FDR at the level $\alpha_0$ is determined by
\begin{equation}\label{threshold}
\begin{split}
t_{\alpha_0} &= arg\underset{t}{\operatorname{min}}\Big\{\frac{\sum_{i=1}^{m}\delta_i(t)(1-v_i)}{\max(1, D(t))}\leq\alpha_0\Big\} \\
\end{split}
\end{equation} 
{As stated above, the BFDR procedure requires an estimate of $v_i$, which in turn depends on $\pi_0$, the proportion of true null hypotheses (i.e, the proportion of true non associated genes/regions). The two main approaches to estimate $v_i$ are based on either setting a prior distribution for $\pi_0$ \citep{Scott2006} or estimating $\pi_0$ from the empirical distribution of the test statistic \citep{Wen2017,Storey2004,Efron2010}, i.e. the BF in our situation. To have a general BFDR control procedure, applicable to both the BF with and without informative prior, we propose to combine these two approaches to estimate $v_i$ in a two-step procedure. The performance of the two-step BFDR procedure in the context of genome-wide inference is assessed by simulation studies and compared to competing methods with frequentist FDR approach. A detailed description of this new BFDR approach and the simulation results are given in the Web Appendix I.}

\section{Simulation study}
\indent\indent  {\it Principle.} We simulated RVs using the $R$ package sim1000G \citep{Dimitromanolakis2019}, which has the advantage of mimicking MAF distributions and short and long-range linkage disequilibirum (LD) across RVs similar to the 1,000 genomes data. We simulated one single region with total number of RV 22, 45, 72 and 145. The MAF of the RVs was in the range $[10^{-6}, 0.01]$. The binary phenotype  (case-control status) was generated from a logistic regression where we assumed that 1/2, 1/3, 1/4 and 1/5 of RVs are causal and randomly selected. The effect size for each causal variant is inversely proportional to its MAF \citep{Wu2011} and is in the range  $[2.23,4.25]$. For each combination of gene size and number of causal variants, we assumed that $100\%$, $90\%$ and $75\%$ of the causal variants are deleterious (ie. increase disease risk) and that the rest are protective. We used sample sizes of 250, 500 and 1000, respectively, for each group of cases and controls, and generated 1000 replicates of simulated data for each scenario. 

{\it Simulations under the null hypothesis.} To check the asymptotic null distribution of the BF with informative prior, we compared the simulated BF distribution to a $\chi^2(3)$ distribution using QQ-plots (See Web Appendix J) under various scenarios. Under most scenarios, the BF using both beta and mixture priors fits well a $\chi^2(3)$ distribution under the null hypothesis, except that conservative results (i.e., empirical type I error lower than nominal type I error) are shown when sample size is equal to 500 or 1000 and number of sites of the gene is equal to 45. In fact, this conservative result is what we expect when using KS test p-values as informative prior. That is because when the number of valid p-values calculated within each gene used for KS test is less than or equal to 30, exact p-value is computed. Otherwise, asymptotic distributions are used and the resulting p-values are known to be conservative when sample size (the number of p-values within the gene for the KS test) is small \citep{Conover1972}. 

{\it Methods comparison.} We compared the statistical power of BF to three standard region-based statistical approaches, Burden test, SKAT and SKAT-O \citep{Lee2014}. The power is estimated using the proportion of replicates with $p$ value less than 0.05, assuming $\alpha=0.05$ for all methods.
The Burden test summarizes the genotype information of multiple RVs in the region into a single genetic score and test for association between this score and a phenotype. Burden tests assume all the variants within a region have a same direction of effect on the disease outcome and the test statistic can be written as the square of the weighted sum of single-variant score statistics $Q = \big(\sum_{j=1}^{n}w_j S_j\big)^2$, where $S_j$ is score statistic of marginal model for variant $j$ and $w_j$ is the weight. The sequence kernel association test (SKAT) is based on a variance component test within a random-effect model. SKAT can overcome the limitation of burden test, assuming the multiple variants in one region can have opposite directions of effect. The SKAT test statistic is a weighted sum of squares of single-variant score statistics $Q=\sum_{j=1}^{n}w_j^2S_j^2$. Since in real data the true genetic model is unknown, SKAT-O uses the data to adaptively weight the SKAT and Burden test statistics $Q= (1-\rho)Q_{skat}+\rho Q_{burden}$. The parameter $\rho$ in the SKAT-O statistic is estimated based on a minimum $p-$value criteria, calculated over a grid of $\rho$'s. 

{\it Type I error results.} The simulation results confirmed that the BF null distribution does not depend on the sample size and gene size. We also assessed the empirical type I error at nominal level $\alpha=0.05$ (Table \ref{tab:typeI.err} and Web Appendices K). The type I error rate is under control under most scenarios, we only noted a slight inflation of the type I error for the BF with informative prior, when sample sizes $\geq 1000$ and gene sizes $=145$. The slight inflation was also confirmed for extremely large genes (i.e. with 5000 sites) (Web Appendix L).\\
\begin{table}[htbp]
	\centering
	\caption{Type I error rate for the informative prior BF and other competing methods, at $\alpha=0.05$ level, for different gene sizes and sample sizes with 1000 replicates}
	\begin{tabular}{lrrrr}
		\hline
		& \multicolumn{1}{p{5em}}{sites=22} & \multicolumn{1}{p{5em}}{sites=45} & \multicolumn{1}{p{5em}}{sites=72} & \multicolumn{1}{p{5em}}{sites = 145} \\
		\hline
		\textbf{250 cases, 250 controls} &       &       &       &  \\
		BF (beta informative prior) & 5.85\% & 2.45\% & 3.51\% & 4.47\% \\
		BF (mixture informative prior) & 5.35\% & 2.45\% & 3.40\% & 5.43\% \\
		BF (beta non-informative prior) & 5.43\% & 4.04\% & 3.30\% & 5.85\% \\
		BF (mixture non-informative prior) & 4.18\% & 2.98\% & 4.89\% & 5.74\% \\
		SKAT  & 5.53\% & 3.09\% & 5.43\% & 3.09\% \\
		Burden & 5.21\% & 3.94\% & 4.26\% & 4.89\% \\
		SKAT-O & 6.60\% & 3.40\% & 5.11\% & 3.62\% \\
		\hline
		\textbf{500 cases, 500 controls} &       &       &       &  \\
		BF (beta informative prior) & 6.91\% & 4.04\% & 5.00\% & 6.49\% \\
		BF (mixture informative prior) & 6.06\% & 4.04\% & 5.96\% & 7.23\% \\
		BF (beta non-informative prior) & 4.89\% & 4.89\% & 4.04\% & 4.57\% \\
		BF (mixture non-informative prior) & 3.72\% & 4.15\% & 4.68\% & 5.85\% \\
		SKAT  & 7.55\% & 2.13\% & 5.43\% & 4.15\% \\
		Burden & 4.89\% & 5.21\% & 3.94\% & 3.94\% \\
		SKAT-O & 6.06\% & 3.94\% & 4.68\% & 3.40\% \\
		\hline
		\textbf{1000 cases, 1000 controls} &       &       &       &  \\
		BF (beta informative prior) & 5.53\% & 5.21\% & 5.00\% & 6.06\% \\
		BF (mixture informative prior) & 4.79\% & 4.79\% & 5.32\% & 7.23\% \\
		BF (beta non-informative prior) & 4.15\% & 3.19\% & 3.62\% & 4.57\% \\
		BF (mixture non-informative prior) & 2.66\% & 2.98\% & 3.51\% & 7.23\% \\
		SKAT  & 3.72\% & 2.87\% & 3.09\% & 5.32\% \\
		Burden & 3.83\% & 3.51\% & 3.62\% & 5.74\% \\
		SKAT-O & 4.04\% & 2.87\% & 2.98\% & 5.53\% \\
		\hline
	\end{tabular}
	\label{tab:typeI.err}
\end{table}

{\it Power results.} Simulation results under the alternative hypothesis are given in Table \ref{tab:power} (see also Web Appendix K). The BF with informative prior is always more powerful than using non-informative prior. The power of BF with beta prior is usually slightly higher than the BF with mixture prior. The BF with informative prior under beta or mixture prior outperforms all three competing methods under all scenarios for small sample size (N=500). For medium and large sample sizes (N=1000, N=2000), under most of situations, SKAT-O performed the best. However, BF with informative prior sometimes can beat SKAT-O, both of which have very close power when all causal variants have same direction of effect or when the proportion of opposite direction variants is low ($10\%$).
\begin{sidewaystable}
	\centering
	\caption{Statistical power of different gene-based tests for different gene sizes and sample sizes with 1000 replicates. The second row of the header indicates $\#$(sites) and proportion of causal RVs within the genes simulated under each scenario.  (Reject null hypothesis when $p<5\%$. The best method is highlighted in bold.)}
	\scalebox{0.8}{
		\begin{tabular}{lrrrrrrrrrrrr}
			\hline
			$\#$(sites), & \multicolumn{4}{c}{All causal RVs are deleterious} & \multicolumn{4}{c}{10\% of causal RVs are protective} & \multicolumn{4}{c}{25\% of causal RVs are protective} \\
			proportion of causal RVs	& \multicolumn{1}{p{3.5em}}{22, 1/2} & 
			\multicolumn{1}{p{3.5em}}{45, 1/3} & 
			\multicolumn{1}{p{3.5em}}{72, 1/4} & 
			\multicolumn{1}{p{3.5em}}{145, 1/5} & 
			\multicolumn{1}{p{3.5em}}{22, 1/2} & 
			\multicolumn{1}{p{3.5em}}{45, 1/3} & 
			\multicolumn{1}{p{3.5em}}{72, 1/4} & 
			\multicolumn{1}{p{3.5em}}{145, 1/5} & 
			\multicolumn{1}{p{3.5em}}{22, 1/2} & 
			\multicolumn{1}{p{3.5em}}{45, 1/3} & 
			\multicolumn{1}{p{3.5em}}{72, 1/4} & 
			\multicolumn{1}{p{3.5em}}{145, 1/5} \\
			\hline
			{\bf 250 cases, 250 controls} &       &       &       &       &       &       &       &       &       &       &       &  \\
			{\bf BF (beta informative prior) }& \textbf{57.77\%} & \textbf{62.23\%} & 56.81\% & 88.09\% & \textbf{45.43\%} & \textbf{55.21\%} & 48.94\% & \textbf{82.34\%} & \textbf{36.49\%} & \textbf{42.13\%} & 35.96\% & 66.38\% \\
			BF (mixture informative prior) & 55.20\% & {58.94\%} & \textbf{57.34\%} & \textbf{89.04\%} & 42.84\% & 54.26\% & \textbf{50.43\%} & 81.91\% & 33.97\% & 40.96\% & \textbf{36.28\%} & \textbf{67.98\%} \\
			BF (beta noninformative prior) & 55.11\% & 57.13\% & 48.19\% & 67.23\% & 41.17\% & 47.55\% & 38.30\% & 57.23\% & 27.77\% & 27.55\% & 23.51\% & 23.72\% \\
			BF (mixture noninformative prior) & 51.02\% & 52.87\% & 50.96\% & 68.83\% & 36.36\% & 44.04\% & 40.11\% & 57.77\% & 25.24\% & 25.64\% & 23.83\% & 26.70\% \\
			SKAT  & 34.15\% & 34.50\% & 32.70\% & 41.70\% & 31.28\% & 37.77\% & 30.74\% & 44.15\% & 29.79\% & 37.02\% & 26.28\% & 38.40\% \\
			Burden & 57.02\% & 57.60\% & 50.30\% & 70.00\% & 39.57\% & 48.19\% & 42.23\% & 57.77\% & 28.40\% & 27.55\% & 22.66\% & 26.38\% \\
			SKAT-O & 55.11\% & 55.50\% & 50.40\% & 70.40\% & 42.02\% & 51.06\% & 44.15\% & 62.23\% & 35.11\% & 38.40\% & 27.98\% & 40.21\% \\
			\hline
			{\bf 500 cases, 500 controls} &       &       &       &       &       &       &       &       &       &       &       &  \\
			BF (beta informative prior) & 83.09\% & 80.00\% & 82.34\% & \textbf{96.91\%} & 71.60\% & \textbf{76.28\%} & 74.04\% & \textbf{92.45\%} & 60.00\% & 52.87\% & 52.02\% & 73.83\% \\
			BF (mixture informative prior) & 80.49\% & 77.87\% & 80.21\% & 96.17\% & 67.41\% & 72.87\% & 72.77\% & 91.49\% & 56.98\% & 52.77\% & 52.34\% & 75.85\% \\
			BF (beta noninformative prior) & 80.53\% & 79.36\% & 81.91\% & 94.36\% & 64.36\% & 72.98\% & 67.13\% & 85.96\% & 47.66\% & 37.34\% & 36.28\% & 46.49\% \\
			BF (mixture noninformative prior) & 76.33\% & 77.55\% & 79.36\% & 94.57\% & 57.40\% & 69.36\% & 65.64\% & 86.81\% & 43.77\% & 38.30\% & 36.70\% & 51.70\% \\
			SKAT  & 66.70\% & 65.50\% & 66.40\% & 83.70\% & 64.57\% & 62.87\% & 68.09\% & 81.28\% & 59.89\% & 63.40\% & 60.53\% & \textbf{78.83\%} \\
			Burden & 81.91\% & 80.90\% & 81.50\% & 95.20\% & 63.94\% & 72.87\% & 69.26\% & 87.45\% & 47.45\% & 40.21\% & 36.81\% & 50.64\% \\
			{\bf SKAT-O}& \textbf{83.30\%} & \textbf{82.90\%} & \textbf{85.90\%} & 96.40\% & \textbf{73.40\%} & 76.17\% & \textbf{79.26\%} & 90.85\% & \textbf{63.83\%} & \textbf{64.04\%} & \textbf{60.74\%} & 78.40\% \\
			\hline
			{\bf 1000 cases, 1000 controls} &       &       &       &       &       &       &       &       &       &       &       &  \\
			BF (beta informative prior) & \textbf{98.40\%} & 97.87\% & 99.04\% & 99.89\% & 94.57\% & 95.32\% & 97.87\% & 99.36\% & 86.91\% & 87.23\% & 87.66\% & 92.02\% \\
			BF (mixture informative prior) & 98.19\% & 97.23\% & 98.83\% & 99.89\% & 92.98\% & 94.26\% & 96.38\% & 99.36\% & 85.32\% & 85.85\% & 87.66\% & 93.09\% \\
			BF (beta noninformative prior) & 96.49\% & 96.60\% & 97.23\% & 99.89\% & 87.77\% & 90.11\% & 93.62\% & 97.02\% & 69.04\% & 60.11\% & 65.00\% & 62.66\% \\
			BF (mixture noninformative prior) & 95.00\% & 94.79\% & 97.02\% & 99.15\% & 85.11\% & 87.55\% & 91.60\% & 97.23\% & 65.43\% & 59.15\% & 62.34\% & 64.36\% \\
			SKAT  & 93.09\% & 95.40\% & 97.80\% & 98.90\% & 90.32\% & 94.47\% & 95.96\% & 97.55\% & 89.89\% & \textbf{94.68\%} & \textbf{95.96\%} & \textbf{97.55\%} \\
			Burden & 97.23\% & 97.00\% & 98.30\% & 99.90\% & 87.98\% & 91.49\% & 93.40\% & 98.09\% & 69.04\% & 62.98\% & 64.26\% & 63.40\% \\
			{\bf SKAT-O} & 98.30\% & \textbf{98.50\%} & \textbf{99.40\%} & \textbf{100.00\%} & \textbf{95.53\%} & \textbf{97.45\%} & \textbf{98.51\%} & \textbf{99.68\%} & \textbf{91.28\%} & 94.04\% & 95.53\% & 97.13\% \\
			\hline
		\end{tabular}
	}
	\label{tab:power}
\end{sidewaystable}

{\it BFDR results.} We evaluated the BFDR control procedure in the context of genome-wide inference and compared it to three competing methods (SKAT, SKAT-O and the Burden test) where {Benjamini-Hochberg (BH) FDR procedure} \citep{Benjamini1995} was used for multiple test comparisons. The principle of these simulations are described in Web Appendix M.  The BFDR procedure controls well the FDR nominal level of 0.05 under all scenarios as the other methods do (Web Appendix Figure 20) except the burden test, which fails when the sample size is small (N1=N2=250) and both deleterious and protective variants are present in the gene (Web Appendix Figure 21). For small sample sizes, $N1=N2=250$, the BF with BFDR procedure was superior to all other methods based on true discovery rate (TDR) values. For intermediate sample sizes, $N1=N2=500$, the BF with BFDR procedure was superior to the Burden test and the performances were similar to SKAT and SKAT-O. For {large} sample sizes, $N1=N2=1000$, all methods achieved similar TDR levels.

\section{Application}	
\textit{Design. }
We apply the BF approach to a whole-exome sequencing (WES) study on lung cancer, which includes 262 patients affected by lung cancer and 261 healthy controls all recruited in Toronto. In the original variant call format (VCF) file of the study, there were totally 2,017,458 sites, including not only DNA polymorphism data but also rich annotations. We performed quality control procedures before conducting the BF analysis (See Web Appendix N).

\begin{figure}
	\centering
	\includegraphics[scale=0.7]{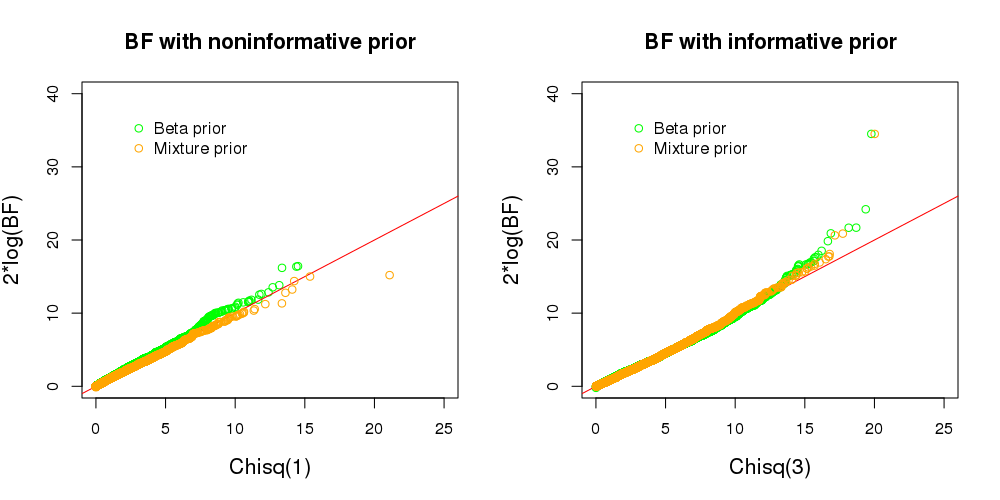}
	\caption{QQ plot of the BF test in the lung cancer WES study. The left panel shows non-informative prior results on 13,738 genes. The green circles represent the BF results with beta prior and the orange circles with the mixture prior. The right panel shows results on 11,721 genes (with at least 5 valid p-values within each gene for KS test). Similarly, green and orange circles correspond to the beta and mixture prior results, respectively. }
	\label{QQ}
\end{figure}

\textit{Results.}
We calculated the BF with non-informative prior assuming either the beta or mixture prior for 13,738 genes ($>=$20 sites) and the BF with informative prior for 11,721 genes, {which all contain at least 5 RVs for the KS test after excluding RVs that have total standardized counts less than 5 and $MAF<0.1\%$.} 
We compared the distribution of $2\times\log(BF)$ assuming non-informative and informative priors, which under the null are $\chi^2(1)$ and $\chi^2(3)$, respectively, using QQ plots (See Figure \ref{QQ}). Despite the increase in degrees of freedom, the distribution of BF with informative prior deviates more from the null distribution compared with the BF with non-informative prior and leads to increased gene discovery. 

For instance, for the top 20 genes identified by the BF assuming an informative beta prior, the $p$-value varies between $4.88 \times 10^{-10}$ and $8.32 \times 10^{-4}$ while it ranges from $5.07 \times 10^{-5}$ to $0.92$ assuming a BF with non-informative beta prior (Table \ref{Topgenes_betaprior}). The difference between these 2 BFs is explained by the contribution of the KS test, which is significant at the 5\% level for 16 of the 20 top genes. Among these 20 genes, 6 of them are also ranked in the top 20 according to the BF with non-informative beta prior (i.e. $REG4$, $TLR6$, $ERAP2$, $PLEKHG7$, $ANKRD44$, and $INTS7$). Among the top genes identified by the BF with an informative prior, the two highest BFs correspond to the genes $KCNIP4$ (p=$4.88 \times 10^{-10}$ and p=$1.55 \times 10^{-7}$) and $GPC5$  (p=$5.12 \times 10^{-10}$ and p=$1.55 \times 10^{-7}$) assuming either a beta prior (Table \ref{Topgenes_betaprior}) or a mixture prior (Web Appendix Table 4).  The KS test is very significant for these 2 genes (p=$9.54 \times 10^{-11}$ and p=$3.23 \times 10^{-8}$), which shows the interest of using informative prior in our BF framework. Besides, 11 out of the top 20 genes identified by the BF with informative beta prior overlap with those identified by the BF with mixture informative prior. Among the top 20 genes detected by the BF with informative beta prior, 12 and 2 overlap with the top 100 genes ranked according to the Burden and SKAT tests, respectively. Of note, the top 2 genes selected by the BF ranks very low with the Burden and SKAT tests, i.e. 10745 and 459 for  $KCNIP4$, respectively, and 8601 and 5955 for $GPC5$.  The difference between the informative BF and the Burden test might be explained by the contribution of the KS component to the BF. Finally, comparing the top gene list identified by different approaches, we did not notice any excess of large genes ($\ge$ 100 sites) with the BF with informative prior compared to the competing methods (Web Appendix Table 6). For SKAT results, we sought to better understand the difference with the BF below.

\textit{Difference between BF and SKAT.}
The top 20 genes identified by each competing method are given in Web Appendix Table 5. Four of the top genes identified by SKAT-O overlap with the BF with informative prior and most of the genes identified by the Burden test are also ranked among the top with the BF. In Web Appendix Figure 22, we have presented 2 genes $KCNIP4$ and $YAP1$ that show a large discrepancy in their rank according to the BF and SKAT. The top gene identified by BF, $KCNIP4$, exhibits a large difference in individual-RV associated $p$-values when compared the the null distribution, especially in the lower tail of the distribution (i.e. small $p$-values). This difference is likely to contribute to the significance of the KS test used as prior in the BF with informative prior. Such difference is not observed for the gene $YAP1$, the top fifth gene identified by SKAT. On the other hand, the distribution of RV counts between cases and control shows a systematic difference for the gene $YAP1$, with very few zeros, which is likely to contribute to the significance of the SKAT test. \\

\begin{sidewaystable}[ht]
	\centering
	\caption{Top 20 genes identified by the BF with beta informative prior }
	\label{Topgenes_betaprior}
	\begin{tabular}{rlrrrrrrr}
		\hline
		Chromosome & Genes & $\#$ of sites & KS test$^1$  & BF (info)$^2$ & SKAT$^3$ & Burden$^4$ & SKAT-O$^5$ & BF (noninfo) $^6$ \\ 
		& & in the gene & p-value & p-value &  rank & rank & rank & rank \\	\hline
		4 & KCNIP4 & 15904 & 9.54E-11 & 4.88E-10 & 459 & 10745 & 796 & 9305 \\ 
		13 & GPC5 & 17067 & 3.23E-08 & 1.55E-07 & 5955 & 8601 & 8813 & 12210 \\ 
		17 & RAB40B & 39 & 2.40E-05 & 2.27E-05 & 814 & 1351 & 1133 & 1286 \\ 
		16 & PAQR4 & 37 & 4.43E-04 & 7.58E-05 & 292 & 262 & 265 & 193 \\ 
		12 & PLEKHG7 & 51 & 1.97E-02 & 7.62E-05 & 233 & 10 & 2 & 4 \\ 
		1 & REG4 & 41 & 1.06E-01 & 1.10E-04 & 15 & 40 & 14 & 1 \\ 
		4 & TLR6 & 30 & 1.76E-01 & 1.83E-04 & 1607 & 5 & 10 & 2 \\ 
		3 & ERC2 & 78 & 6.60E-03 & 3.44E-04 & 852 & 82 & 144 & 68 \\ 
		10 & ZFYVE27 & 61 & 1.91E-02 & 4.46E-04 & 1807 & 90 & 224 & 34 \\ 
		8 & OTUD6B & 20 & 1.24E-02 & 5.24E-04 & 57 & 72 & 57 & 63 \\ 
		17 & ATP6V0A1 & 75 & 5.07E-03 & 5.25E-04 & 1094 & 71 & 150 & 125 \\ 
		2 & ANKRD44 & 110 & 8.96E-02 & 5.70E-04 & 8511 & 57 & 135 & 7 \\ 
		19 & ZNF556 & 42 & 2.58E-02 & 6.63E-04 & 853 & 38 & 68 & 41 \\ 
		10 & SLF2 & 56 & 2.58E-02 & 7.22E-04 & 778 & 47 & 78 & 45 \\ 
		1 & INTS7 & 66 & 4.89E-02 & 7.38E-04 & 795 & 34 & 54 & 19 \\ 
		16 & SDR42E1 & 26 & 4.62E-03 & 7.80E-04 & 311 & 190 & 253 & 216 \\ 
		16 & IL17C & 38 & 1.45E-03 & 8.16E-04 & 2923 & 811 & 1756 & 870 \\ 
		5 & ERAP2 & 79 & 7.85E-01 & 8.24E-04 & 3096 & 1 & 3 & 3 \\ 
		17 & TRIM47 & 33 & 3.23E-03 & 8.29E-04 & 1560 & 613 & 1151 & 336 \\ 
		17 & BRCA1 & 102 & 5.07E-03 & 8.32E-04 & 3305 & 563 & 1341 & 207 \\ 
		\hline
		\multicolumn{9}{l}{ {\small 1. KS p-value used as the p random variable in equation (15) of Web Appendix D;}}\\
		\multicolumn{9}{l}{ {\small 2. The p-value of BF with informative prior is calculated based on the null distribution of $2log(BF)$ of $\chi^2(3)$;}}\\
		\multicolumn{9}{l}{{\small 3. Ranking of genes according to SKAT (see Section 5);}}\\
		\multicolumn{9}{l}{ {\small 4. Ranking of genes according to the Burden test (see Section 5);}}\\
		\multicolumn{9}{l}{ {\small 5. Ranking of genes according to SKAT-O (see Section 5);}}\\	
		\multicolumn{9}{l}{ {\small 6. Ranking of genes according to BF with non-informative prior as shown in equation (15) in Web Appendix D.}}\\		
	\end{tabular}
\end{sidewaystable}

\textit{BFDR evaluation.}
{We also applied the BFDR control procedure using the upper bound estimates of $\pi_0$. Given $\gamma=0.999$, $q_\gamma^*=16.27$, with the number of genes (hypotheses) to test using beta and mixture informative prior BF are 11,721 and  9,368 respectively, the corresponding $\hat{\pi}_0$ using these two methods are $99.90\%$ and $99.97\%$. The BFDR level of gene \textit{KCNIP4} and \textit{GPC5} are $6.09\times 10^{-5}$ and $1.07\times 10^{-2}$ with beta prior and $7.16\times 10^{-4}$ and $8.84\times 10^{-2}$ with mixture prior.}

\section{Conclusion}
\label{sec:conc}
Our BF method with informative priors provides a sensitive approach to detect RV associations with complex diseases based on NGS technology. It has greater power than competing approaches such as SKAT-O and Burden Test in designs with moderate sample sizes (number of cases and number of controls each $<$ 500), which is an important result since many NGS applications face this situation including our real data application. The BF provided better power performances when genes had a higher proportion of RVs with the same direction of effects (ratio of protective RVs $\le 10\%$) in contrast to SKAT-O, which performed better when genes had a higher proportion of RVs with opposite direction of effects (ratio of protective RVs $\ge 25\%$). The problem of protective variants might be less critical in WES where variants are located within protein-coding genes and have predominantly deleterious effects on the disease of interest. 

\indent Another advantage of our BF approach is the incorporation of informative priors in the form of a prior probability of association given by a KS test. The BF can then be thought of as a {composite test} where the first component is a ratio of marginal likelihoods comparing the distribution of RV counts among cases vs. controls for a particular gene using the beta-binomial distribution {and the second (independent) component, the KS test, compares the $p$-value distribution across RVs in that gene to an empirical null distribution}. It is therefore sensitive to either overall RV counts difference between cases and controls or allelic distribution differences (or both). Combining different gene-based RV tests has also been found to be an efficient strategy to detect associations in the recent literature \citep{Lee2012, Porsch2018}. 

\indent An empirical Bayesian approach for genetic associations based on the beta-binomial distribution has also been proposed by \cite{McCallum2015}. In the framework, the hyperprior parameter of the beta distribution corresponds to the mean RV probability across individuals unlike our approach which uses the mean RV probability across a genomic region. A great advantage of our approach over this paper is that the asymptotic distribution of the BF under the null hypothesis of no association is provided and a Bayesian FDR procedure is also proposed for genome-wide inference. 


\indent Our real data application illustrates the interest of our approach. Using the BF with informative prior improves the overall gene discovery compared to the BF with non-informative prior. Several important genes have been identified by the BF approach including the 2 top genes $KCNIP4$ and $GPC5$ 
and 4 known cancer genes. Interestingly, the gene $KCNIP4$ has been previously associated with lung cancer in a large fine-mapping study using common genetic variants and a Bayesian prioritization approach \citep{Brenner2015}. This gene encodes a member of the family of voltage-gated potassium (Kv) channel-interacting proteins ($KCNIP$s), which are small calcium binding proteins. This gene is believed to play a role in regulation of Wnt/$\beta$-catenin signaling and target gene transcription pathway \citep{Pruunsild2005}. The gene $GPC5$, the glypican Proteoglycan 5, is a member of the glypican-related integral membrane proteoglycan family (GRIPS) proteins. It has recently been shown that $GPC5$ is a novel epigenetically silenced tumor suppressor, which inhibits tumor growth by suppressing Wnt/$\beta$-catenin signaling in lung adenocarcinoma \citep{Yuan2016}. 

\indent Finally, we only have considered relatively equal sample sizes in our simulations and application but our method is also applicable to unbalanced case-control designs. Our framework could also be extended to include covariates, either individual-level or variant-level covariates.  We have not  addressed specifically the problem of population structure in this paper, which arises when cases and controls are sampled at differential rates from genetically divergent populations \citep{Devlin1999}. In our real data analysis, we did not notice any inflation of the BF due to population stratification, probably because we used a relatively homogenous population of European descent. 	
\section*{Acknowledgements}

The authors also thank Drs Rayjean Hung and Geoffrey Liu for providing the Lung Cancer whole-exome sequencing data, Apostolos Dimitromanolakis for creating R packages sim1000G and rareBF, professors H\'el\`ene Massam and Michael Escobar for constructive discussions about the method development.
\vspace*{-8pt}

  \bibliographystyle{biom}  \bibliography{ms}

\end{document}


\maketitle

\section{Web Appendix A: Rare variant count distribution for the gene $CHEK2$}
\begin{figure}
\centering
\includegraphics[scale=0.7]{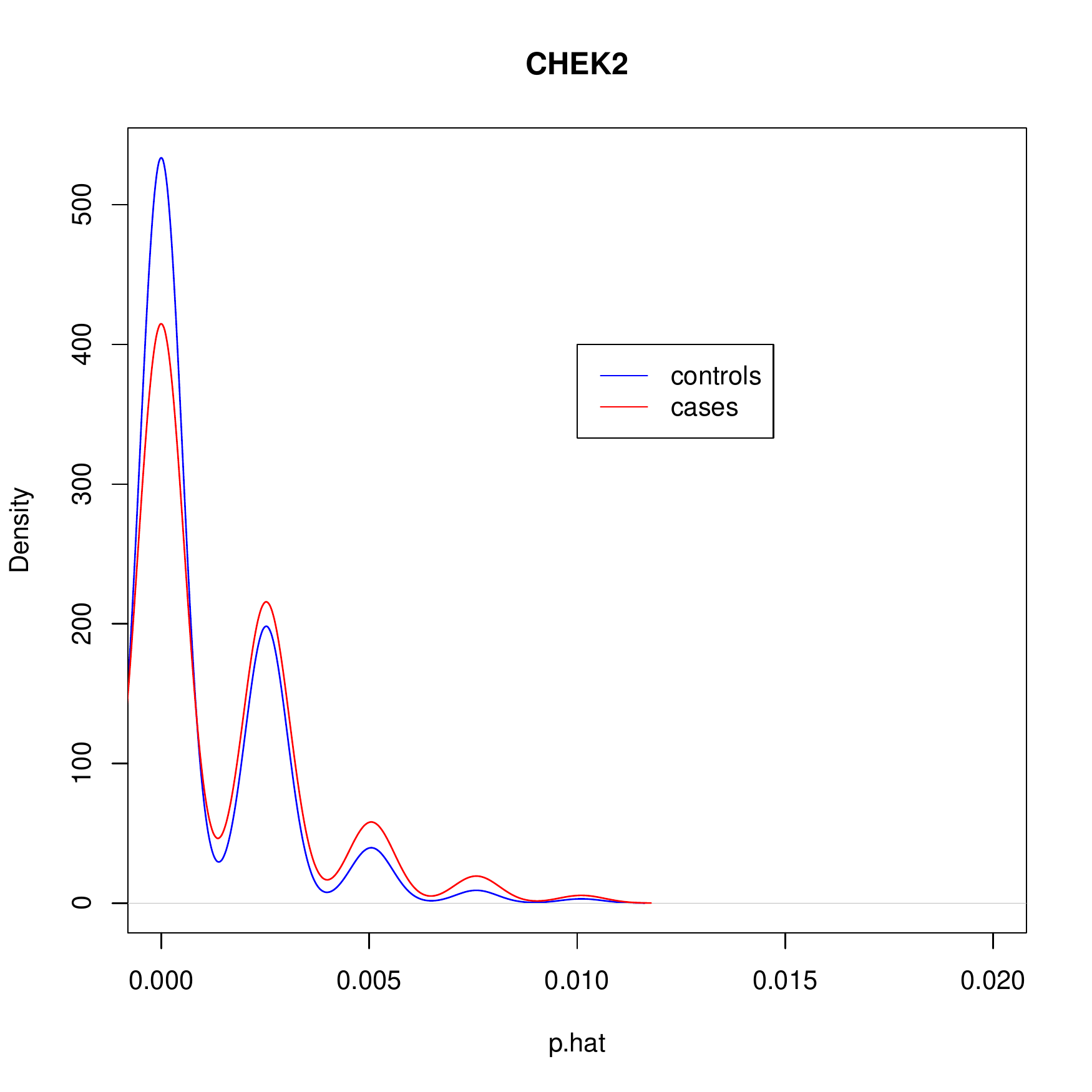}
\caption{\label{Fig_Check2} Distribution of RV proportion in cases and controls for the gene $CHEK2$. Proportion of RV $\hat{p}=\frac{x}{n}$ is calculated for each individual, where $x$ is the total RV count, $n$ is the number of sites in the region.}
\end{figure}

\section{Web Appendix B: Fit of the the beta-binomial distribution to the simulated and WES lung cancer data}
\begin{figure}[h!]
\centering	
\includegraphics[scale=0.8]{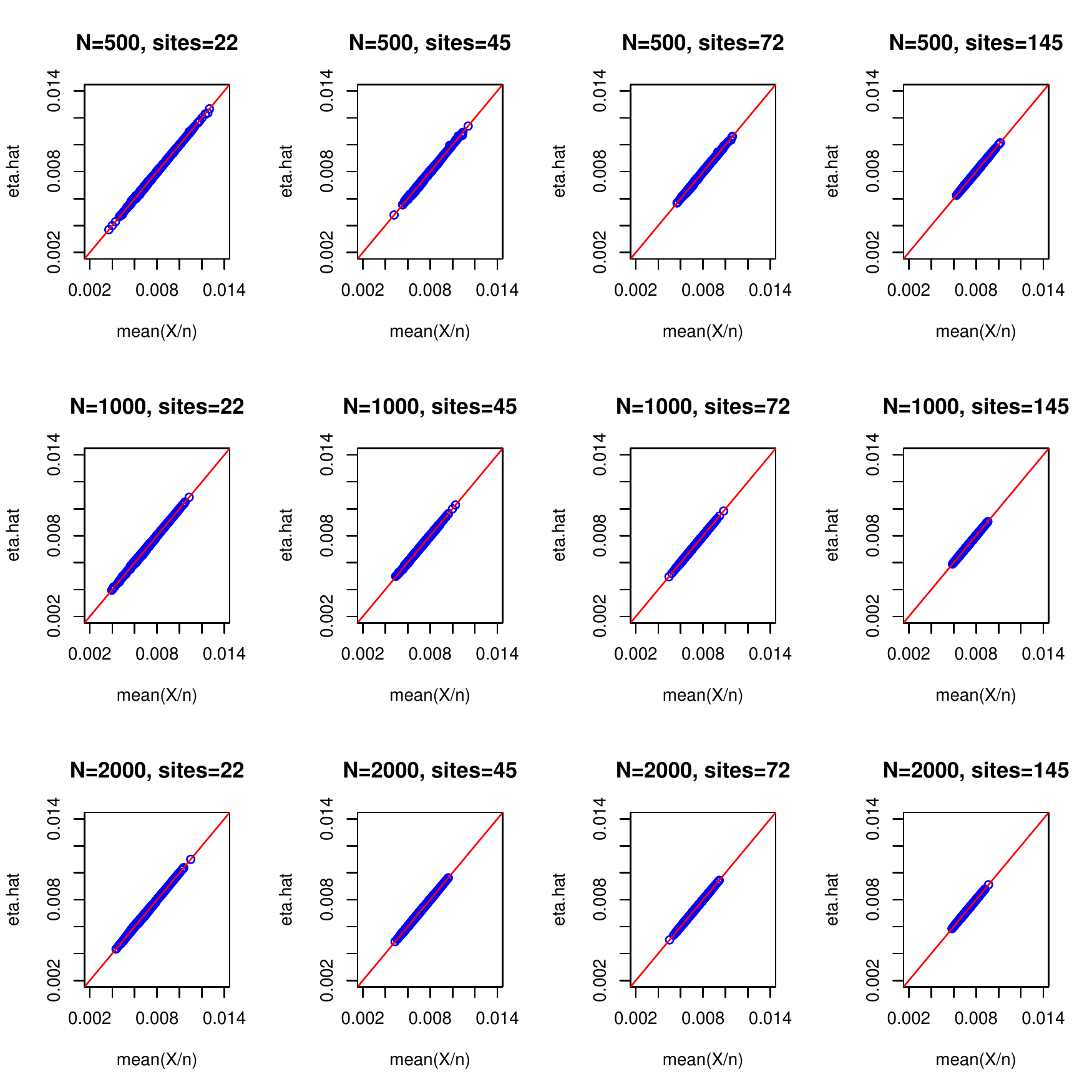}
\caption{Scatter plots comparing $\hat{\eta}$ estimated by the beta-binomial model used in the BF derivation (y-axis) and the mean of RV rates (x/n) across individuals (x-axis). Under each simulation scenario with different gene sizes and sample sizes, the plot is based on 1000 genes simulated under $H_0$ (See Section 5 in the main manuscript). The scatter points close to 45 degree line indicate that the model-based RV rate estimate and mean RV rate are consistent across genes. Besides, it shows that the mean RV rate has larger variation when the gene size is small, compared with scenario where gene size is large.}
\label{RV_rate_est_sim}
\end{figure}

\begin{figure}[h!]
\centering	
\includegraphics[scale=0.7]{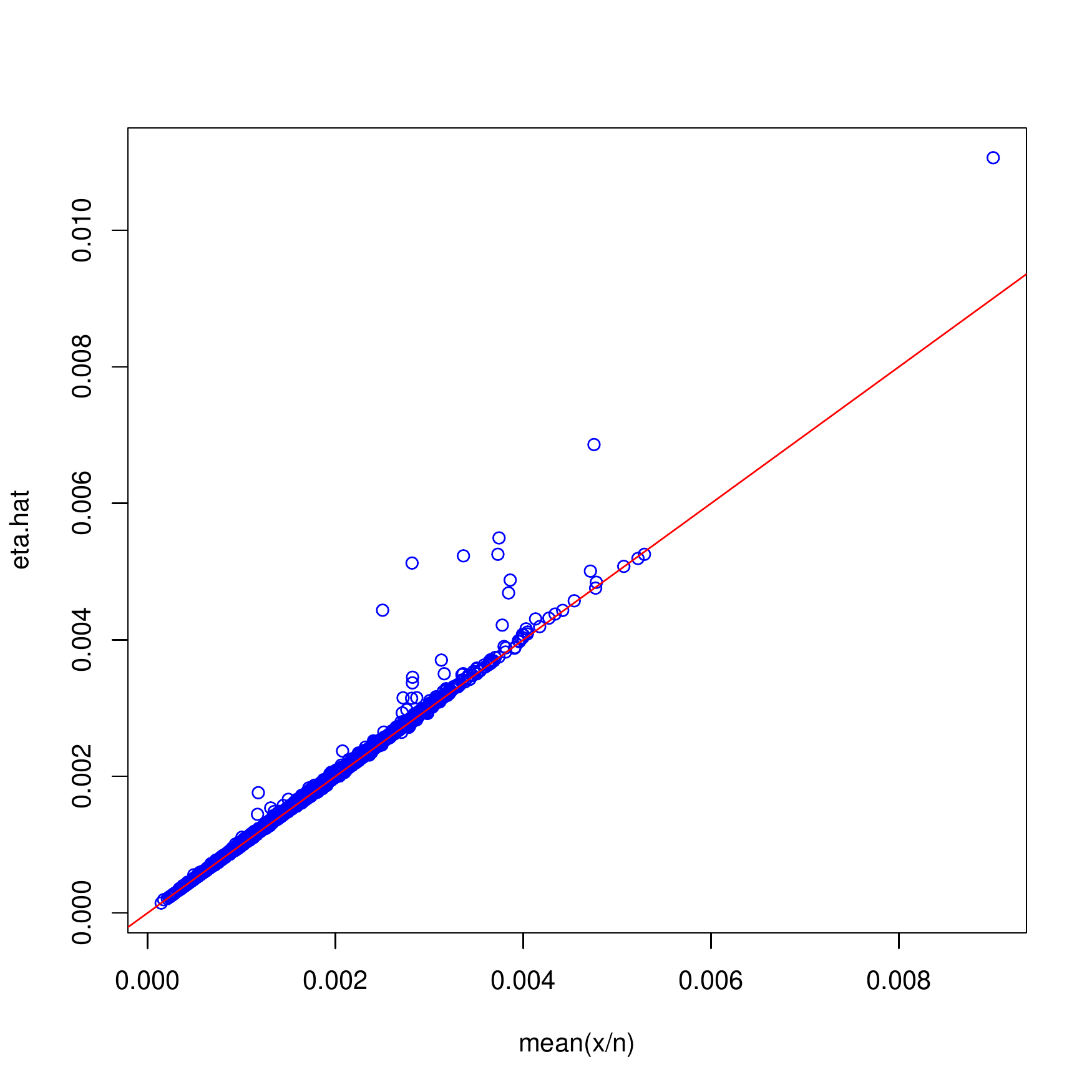}
\caption{Scatter plot comparing $\hat{\eta}$ estimated by the beta-binomial model used in the BF derivation (y-axis) and the mean of RV rates (x/n) across individuals (x-axis). This result is based on the Toronto lung cancer WES data. The scatter points close to 45 degree line indicate that the model-based RV rate estimate and mean RV rate are consistent across genes.}
\label{RV_rate_est}
\end{figure}

To check the fit of the beta-binomial distribution, we compare the model-based overall RV rate $\hat{\eta}$ to the mean RV rate (across all individuals) $\frac{1}{N}\sum_{j,k}\frac{x_{jk}}{n}$ using both the simulated data and Toronto WES lung cancer data (see Figures \ref{RV_rate_est_sim} and \ref{RV_rate_est}). Our simulation studies and real data analysis showed that the beta-binomial distribution provided a very good fit of the sequencing RV data.

\section{Web Appendix C: Estimation of correlation between RVs within the same gene/region.}

\begin{figure}[h!]
\centering	
\includegraphics[scale=0.7]{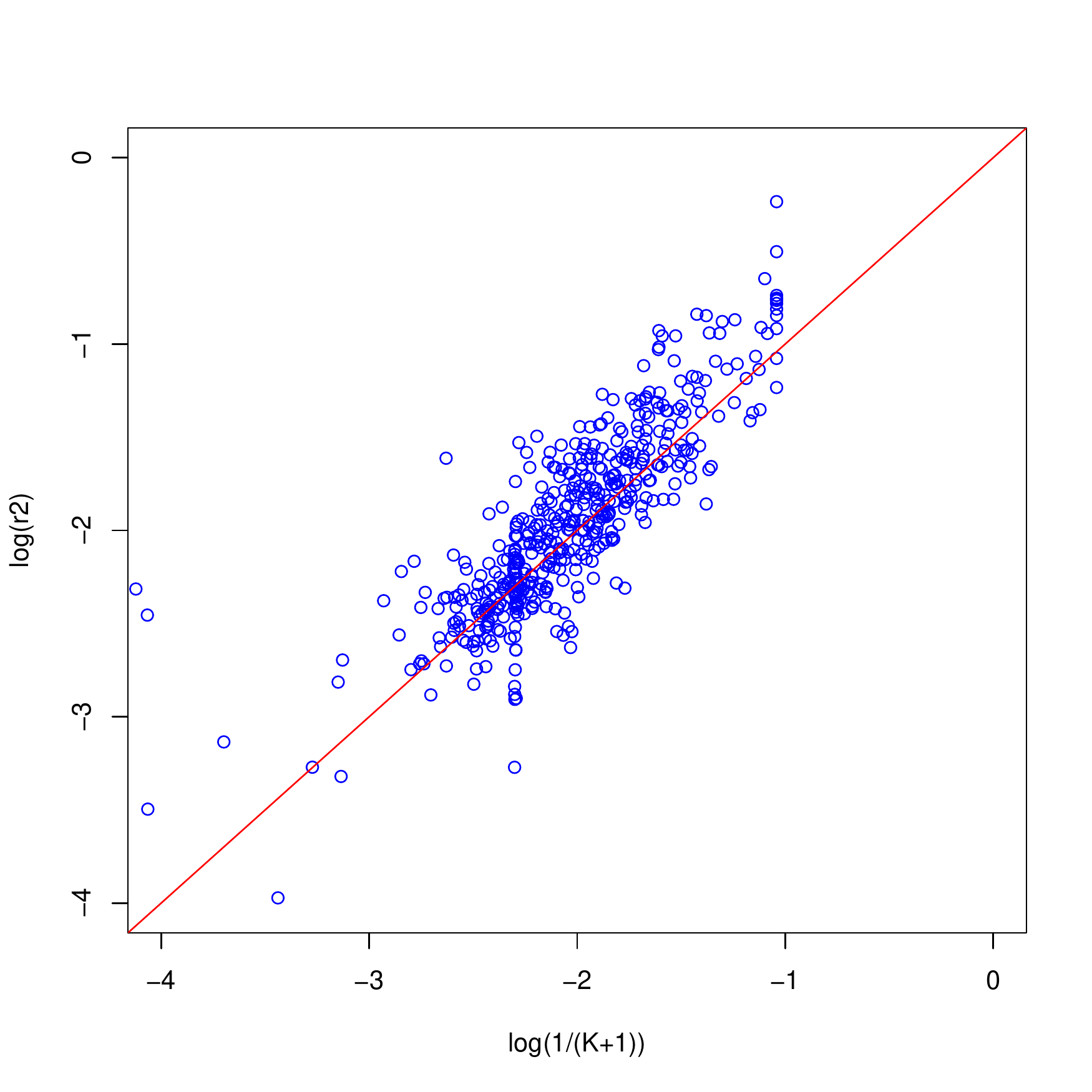}
\caption{Comparison between the intra-class correlation (ICC) estimated by the beta-binomial distribution used in the BF derivation (log scale on x-axis) and the mean Pearson correlation ($r^2$) between pairs of variants within the same gene (log scale on y-axis). This result is based on 500 simulated genes, with a number of sites varying from 20 to 974. The model-based ICC is estimated by $\frac{1}{\hat{K}+1}$, where $\hat{K}$ is the MLE of the precision parameter $K$ in the beta prior distribution based on the marginal likelihood (see section 3.3 in the main manuscript). The scatter points close to 45 degree line indicate that the model-based ICC and Pearson correlation coefficient are consistent across genes.}
\label{correlation_est}
\end{figure}

The results showed in Figure \ref{correlation_est} below demonstrate that the beta-binomial distribution used in the derivation of the BF accounts for correlation between RVs within the same gene/region and that this model-based estimate of the intra-class correlation is close to some average correlation estimate between pairs of variants obtained with a Pearson correlation coefficient.

\section{Web Appendix D: Proof of Theorem 1}
In equation (\ref{m0_beta}) of the main manuscript, we showed that assuming a beta prior, the marginal likelihood of the data under $H_0$ is proportional to the integral function
\begin{equation}\label{Laplace_beta}
\begin{split}
m_0({\bf X}| \tilde{K}, \eta^{*}, K^{*}) &\propto   \;\; I=\int_{0}^{1}\frac{\prod_{j=1}^{2}\prod_{k=1}^{N_j}B\{x_{jk}+\tilde{K}\eta,n_{jk}-x_{jk}+\tilde{K}(1-\eta)\}}{B\{\tilde{K}\eta,\tilde{K}(1-\eta)\}^N}\pi(\eta|\eta^*,K^*)d\eta.\\
\end{split}
\end{equation}

According to equation (5) in \cite{Kass1995}, this integral can be approximated using Laplace's method by
\begin{eqnarray}\label{Laplace_app}
\begin{split}
\hat{I}_{MLE} &=& (2\pi)^{1/2}\hat{\Sigma}^{1/2}Pr({\bf X}|\hat{\eta})\pi(\hat{\eta}|\eta^*,K^*),\\
\end{split}
\end{eqnarray}

where  $Pr({\bf X}|\eta)=\frac{\prod_{j=1}^{2}\prod_{k=1}^{N_j}B\{x_{jk}+\tilde{K}\eta,n_{jk}-x_{jk}+\tilde{K}(1-\eta)\}}{B\{\tilde{K}\eta,\tilde{K}(1-\eta)\}^N}$ is the likelihood function, $\hat{\eta}$ is the MLE of $\eta$ based on this likelihood and $\hat{\Sigma}$ is the estimated variance of $\hat{\eta}$ with $\hat{\Sigma}=\{-\frac{d^2\log Pr({\bf X}|\eta)}{d \eta^2}\}^{-1}|_{\eta=\hat{\eta}}$, where $\hat{\eta} \sim N(\eta_0,\hat{\Sigma}).$ As $N \rightarrow \infty$,
$$ I= \hat{I}_{MLE}\{1+\mathcal{O}(N^{-1})\}.$$

When $K^*\eta^* \rightarrow \infty$ and $\eta^* \rightarrow 0$, 
we have $K^*(1-\eta^*) \rightarrow \infty$, $K^* \rightarrow \infty$ and $\log(1-{\eta^*}) \rightarrow 0$, therefore, by using stirling's series of $\log\Gamma(K^*\eta^*)$, $\log\Gamma\{K^*(1-\eta^*)\}$ and $\log\Gamma(K^*)$ \citep{Impens2003}, $\log \pi({\eta}|\eta^*,K^*)\Big|_{\eta = \hat{\eta},\eta^*=\hat{\eta}}$ can be written as 
\begin{eqnarray*}
\log \pi({\eta}|\eta^*,K^*)\Big|_{\eta = \hat{\eta},\eta^*=\hat{\eta}} &=& 
\log\frac{\eta^{K^*\eta^*-1}(1-\eta)^{K^*(1-\eta^*)-1}}{B\{K^*\eta^*,K^*(1-\eta^*)\}}\Big|_{\eta = \hat{\eta},\eta^*=\hat{\eta}}\\
& = & (K^*\eta^*-1)\log\eta + \{K^*(1-\eta^*)-1\}\log(1-\eta) - \log\sqrt{2\pi} \\
&& - (K^*\eta^*-\frac{1}{2})\log(K^*\eta^*) - (K^*-K^*\eta^*-\frac{1}{2})\log(K^*-K^*\eta^*) \\
&& + (K^*-\frac{1}{2})\log K^* - \mathcal{O}(\frac{1}{K^*\eta^*}) - \mathcal{O}\{\frac{1}{K^*(1-\eta^*)}\} + \mathcal{O}(\frac{1}{K^*})\Big|_{\eta = \hat{\eta},\eta^*=\hat{\eta}}\\
&=& -\frac{1}{2}\log\hat{\eta} + \frac{1}{2}\log K^* - \log\sqrt{2\pi} - \frac{1}{2}\log(1-\hat{\eta}),\\
&=& -\frac{1}{2}\log\hat{\eta} + \frac{1}{2}\log K^* - \log\sqrt{2\pi}.
\end{eqnarray*}
Thus, by taking $\eta^* = \hat{\eta}$,
\begin{eqnarray}\label{Int}\nonumber
\log \hat{I}_{MLE} &\approx& \frac{1}{2}\log(2\pi) + \frac{1}{2}\log\hat{\Sigma}+\log Pr({\bf X}|\hat{\eta}) -\frac{1}{2}\log\hat{\eta} + \frac{1}{2}\log K^* - \log\sqrt{2\pi},\\ 
&\approx& \frac{1}{2}\log\hat{\Sigma}+\log Pr({\bf X}|\hat{\eta}) -\frac{1}{2}\log\hat{\eta} + \frac{1}{2}\log K^*.\\ \nonumber
\end{eqnarray}

Besides, in equation (\ref{m1_beta}) of the main manuscript, it is shown that under a beta prior, the marginal likelihood function under $H_1$ is proportional to a product of two integral functions:

\begin{equation}
\begin{split}
m_1(X|\tilde{K},\eta_1^*,K_1^*,\eta_2^*,K_2^*) \propto &  \prod_{j=1}^{2}\displaystyle\int_{0}^{1}\frac{\prod_{k=1}^{N_j}B\{x_{jk}+\tilde{K}\eta,n_{jk}-x_{jk}+\tilde{K}(1-\eta)\}}{B\{\tilde{K}\eta,\tilde{K}(1-\eta)\}^{N_j}}\\ &\frac{\eta^{K_j^*\eta_j^*-1}(1-\eta)^{K_j^*(1-\eta_j^*)-1}}{B\{K_j^*\eta_j^*,K_j^*(1-\eta_j^*)\}} d\eta,\\
= & I_1 \times I_2. \\
\end{split}
\end{equation}		

%
%

Similar to $\hat{I}_{MLE}$, by assuming $\eta_1^* = \hat{\eta}_1$, $\eta_2^* = \hat{\eta}_2$ and using Laplace's approximation, we have
\begin{eqnarray*}
\hat{I}_{1_{MLE}} &=& (2\pi)^{1/2}\hat{\Sigma_1}^{1/2}Pr({\bf X_1}|\hat{\eta_1})\pi(\hat{\eta_1}|\hat{\eta}_1,K_1^*),
\end{eqnarray*}
and
\begin{eqnarray*}
\hat{I}_{2_{MLE}} &=& (2\pi)^{1/2}\hat{\Sigma_2}^{1/2}Pr({\bf X_2}|\hat{\eta_2})\pi(\hat{\eta_2}|\hat{\eta}_2,K_2^*),
\end{eqnarray*}

with $\hat\Sigma_1=\{-\frac{d^2\log Pr({\bf X_1}|\eta_1)}{d \eta_1^2}\}^{-1}|_{\eta_1=\hat\eta_1}$ and $\hat\Sigma_2=\{-\frac{d^2\log Pr({\bf X_2}|\eta_2)}{d \eta_2^2}\}^{-1}|_{\eta_2=\hat\eta_2}$.\\

As $N_1 \rightarrow \infty,  I_1= \hat{I}_{1_{MLE}}\{1+\mathcal{O}(N_1^{-1})\}$ and as $N_2 \rightarrow \infty,  I_2= \hat{I}_{2_{MLE}}\{1+\mathcal{O}(N_2^{-1})\}.$ \\

Therefore, similar to equation (\ref{Int}), we have
\begin{eqnarray*}
\log \hat{I}_{1_{MLE}} \approx \frac{1}{2}\log\hat{\Sigma}_1+\log Pr({\bf X_1}|\hat{\eta}_1) -\frac{1}{2}\log\hat{\eta}_1 + \frac{1}{2}\log K_1^*,
\end{eqnarray*}
and
\begin{eqnarray*}
\log \hat{I}_{2_{MLE}} \approx \frac{1}{2}\log\hat{\Sigma}_2+\log Pr({\bf X_2}|\hat{\eta}_2) -\frac{1}{2}\log\hat{\eta}_2 + \frac{1}{2}\log K_2^*.
\end{eqnarray*}

The BF is then given by
\begin{eqnarray}
BF = \frac{m_1({\bf X}| \tilde{K}, \eta_1^{*}, K_1^{*}, \eta_2^{*}, K_2^{*})}{ m_0({\bf X}| \tilde{K}, \eta^{*}, K^{*})} \approx \frac{\hat{I}_{1_{MLE}}  \times \hat{I}_{2_{MLE}} }{\hat{I}_{MLE} }.
\end{eqnarray}

The last equality comes from the fact that the term before the integral in equation (\ref{m0_beta}) and (\ref{m1_beta}) of the main manuscript cancels out.\\

Let the parameters 
$K^* = p^2\hat\eta  \hat\Sigma^{-1}$, $K_1^* =   \hat\eta_1 \hat\Sigma_1^{-1}$, $K_2^*=\hat\eta_2 \hat\Sigma_2^{-1}$, then we have
\begin{equation}\label{log-BF}
\begin{split}
\log BF \approx & \log \hat{I}_{1_{MLE}}  + \log \hat{I}_{2_{MLE}}  -\log \hat{I}_{MLE}, \\	
\approx & \frac{1}{2}\Big(-\log\frac{\hat{\eta}_1}{\hat{\Sigma}_1}-\log\frac{\hat{\eta}_2}{\hat{\Sigma}_2} +\log\frac{\hat{\eta}}{\hat{\Sigma}}\Big)+\frac{1}{2}\log\frac{K_1^*K_2^*}{K^*} - \log Pr({\bf X_1}|\hat{\eta}) -\log Pr({\bf X_2}|\hat{\eta}) \\
& + \log Pr({\bf X_1}|\hat{\eta}_1) + \log Pr({\bf X_2}|\hat{\eta}_2)-log(p),\\
\approx & \log Pr({\bf X_1}|\hat{\eta}_1) + \log Pr({\bf X_2}|\hat{\eta}_2) - \log Pr({\bf X_1}|\hat{\eta}) -\log Pr({\bf X_2}|\hat{\eta})-log(p).
\end{split}
\end{equation}

Expanding the log-likelihood function around $\eta_0$, the true value of $\eta$, according to Taylor's expansion, we have	
\begin{equation}\label{log-lik}
\begin{split}
\log Pr({\bf X_1}|\hat{\eta}_1) = \log Pr({\bf X_1}|\eta_0) + ({\hat\eta_1}-\eta_0) \ell_1'({\eta_0})+\frac{1}{2} ({\hat\eta_1}-\eta_0)^2\ell_1''({\eta_0})+o({\hat\eta_1}-\eta_0)^2,\\
\log Pr({\bf X_2}|\hat{\eta}_2) = \log Pr({\bf X_2}|\eta_0) + ({\hat\eta_2}-\eta_0) \ell_2'({\eta_0})+\frac{1}{2} ({\hat\eta_2}-\eta_0)^2\ell_2''({\eta_0})+o({\hat\eta_2}-\eta_0)^2,\\
\log Pr({\bf X_1}|\hat{\eta}) = \log Pr({\bf X_1}|\eta_0) + ({\hat\eta}-\eta_0) \ell_1'({\eta_0})+\frac{1}{2} ({\hat\eta}-\eta_0)^2\ell_1''({\eta_0})+o({\hat\eta}-\eta_0)^2,\\
\log Pr({\bf X_2}|\hat{\eta}) = \log Pr({\bf X_2}|\eta_0) + ({\hat\eta}-\eta_0) \ell_2'({\eta_0})+\frac{1}{2} ({\hat\eta}-\eta_0)^2\ell_2''({\eta_0})+o({\hat\eta}-\eta_0)^2,
\end{split}	
\end{equation}
where $\ell'({\eta_0}) \equiv \frac{d\log Pr({\bf X}|{\eta})}{d \eta}|_{\eta=\eta_0}$, $\ell_1'({\eta_0}) \equiv \frac{d\log Pr({\bf X_1}|{\eta})}{d \eta}|_{\eta=\eta_0}$, $\ell_2'({\eta_0}) \equiv \frac{d\log Pr({\bf X_2}|{\eta})}{d \eta}|_{\eta=\eta_0}$, $\ell''({{\eta_0}}) \equiv \frac{d^2\log Pr({\bf X}|{\eta})}{d \eta^2}|_{\eta=\eta_0}$,
$\ell_1''({{\eta_0}}) \equiv \frac{d^2\log Pr({\bf X_1}|{\eta})}{d \eta^2}|_{\eta=\eta_0}$ and  $\ell_2''({{\eta_0}}) \equiv \frac{d^2\log Pr({\bf X_2}|{\eta})}{d \eta^2}|_{\eta=\eta_0}$
and the approximation error term $o()$ can be ignored.

From equation (\ref{log-like-der}) in Web Appendix H, we have the following approximation   \\
$(\hat{\eta}_1-\eta_0)\ell^{''}_1(\eta_0) \approx -(\hat{\eta}_1-\eta_0)\Sigma^{-1}_1 \approx  -\ell'_1(\eta_0)$ and  $(\hat{\eta}_2-\eta_0)\ell^{''}_2(\eta_0) \approx -(\hat{\eta}_2-\eta_0)\Sigma^{-1}_2 \approx  -\ell'_2(\eta_0)$. In equation (\ref{log-lik}) we then have
\begin{equation*}
\begin{split}
& ({\hat\eta_1}-\eta_0) \ell_1'({\eta_0}) + ({\hat\eta_2}-\eta_0) \ell_2'({\eta_0}) -  ({\hat\eta}-\eta_0) \ell_1'({\eta_0}) - ({\hat\eta}-\eta_0) \ell_2'({\eta_0}) \\
=& ({\hat\eta_1}-\hat\eta) \ell_1'({\eta_0}) + ({\hat\eta_2}-\hat\eta) \ell_2'({\eta_0}), \\
\approx & ({\hat\eta_1}-\hat\eta)({\hat\eta_1}-\eta_0) \Sigma_1^{-1} +  ({\hat\eta_2}-\hat\eta)({\hat\eta_2}-\eta_0)  \Sigma_2^{-1},\\
\end{split}
\end{equation*}
using $ \hat\eta \approx  \frac{\hat\eta_1\Sigma_2+\hat\eta_2\Sigma_1}{\Sigma_1+\Sigma_2}$ from Proposition 1, this yields
\begin{eqnarray}\label{log-BF-1}
\begin{split}
& \approx \frac{({\hat\eta_1}-\hat\eta_2)}{({\Sigma}_1+{\Sigma}_2)}({\hat\eta_1}-\eta_0) -  \frac{({\hat\eta_1}-\hat\eta_2)}{({\Sigma}_1+{\Sigma}_2)}({\hat\eta_2}-\eta_0),  \\
& \approx \frac{(\hat{\eta}_1-\hat{\eta}_2)^2}{({\Sigma}_1+{\Sigma}_2)}.
\end{split}
\end{eqnarray}

Based on a similar derivation using proposition 1, we have
\begin{eqnarray}\label{log-BF-2}
\begin{split}
& \frac{1}{2} ({\hat\eta_1}-\eta_0)^2\ell_1''({\eta_0}) + \frac{1}{2} ({\hat\eta_2}-\eta_0)^2\ell_2''({\eta_0}) - \frac{1}{2} ({\hat\eta}-\eta_0)^2\ell_1''({\eta_0}) - \frac{1}{2} ({\hat\eta}-\eta_0)^2\ell_2''({\eta_0}) \\
\approx & - \frac{(\hat{\eta}_1-\hat{\eta}_2)^2}{2({\Sigma}_1+{\Sigma}_2).}
\end{split}
\end{eqnarray}

From equations (\ref{log-BF}) and (\ref{log-lik}), we know that $2\log BF$ can be written as
\begin{eqnarray}\label{final_form}
2\log BF &\approx&  \frac{(\hat{\eta}_1-\hat{\eta}_2)^2}{{\Sigma}_1+{\Sigma}_2} - 2log(p),
\end{eqnarray}
where $\frac{(\hat{\eta}_1-\hat{\eta}_2)^2}{{\Sigma}_1+{\Sigma}_2} \xrightarrow{d} \chi^2(1)$.\\
Since under $H_0$, $p \sim Unif(0,1)$, thus $-2log(p) \sim \chi^2(2)$ , {which we assumed independent of $\frac{(\hat{\eta}_1-\hat{\eta}_2)^2}{{\Sigma}_1+{\Sigma}_2}$. Indeed, the correlation coefficient between these two components is empirically evaluated based on the lung cancer WES data (see Figures \ref{correlation} and  \ref{corr_real}) and the simulated data (see Figure \ref{corr_sim}), and was found very small}. Therefore, 
\begin{eqnarray}
2\log BF \xrightarrow{d} \chi^2(3).
\end{eqnarray}

When the number of $p$-values from the KS test was less than 5, we assumed a non-informative prior probability $p=1$ for the BF.\\

{The derivation of the asymptotic distribution of the BF above requires the assumptions that $K^{*}$ and $K^{*}\eta^{*}$ be large. In Web Appendix O, we checked that these assumptions are realistic based on our simulation results.}

\section{Web Appendix E: Assessment of the Laplace approximation}
In this section, the performance of the Laplace's approximation is assessed from the simulated data. First, we compared the integrand function to the approximated normal density function; Second, we compared the Laplace's estimate to a Monte Carlo estimate.

\subsection{Comparison of the integrand function}

According to equation (\ref{Laplace_beta}) of the Web Appendix, we have

\begin{equation}
\begin{split} I=&\int_{0}^{1}\frac{\prod_{j=1}^{2}\prod_{k=1}^{N_j}B\{x_{jk}+\tilde{K}\eta,n_{jk}-x_{jk}+\tilde{K}(1-\eta)\}}{B\{\tilde{K}\eta,\tilde{K}(1-\eta)\}^N}\pi(\eta|\eta^*,K^*)d\eta.\\
= &  \displaystyle\int_{-\infty}^{+\infty}\frac{\prod_{j=1}^{2}\prod_{k=1}^{N_j}B(x_{jk}+\tilde{K}\eta,n_{jk}-x_{jk}+\tilde{K}(1-\eta))}{B(\tilde{K}\eta,\tilde{K}(1-\eta))^N}\frac{\eta^{K^*\eta^*-1}(1-\eta)^{K^*(1-\eta^*)-1}}{B(K^*\eta^*,K^*(1-\eta^*))}\\ &\eta(1-\eta)\bigg|_{\eta = \frac{e^\theta}{1+e^\theta}}  d\theta, \\
= & \displaystyle\int_{-\infty}^{+\infty} f(\theta)  d\theta.
\end{split}
\end{equation}

\begin{figure}[ht!]
\centering
\includegraphics[scale=0.5]{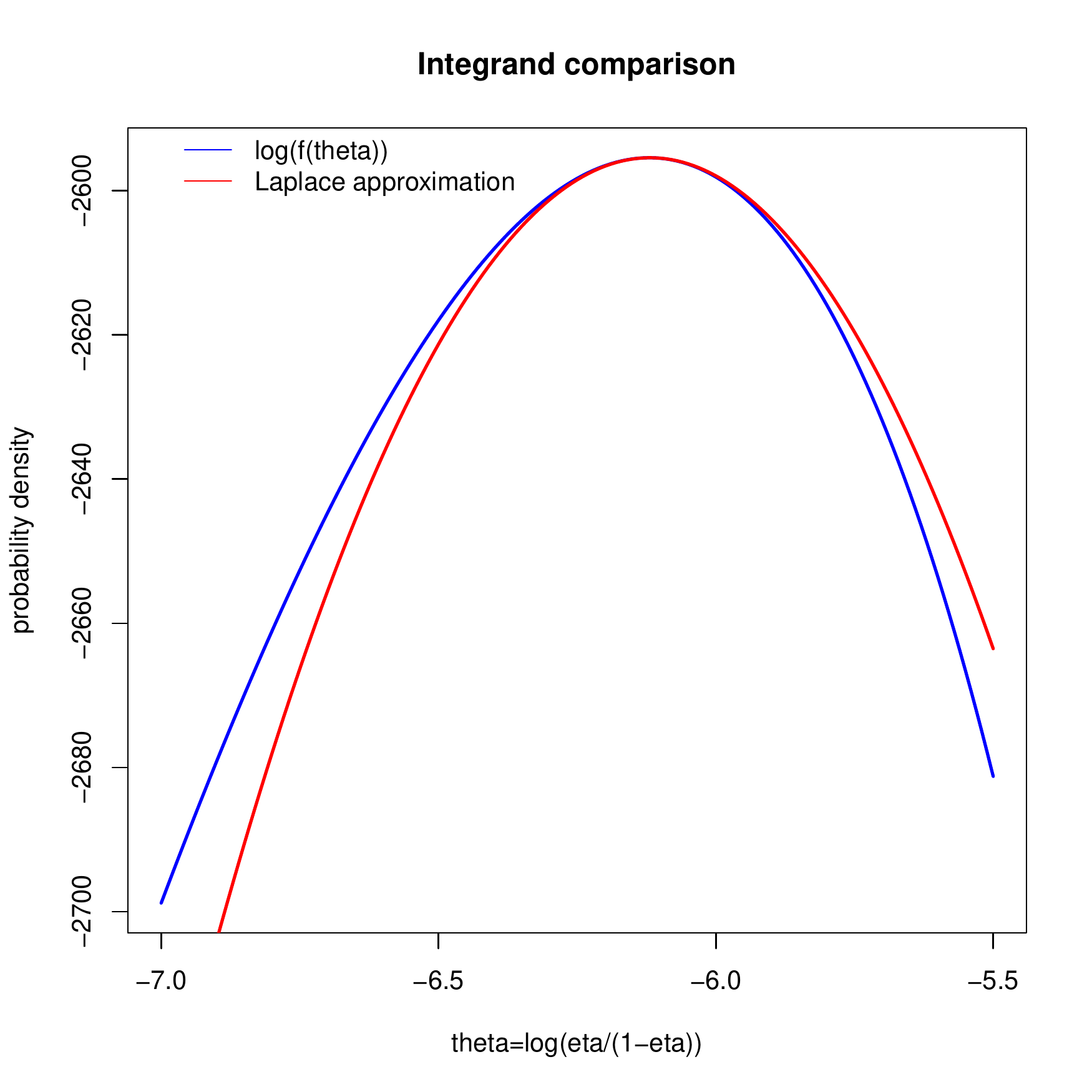}
\caption{Integrand function approximation. Red curve represents $\log g(\theta)$, and blue curve represents $\log f(\theta)$.}
\label{log_laplace_compare}
\end{figure}

Applying Laplace approximation, we first find $\hat{\theta}, s.t. f(\hat{\theta})=\max f(\theta)$,
\begin{equation*}
\begin{split}
f(\theta) \approx g(\theta) = f(\hat{\theta})\sigma \sqrt{2\pi}\phi (\theta;\mu,\sigma^2)
\end{split}
\end{equation*}

where $\phi(.)$ is Normal density function with parameter $\mu = \hat{\theta}, \sigma^2 = -\frac{1}{\ell''(\hat{\theta})}$. We then draw two curves: $\log f(\theta)$ and $\log g(\theta)$ {in Figure \ref{log_laplace_compare}} to examine the similarity of these two functions. 

Figure \ref{log_laplace_compare} shows that we have a very good approximation of the likelihood function for values of $\theta=\log(\eta / (1- \eta))$ in the range of -6.9 to -5.5, which corresponds to $\eta$ values in the range of 0.001 to 0.004. Most of the estimated $\eta$ values in our real data application lie in this interval (see Figure \ref{RV_rate_est} of the Web Appendix). 

\subsection{Monte Carlo Estimate of the integral function}
According to equation (\ref{Laplace_beta}), we have

\begin{equation}
\begin{split} I=&\int_{0}^{1}\frac{\prod_{j=1}^{2}\prod_{k=1}^{N_j}B\{x_{jk}+\tilde{K}\eta,n_{jk}-x_{jk}+\tilde{K}(1-\eta)\}}{B\{\tilde{K}\eta,\tilde{K}(1-\eta)\}^N}\pi(\eta|\eta^*,K^*)d\eta.\\
\end{split}
\end{equation}
where $\pi(\eta;k^*,\eta^*)$ is the density function of Beta distribution.

The simplest Monte Carlo integration estimate of the above formula is
\begin{equation*}
\begin{split}
&\int_{0}^{1}\frac{\prod_{j=1}^{2}\prod_{k=1}^{N_j}B\{x_{jk}+\tilde{K}\eta,n_{jk}-x_{jk}+\tilde{K}(1-\eta)\}}{B\{\tilde{K}\eta,\tilde{K}(1-\eta)\}^N}\pi(\eta|\eta^*,K^*)d\eta \\
\approx& \frac{1}{L}\sum_{l=1}^{L}\frac{\prod_{j=1}^{2}\prod_{k=1}^{N_j}B(x_{jk}+\tilde{K}\eta_l,n-x_{jk}+\tilde{K}(1-\eta_l))}{B(\tilde{K}\eta_l,\tilde{K}(1-\eta_l))^N},\\
\end{split}
\end{equation*}\
where $\{\eta_l,l=1,...,L\}$ are Monte Carlo samples generated from the distribution of $\eta$, $Beta(\eta^*,K^*)$.

\begin{figure}
\centering
\includegraphics[scale=0.45]{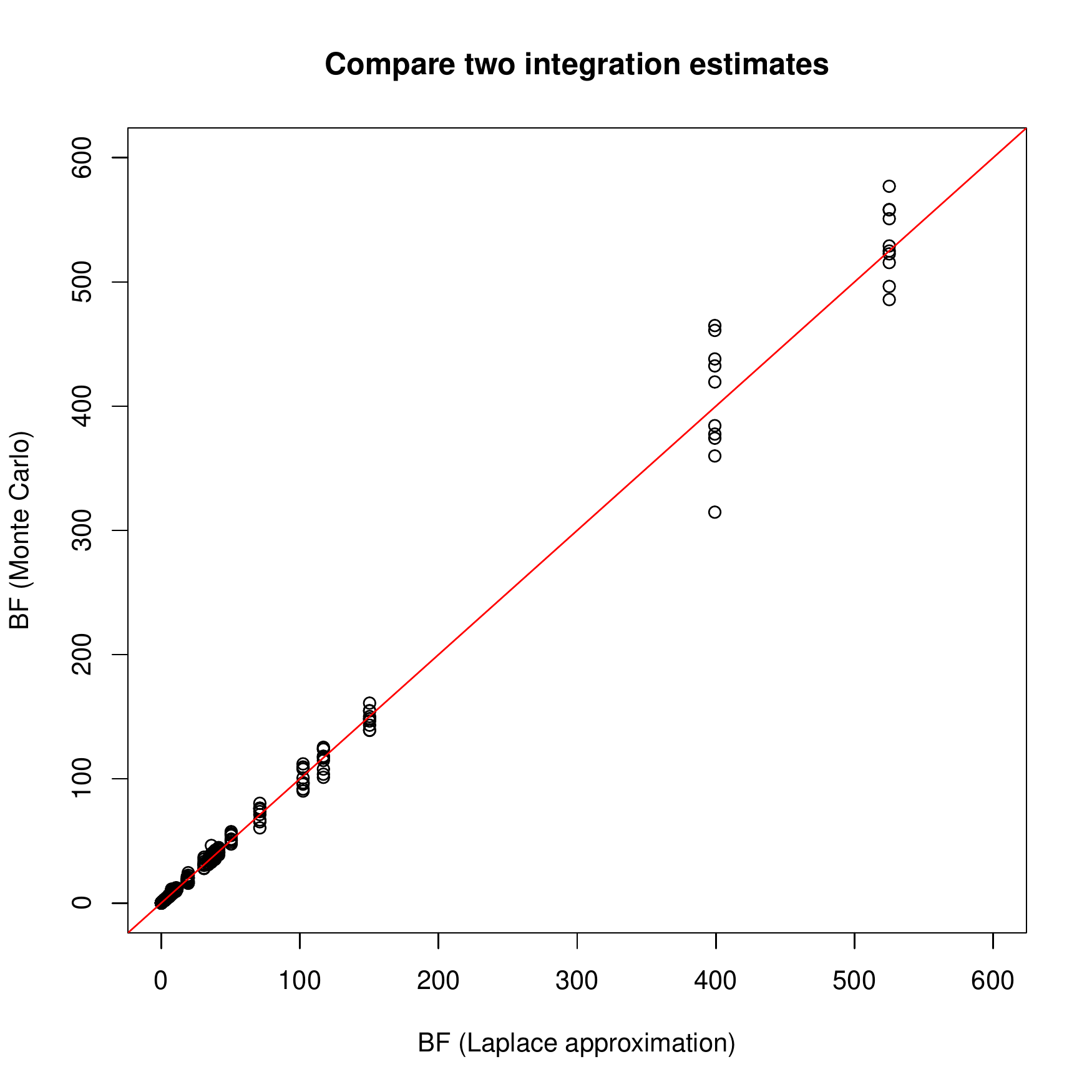}
\caption{Compare BF computations based on two approximations: Laplace method and Monte Carlo estimate}
\label{Laplace_MC}
\end{figure}

\begin{figure}
\centering
\includegraphics[scale=0.45]{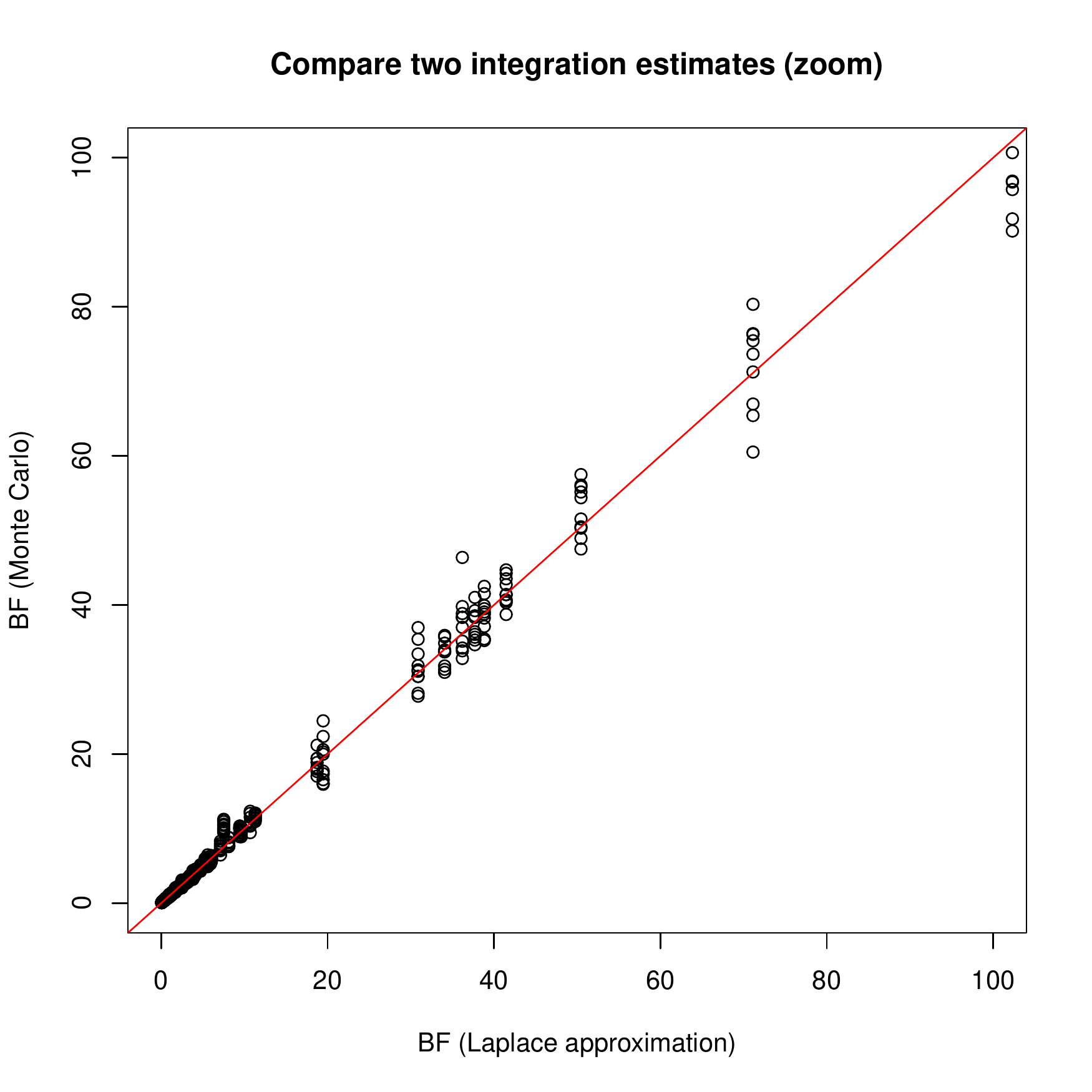}
\caption{Compare BF computations based on two approximations: Laplace method and Monte Carlo estimate (in zoomed-in version)}
\label{Laplace_MC_zoom}
\end{figure}

To assess the accuracy of Laplace method, we compared the Laplace estimate to 10 Monte Carlo estimates. {Figure \ref{Laplace_MC} and \ref{Laplace_MC_zoom} (zoomed-in version)} show that the BF computed by the Laplace method is very close to the mean of the BF computed by the Monte Carlo approximations. The former method is a more efficient and accurate approach compared to the latter.

\section{Web Appendix F:  Derivation of BF with  mixture prior}

\subsection{Compare $\eta_1$ and $\eta_2$ with fixed $w_{0j}$ ($w_{0j} \equiv \tilde{w_0}$) and fixed $K$ ($K\equiv\tilde{K}$)}

The parameters $\tilde{w}_0$ and $\tilde{K}$ are obtained by MLE, from the whole sample {under the null hypothesis}. We can formulate the null and alternative hypotheses as:
$$H_0: \eta_{1}  = \eta_{2} = \eta$$
$$H_1: \eta_{1}  \neq \eta_{2}$$
The marginal likelihood under $H_0$ is
\begin{equation*}
\begin{split}
m_0({\bf X}|\tilde{w_0},\tilde{K},\eta^*,K^*) = 
& \displaystyle\int_{\eta} \displaystyle\int_P f({\bf X}|{\bf P}) g({\bf P}|\eta, \tilde{w_0},\tilde{K}) d{\bf P} \pi(\eta|\eta^*,K^*) d\eta,\\
\end{split}
\end{equation*}

where
\begin{equation}\label{mix_lik}
\begin{split}
\displaystyle\int_P f({\bf X}|{\bf P}) g({\bf P}|\eta,\tilde{w_0},\tilde{K}) d{\bf P} 
= &  \prod_{j=1}^{2}\prod_{k=1}^{N_j} \displaystyle\int_{p_{jk}} f(x_{jk}|p_{jk})g(p_{jk}|\eta, \tilde{w_0},\tilde{K}) dp_{jk},\\
= & \prod_{x_{jk}=0}h_0(\tilde{w_0},\eta,\tilde{K}) \prod_{x_{jk}>0}h_1(x_{jk},\tilde{w_0},\eta,\tilde{K}).
\end{split}
\end{equation}

We have
\begin{equation*}
\begin{split}
h_0(\tilde{w_0},\eta,\tilde{K})
= & h_0(\tilde{w_0},\eta,\tilde{K}|p_{jk}=0)P(p_{jk}=0) + h_0(\tilde{w_0},\eta,\tilde{K}|p_{jk}>0)P(p_{jk}>0),\\
= & E\{f(x_{jk}=0|p_{jk}=0)\}P(p_{jk}=0) + (1-\tilde{w_0})\displaystyle\int_0^1 p_{jk}^0(1-p_{jk})^{n_{jk}}g(p_{jk}|\eta,\tilde{K})dp_{jk},\\
= & 1 \times P(p_{jk}=0) + (1-\tilde{w_0})\frac{B\{\eta \tilde{K},\tilde{K}(1-\eta)+n_{jk}\}}{B\{\eta \tilde{K},\tilde{K}(1-\eta)\}} \displaystyle\int_0^1 \frac{p_{jk}^{\eta \tilde{K}-1}(1-p_{jk})^{\tilde{K}(1-\eta)+n_{jk}-1}}{B\{\eta \tilde{K},\tilde{K}(1-\eta)+n_{jk}\}}dp_{jk},\\
= & \tilde{w_0} + (1-\tilde{w_0})\frac{B\{\eta \tilde{K},\tilde{K}(1-\eta)+n_{jk}\}}{B\{\eta \tilde{K},\tilde{K}(1-\eta)\}}, \\
\end{split}
\end{equation*}

and
\begin{equation*}
\begin{split}
h_1(x_{jk},\tilde{w_0},\eta,\tilde{K}) =& h_1(x_{jk},\tilde{w_0},\eta,\tilde{K}|p_{jk}=0)\tilde{w_0} + h_1(x_{jk},\tilde{w_0},\eta,\tilde{K}|p_{jk}>0)(1-\tilde{w_0}),\\
= & (1-\tilde{w_0}) \displaystyle \int_0^1 \binom{n_{jk}}{x_{jk}}p_{jk}^{x_{jk}}(1-p_{jk})^{n_{jk}-x_{jk}}\frac{p_{jk}^{\eta \tilde{K}-1}(1-p_{jk})^{\tilde{K}(1-\eta)-1}}{B\{\eta \tilde{K},\tilde{K}(1-\eta)\}}dp_{jk},\\
= &(1-\tilde{w_0})\binom{n_{jk}}{x_{jk}}\frac{B(\eta \tilde{K}+x_{jk},\tilde{K}(1-\eta)+n_{jk}-x_{jk})}{B\{\eta \tilde{K},\tilde{K}(1-\eta)\}}\\
& \displaystyle\int_0^1 \frac{p_{jk}^{\eta \tilde{K}+x_{jk}-1}(1-p_{jk})^{\tilde{K}(1-\eta)+n_{jk}-x_{jk}-1}}{B\{\eta \tilde{K}+x_{jk},\tilde{K}(1-\eta)+n_{jk}-x_{jk}\}}dp_{jk},\\
= &(1-\tilde{w_0})\binom{n_{jk}}{x_{jk}}\frac{B\{\eta \tilde{K}+x_{jk},\tilde{K}(1-\eta)+n_{jk}-x_{jk}\}}{B\{\eta \tilde{K},\tilde{K}(1-\eta)\}}. \\
\end{split}
\end{equation*}


Therefore, 
\begin{equation}\label{m0_mix}
\begin{split}
m_0({\bf X}|\tilde{w_0},\tilde{K},\eta^*,K^*) = & \displaystyle\int_{\eta} f({\bf X}|\tilde{w_0},\eta,\tilde{K}) \pi(\eta|\eta^*,K^*) d\eta,\\
= & \displaystyle\int_{\eta} \prod_{x_{jk}=0} \big[\tilde{w_0} + (1-\tilde{w_0})\frac{B\{\eta \tilde{K},\tilde{K}(1-\eta)+n_{jk}\}}{B\{\eta \tilde{K},\tilde{K}(1-\eta)\}}\big]\\
& \times \prod_{x_{jk}>0} \Big[(1-\tilde{w_0})\binom{n_{jk}}{x_{jk}}\frac{B\{\eta \tilde{K}+x_{jk},\tilde{K}(1-\eta)+n_{jk}-x_{jk}\}}{B\{\eta \tilde{K},\tilde{K}(1-\eta)\}} \Big]\\
& \pi(\eta|\eta^*,K^*)d\eta,
\end{split}
\end{equation}

and 
\begin{equation}
\begin{split}
m_1({\bf X}|\tilde{w_0},\tilde{K},\eta_1^*,K_1^*,\eta_2^*,K_2^*) = & \prod_{j=1}^2\displaystyle\int_{\eta} f({\bf X}|\tilde{w_0},\eta,\tilde{K}) \pi(\eta|\eta^*,K^*) d\eta,\\
= &\prod_{j=1}^2 \displaystyle\int_{\eta} \prod_{x_{jk}=0} \big[\tilde{w_0} + (1-\tilde{w_0})\frac{B\{\eta \tilde{K},\tilde{K}(1-\eta)+n_{jk}\}}{B\{\eta \tilde{K},\tilde{K}(1-\eta)\}}\big]\\
& \times \prod_{x_{jk}>0} \Big[(1-\tilde{w_0})\binom{n_{jk}}{x_{jk}}\frac{B\{\eta \tilde{K}+x_{jk},\tilde{K}(1-\eta)+n_{jk}-x_{jk}\}}{B\{\eta \tilde{K},\tilde{K}(1-\eta)\}} \Big]\\
& \pi(\eta|\eta_j^*,K_j^*)d\eta.
\end{split}
\end{equation}
The  marginal likelihood is computed using Laplace approximation, see equation (\ref{Laplace_app}).  
The MLE of $\eta$ can be expressed as $\hat{\eta}=arg\underset{\eta}{\operatorname{max}} \{ \log f({\bf X}|\eta, \tilde{w_0}, \tilde{K}) \}$, where $f({\bf X}|\eta, \tilde{w_0}, \tilde{K}) =\displaystyle\int_P f({\bf X}|{\bf P}) g({\bf P}|\eta,\tilde{w_0},\tilde{K}) d{\bf P}$. 
The estimates of ${\eta}_1$ and ${\eta}_2$ are obtained in a similar way by the MLE computed separately in controls and cases, but assuming the same $\tilde{w_0}$ and $\tilde{K}$, i.e. $\hat{\eta}_1=arg\underset{\eta}{\operatorname{max}} \{ \log f({\bf X_1}|\eta, \tilde{w_0}, \tilde{K}) \}$ and $\hat{\eta}_2=arg\underset{\eta}{\operatorname{max}} \{ \log f({\bf X_2}|\eta, \tilde{w_0}, \tilde{K}) \}$. {The parameters in the hyperprior distribution are defined in a similar way as indicated at the end of Section 3.4.1 of the main manuscript under a non-informative prior setting.}

\subsection{ Compare both $\eta_1$,  $w_{01}$ and $\eta_2$, $w_{02}$ with fixed $K$ ($K\equiv\tilde{K}$)}
The parameter $\tilde{K}$ is obtained by MLE, from the whole sample. For this comparison, the null and alternative hypotheses can be written as
$$H_0: \eta_{1}  = \eta_{2} = \eta \;\text{and} \; w_{01}=w_{02}=w_0$$
$$H_1: \eta_{1}  \neq \eta_{2} \;\text{or}\; w_{01} \neq w_{02}$$

In our simulations and real data analysis, this BF formulation did not result in improved power compared to the other versions so the results with this approach are omitted in these sections. \\

To simplify the derivation, we assume $n_{jk}\equiv n$, when $x_{jk}=0$.\\

The marginal likelihood under $H_0$ is
\begin{equation*}
\begin{split}
m_0({\bf X}|\tilde{K}, \eta^*,K^*, \eta^{**},K^{**}) = & \displaystyle\int f({\bf X}|{\bf P})g({\bf P})d{\bf P},\\
= & \displaystyle\int_P f({\bf X}|{\bf P}) \displaystyle\int_{\eta} \displaystyle\int_{w_0}g({\bf P}|\eta, w_0, \tilde{K})\pi(w_0| \eta^{**},K^{**})\pi(\eta|\eta^*,K^*)dw_0\ d\eta\ d{\bf P},\\
= & \displaystyle\int_{\eta}\displaystyle\int_{w_0}\ \displaystyle\int_P f({\bf X}|{\bf P}) g({\bf P}|\eta, w_0,\tilde{K}) d{\bf P}\  \pi(w_0| \eta^{**},K^{**} )\pi(\eta|\eta^*,K^*)dw_0\ d\eta,\\
\end{split}
\end{equation*}
\normalsize	
where $\displaystyle\int_{w_0} \displaystyle\int_P f({\bf X}|{\bf P}) g({\bf P}|w_0) d{\bf P} \pi(w_0) dw_0$ is the integral of equation (\ref{mix_lik}) in terms of $w_0$. Suppose $Z \equiv \frac{B(\eta \tilde{K},\tilde{K}(1-\eta)+n)}{B(\eta \tilde{K},\tilde{K}(1-\eta))}$, $\displaystyle\int_{w_0} \displaystyle\int_P f({\bf X}|{\bf P}) g({\bf P}|w_0) d{\bf P} \pi(w_0) dw_0$ can be written as\\

\small	
\begin{equation*}
\begin{split}
\displaystyle\int_{w_0} \displaystyle\int_P f({\bf X}|{\bf P}) g({\bf P}|w_0) d{\bf P} \pi(w_0) dw_0
= & \prod_{x_{jk}>0} \bigg[\binom{n_{jk}}{x_{jk}}\frac{B\{\eta \tilde{K}+x_{jk},\tilde{K}(1-\eta)+n_{jk}-x_{jk}\}}{B(\eta \tilde{K},\tilde{K}(1-\eta))}\bigg]\frac{1}{B(\eta^{**} K^{**}, K^{**}(1-\eta^{**}))}\\
& \times \displaystyle\int_0^1 \big[ \binom{M}{0}(1-w_0)^MZ^M + \binom{M}{1}w_0(1-w_0)^{M-1}Z^{M-1} + ... + \binom{M}{M}w_0^M \big]\\
& \times w_0^{\eta^{**}K^{**}-1}(1-w_0)^{N-M+K^{**}(1-\eta^{**})-1}dw_0\\
= &\prod_{x_{jk}>0} \bigg[\binom{n_{jk}}{x_{jk}}\frac{B\{\eta \tilde{K}+x_{jk},\tilde{K}(1-\eta)+n_{jk}-x_{jk}\}}{B(\eta \tilde{K},\tilde{K}(1-\eta))}\bigg]\frac{1}{B(\eta^{**} K^{**}, K^{**}(1-\eta^{**}))}\\
&  \sum_{h=0}^{M} \bigg[\binom{M}{h}Z^{M-h}B\{h+\eta^{**}K^{**},-h+N+K^{**}(1-\eta^{**})\} \\
& \times \displaystyle\int_{0}^{1} \frac{w_0^{h+\eta^{**}K^{**}-1}(1-w_0)^{-h+N+K^{**}(1-\eta^{**})-1}}{B\{h+\eta^{**}K^{**},-h+N+K^{**}(1-\eta^{**})\}} dw_0\bigg]\\
= & \prod_{x_{jk}>0} \bigg[\binom{n_{jk}}{x_{jk}}\frac{B\{\eta \tilde{K}+x_{jk},\tilde{K}(1-\eta)+n_{jk}-x_{jk}\}}{B(\eta \tilde{K},\tilde{K}(1-\eta))}\bigg]\\
& \frac{1}{B(\eta^{**} K^{**}, K^{**}(1-\eta^{**}))} \times \sum_{h=0}^{M} \bigg[\binom{M}{h}\frac{B\{\eta \tilde{K},\tilde{K}(1-\eta)+n\}^{M-h}}{B\{\eta \tilde{K},\tilde{K}(1-\eta)\}^{M-h}}\\
& \times B\{h+\eta^{**}K^{**},N-h+K^{**}(1-\eta^{**})\}\bigg].\\
\end{split}
\end{equation*}
\normalsize	
{where $M$ denotes the number of individuals with 0 rare variants for the specific gene (see Section 3.4.2 in the main manuscript).}

Therefore, 
\begin{equation}
\begin{split}
m_0({\bf X}|\tilde{K}, \eta^*,K^*, \eta^{**},K^{**})  
= & \prod_{x_{jk}>0} \binom{n_{jk}}{x_{jk}}\frac{1}{B\{\eta^{**} K^{**}, K^{**}(1-\eta^{**})\}} \\
& \displaystyle\int_{\eta} \prod_{x_{jk}>0} \bigg[\frac{B\{\eta \tilde{K}+x_{jk},\tilde{K}(1-\eta)+n_{jk}-x_{jk}\}}{B\{\eta \tilde{K},\tilde{K}(1-\eta)\}}\bigg]\\
& \times \sum_{h=0}^{M} \bigg[\binom{M}{h}\frac{B\{\eta \tilde{K},\tilde{K}(1-\eta)+n\}^{M-h}}{B\{\eta \tilde{K},\tilde{K}(1-\eta)\}^{M-h}}\\
& B\{h+\eta^{**} {K}^{**},N-h+ {K}^{**}(1-\eta^{**})\}\bigg] \times \pi(\eta|\eta^*,K^*)d\eta\\
\end{split}
\end{equation}
and
\begin{equation}
\begin{split}
m_1({\bf X}|\tilde{K}, \eta_1^*,K_1^*,  \eta_2^*,K_2^*, \eta_1^{**},K_1^{**},\eta_2^{**},K_2^{**})  
= & \prod_{j=1}^2\prod_{x_{jk}>0} \binom{n_{jk}}{x_{jk}}\frac{1}{B\{\eta_j^{**} K_j^{**}, K_j^{**}(1-\eta_j^{**})\}} \times\\
&\prod_{j=1}^2 \displaystyle\int_{\eta} \prod_{x_{jk}>0} \bigg[\frac{B\{\eta \tilde{K}+x_{jk},\tilde{K}(1-\eta)+n_{jk}-x_{jk}\}}{B\{\eta \tilde{K},\tilde{K}(1-\eta)\}}\bigg]\\
& \times \sum_{h=0}^{M} \bigg[\binom{M}{h}\frac{B\{\eta \tilde{K},\tilde{K}(1-\eta)+n\}^{M-h}}{B\{\eta \tilde{K},\tilde{K}(1-\eta)\}^{M-h}}\\
& B\{h+\eta_j^{**} {K}_j^{**},N-h+ {K}_j^{**}(1-\eta_j^{**})\}\bigg] \times \pi(\eta|\eta_j^*,K_j^*)d\eta\\
\end{split}
\end{equation}

As before, $\eta$ is estimated by MLE as $\hat{\eta}=arg\underset{\eta}{\operatorname{max}} \{  \log f({\bf X}|\eta, \tilde{K}, \eta^{**}, K^{**}) \}$ and $\eta_j$ is estimated by $\hat{\eta}_j=arg\underset{\eta}{\operatorname{max}} \{  \log f({\bf X_j}|\eta, \tilde{K}, \eta_j^{**}, K_j^{**}) \}$, where 

\begin{equation} 
\begin{split}
f({\bf X}|\eta, \tilde{K}, \eta^{**}, K^{**})=&\prod_{x_{jk}>0} \bigg[\frac{B\{\eta \tilde{K}+x_{jk},\tilde{K}(1-\eta)+n_{jk}-x_{jk}\}}{B\{\eta \tilde{K},\tilde{K}(1-\eta)\}}\bigg]\\
& \times \sum_{h=0}^{M} \bigg[\binom{M}{h}\frac{B\{\eta \tilde{K},\tilde{K}(1-\eta)+n\}^{M-h}}{B\{\eta \tilde{K},\tilde{K}(1-\eta)\}^{M-h}}\\
& B\{h+\eta^{**} {K}^{**},N-h+ {K}^{**}(1-\eta^{**})\}\bigg].\\
\end{split}
\end{equation}

\section{Web Appendix G: Correlation assessment between the two components of the BF with informative prior}
\begin{figure}
\centering
\includegraphics[scale=0.6]{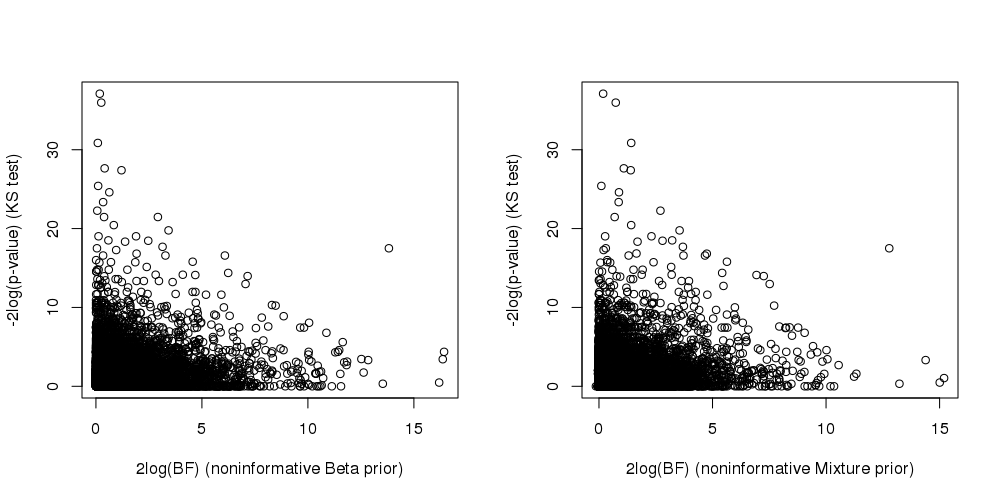}
\caption{Scatter plot of the two BF components with $x$-axis: $\frac{(\hat{\eta}_1-\hat{\eta}_2)^2}{{\Sigma}_1+{\Sigma}_2}$ and $y$-axis: $-2log(p)$ given in equation (\ref{final_form}), based on all the genes from the lung cancer WES study. The Kendall correlation coefficient between the 2 components is 0.03 for beta prior and 0.04 for mixture prior.}
\label{correlation}
\end{figure}

\begin{figure}
\centering
\includegraphics[scale=0.6]{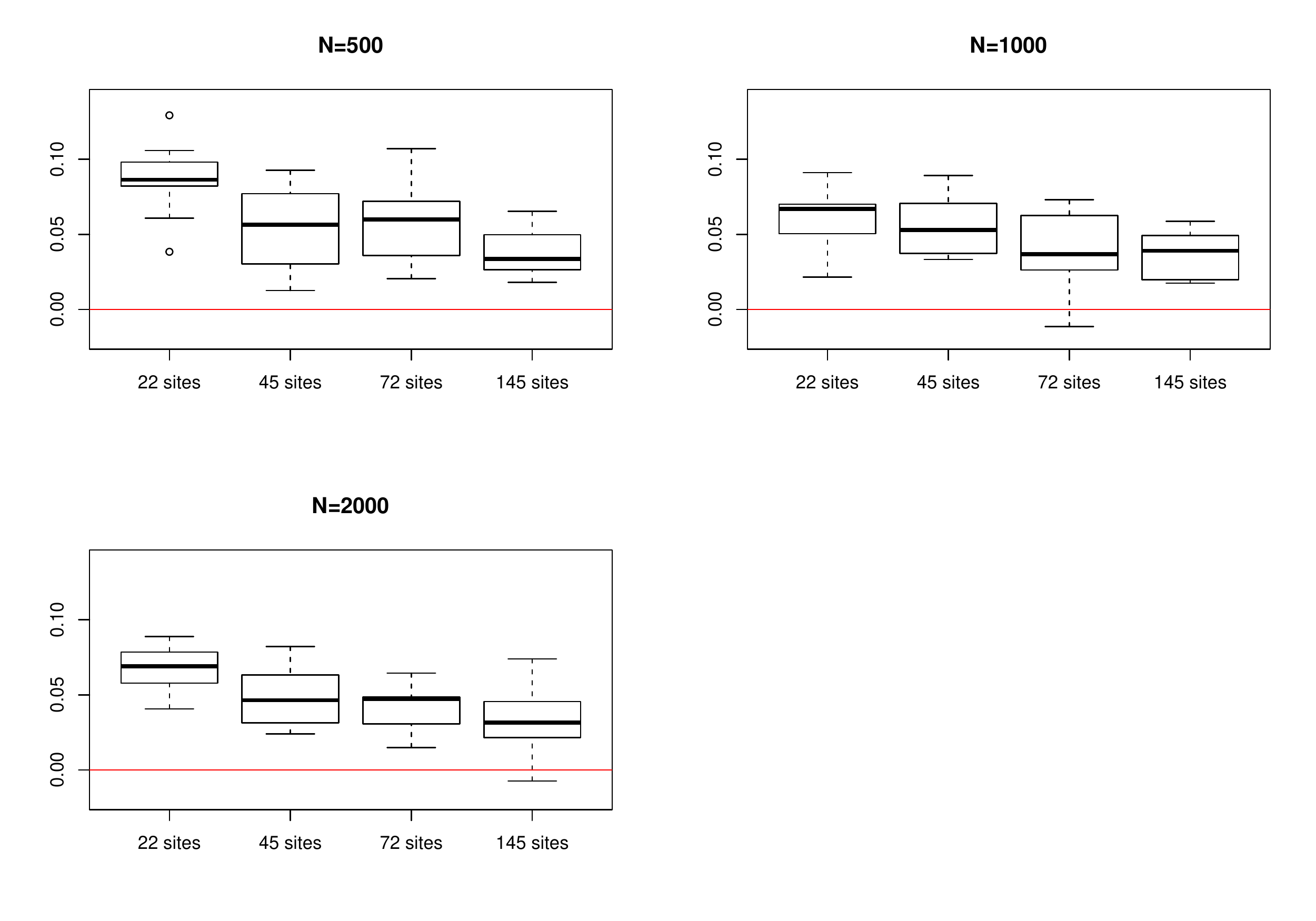}
\caption{Distribution of Kendall correlation coefficient between $\frac{(\hat{\eta}_1-\hat{\eta}_2)^2}{{\Sigma}_1+{\Sigma}_2}$ and $-2log(p)$ under the null hypothesis for the BF with beta prior with respect to different gene sizes and sample sizes. The KS test p-values are computed after pruning highly correlated variants (Pearson correlation coefficient $>$ 0.5). For all simulation scenarios, the Kendall correlation coefficient is close to zero, justifying our assumption of independence between the two components of the informative BF statistic under $H_0$.}
\label{corr_sim}
\end{figure}

\begin{figure}
\centering
\includegraphics[scale=0.6]{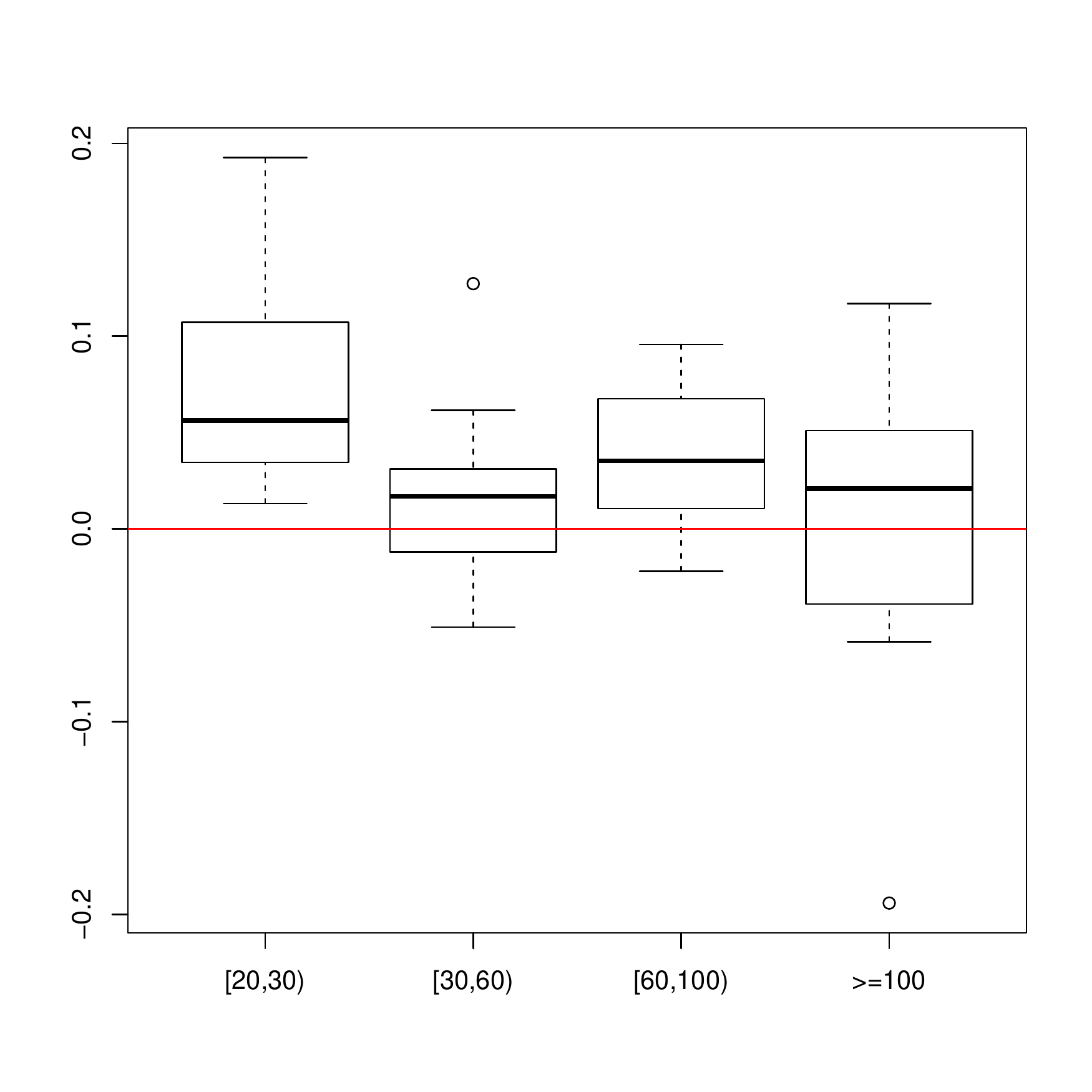}
\caption{Distribution of Kendall correlation coefficient between $\frac{(\hat{\eta}_1-\hat{\eta}_2)^2}{{\Sigma}_1+{\Sigma}_2}$ and $-2log(p)$ for the BF with beta prior with respect to gene sizes in the lung cancer WES study.  The KS test p-values used for this result are computed after pruning highly correlated variants (Pearson correlation coefficient $>$ 0.5).}
\label{corr_real}
\end{figure}

{First, the scatter plots in Figure \ref{correlation} illustrate that the two components of informative BF statistic in the lung cancer WES data are uncorrelated, under both a beta and a mixture prior. Besides, to assess the correlation between the BF with non-informative prior and the KS test $p$-value, we compute the Kendall rank correlation coefficient, which does not rely on the distribution assumption of each component, in the simulated data (Figure \ref{corr_sim}) and lung cancer WES data (Figure \ref{corr_real}). 

It is noteworthy that here we evaluate the correlation between two components of informative BF statistic under $H_0$. It is under this assumption that the majority of genes in the lung cancer WES study are non associated with the disease status. 
To avoid the influence of linkage disequilibrium (LD) across RVs from the same gene on the correlation assumption of the two components, we did prune highly correlated RVs to compute the KS test $p$-values. Even if we notice some slightly higher correlation between the two components of BF statistics for smaller genes, we do not observe an increase in type I error for these genes in our simulation results (Table 1 of the main manuscript). 
Therefore, the independence assumption between the two components of the informative BF can be justified by the fact that they are almost uncorrelated in our simulations and real data. }

\begin{table}
\centering
\caption{Average Kendall rank correlation coefficient estimated from the WES Lung Cancer study based on 1000 selected genes}
\label{table_corr_real}
\begin{tabular}{lrrrrrr}
\hline
& & \multicolumn{4}{c}{$\#(sites)$}\\
& All the genes & $[20,30)$ 	& $[30,60)$ & $[60,100)$& $[100,17067]$ \\
\hline
Beta prior &&&&&\\
$r^2>0.99^*$ & 0.03 & 0.05 & 0.01 & 0.03 & 0.04\\	
$r^2>0.5^*$ & 0.03 & 0.06 & 0.01 & 0.04 & 0.02\\	
\hline	
Mixture prior &&&&&\\
$r^2>0.99^*$ & 0.04 & 0.05 & 0.01 & 0.07 & 0.02\\	
$r^2>0.5^*$ & 0.03 & 0.06 & 0.01 & 0.06 & 0.01\\	
\hline		
\multicolumn{6}{l}{ {\small $^*$ Highly correlated RVs are removed  for the KS test based on either $r^2>0.5 \; \text{or} \; r^2>0.99$.}} \\	
\end{tabular}
\end{table}

\section{Web Appendix H: Proof of Proposition 1}
Expanding the derivatives around $\eta_0$ using Taylor series expansion, assuming $\ell_1$ is 3 times differentiable on the closed interval between $\eta_0$ and $\hat{\eta}_1$ and closed interval between $\eta_0$ and $\hat{\eta}$; $\ell_2$ is 3 times differentiable on the closed interval between $\eta_0$ and $\hat{\eta}_2$ and closed interval between $\eta_0$ and $\hat{\eta}$, we obtain\\
\begin{equation}\label{log-like-der}
\begin{split}
\ell_1'({\hat\eta_1})&=\ell_1'({\eta_0})+({\hat\eta_1}-\eta_0)\ell_1''({\eta_0}) + (\hat{\eta}_1-\eta_0)^2\frac{\ell_1'''(\eta_{10})}{2} \text{ for }\eta_{10}\text{ between }\eta_0 \text{ and } \hat{\eta}_1,\\
\ell_2'({\hat\eta_2})&=\ell_2'({\eta_0})+({\hat\eta_2}-\eta_0)\ell_2''({\eta_0})+ (\hat{\eta}_2-\eta_0)^2\frac{\ell_2'''(\eta_{20})}{2} \text{ for }\eta_{20}\text{ between }\eta_0 \text{ and } \hat{\eta}_2,\\
\ell_1'({\hat\eta})&=\ell_1'({\eta_0})+({\hat\eta}-\eta_0)\ell_1''({\eta_0})+ (\hat{\eta}-\eta_0)^2\frac{\ell_1'''(\eta_{01})}{2} \text{ for }\eta_{01}\text{ between }\eta_0 \text{ and } \hat{\eta}, \\
\ell_2'({\hat\eta})&=\ell_2'({\eta_0})+({\hat\eta}-\eta_0)\ell_2''({\eta_0}) + (\hat{\eta}-\eta_0)^2\frac{\ell_2'''(\eta_{02})}{2} \text{ for }\eta_{02}\text{ between }\eta_0 \text{ and } \hat{\eta}.\\	
\end{split}
\end{equation} 


Since we have $\ell_1'({\hat{\eta_1}})=0$, $\ell_2'({\hat{\eta_2}})=0$ and $\ell'({\hat{\eta}})=\ell_1'({\hat{\eta}})+\ell_2'({\hat{\eta}})=0$, we deduct that 
$$\ell_1'({\eta_0})+\ell_2'({\eta_0})=-({\hat\eta_1}-\eta_0)\ell_1''({\eta_0})-({\hat\eta_2}-\eta_0)\ell_2''({\eta_0})+o(\hat{\eta}_1-\eta_0)+o(\hat{\eta}_2-\eta_0)$$
and
$$\ell_1'({\eta_0})+\ell_2'({\eta_0})=-({\hat\eta}-\eta_0)\ell_1''({\eta_0})-({\hat\eta}-\eta_0)\ell_2''({\eta_0})+o(\hat{\eta}-\eta_0)$$

$\implies ({\hat\eta_1} - {\hat\eta}) \Sigma_1^{-1} + ({\hat\eta_2} - {\hat\eta}) \Sigma_2^{-1} = o(\hat{\eta}_1-\eta_0)+o(\hat{\eta}_2-\eta_0)+o(\hat{\eta}-\eta_0), $

$\implies \hat\eta = \frac{\hat\eta_1\Sigma_2+\hat\eta_2\Sigma_1}{\Sigma_1+\Sigma_2}+o(\hat{\eta}_1-\eta_0)+o(\hat{\eta}_2-\eta_0)+o(\hat{\eta}-\eta_0).$ \\

Since $\Sigma^{-1} = \Sigma_1^{-1}+\Sigma_2^{-1}$, ignoring the approximation error term, we deduct that $\Sigma=\frac{\Sigma_1 \Sigma_2}{\Sigma_1+\Sigma_2}.$

\section{Web Appendix I: BFDR procedure for genome-wide inference}
For the computation of $v_i$ in the BFDR procedure, we need to specify $p(\pi_0|Y)$ as outlined in equation (6) of the main manuscript. Following \cite{Scott2006}, we can assume a beta prior distribution for $\pi_0$, the proportion of non-associated genes, as
\begin{equation}\label{pi0_prior}
\begin{split}
f(\pi_0 | Y) &= (\alpha+1)\pi_0^{\alpha}.\\
\end{split}
\end{equation}	

For this prior distribution of $\pi_0$, we need a specification of $\alpha$ parameter. For $\alpha$, a large parameter value corresponds to a large proportion of genes not associated with the response (case-control status), which leads to a more stringent gene selection at the genome-wide level. We need to emphasize that the BF procedure is sensitive to this hyperparameter specification, in particular, the choice of hyperparameters should be different for the BF with and without informative priors. The reason is that the BF distribution that is needed to compute the component $Pr(Z_i=1 | Y, \pi_0)$ in equation (7) of the main manuscript, is quite different when using informative prior or not.  \\ \ \\


For the specification of the above hyperparameters and thus to estimate $v_i$, we propose the following two-step procedure. 

\begin{enumerate}
\item We first estimate $\pi_0$, based on the approach proposed by \cite{Wen2017} for both the BF with and without informative prior. Both estimates can be different since the distribution of the two BFs {and also their theoretical null distributions can differ.}

\item Based on $\hat{\pi}_0$, estimated in step 1, we decide the hyperparamter ${\alpha}$ in equation (\ref{pi0_prior}). 
The choice of the hyperparameter $\alpha$ also  depends on whether informative prior is incorporated in the BF or not. 

\end{enumerate}

For step 1, \cite{Wen2017} showed that we can obtain an upper bound estimator of $\pi_0$ by
$$ \hat{\pi}_0 = \frac{\sum_{i=1}^{m}\mathnormal{I}(BF_i\leq q_{i,\gamma})}{m\gamma},$$
where $q_{i,\gamma}$ denotes the $\gamma$-quantile of the $BF_i$ under $H_0$, i.e.  $F_i^0(q_{i,\gamma}) =\gamma$, where $F_i^0(x)$ is the {\it c.d.f.} of the BF under $H_0$.
A problem with Wen's procedure, however, is that $F_i^0(x)$ is not known and would need to be estimated empirically, for example from permutation sampling, which could be intractable if the number of tests (i.e., genes) is large. However, from our theoretical results, we know that $2\log BF_i \xrightarrow{d} \chi^2(1)$ or
$2\log BF_i \xrightarrow{d} \chi^2(3)$ under a non-informative or informative prior, respectively. We can then replace the estimator of $\pi_0$ by 

\begin{equation}\label{est_pi}
\begin{split}
\hat{\pi}_0 = \frac{\sum_{i=1}^{m}\mathnormal{I}(2\log BF_i\leq q^*_{\gamma})}{m\gamma},
\end{split}
\end{equation} 
where $q^*_{\gamma}$ is the $\gamma$-quantile of a $\chi^2(1)$ distribution for non-informative prior and $\chi^2(3)$ distribution for informative prior. {The choice of $\gamma$ results from a tradeoff between bias and variance of $\hat{\pi}_0$ \citep{Storey2003,Efron2010,Wen2017}. In the genome-wide analysis, there is a high proportion of non associated genes with BF values close to 1 (more than expected under the null), which could be due to the high proportion of small genes (20-50 sites per gene) that can be sensitive to gentoyping error and data pre-processing. As the number of genes is large in a WES study ($>10,000$), the proportion of truly associated genes $(1-\hat\pi_0)$ is expected to be small ($\approx 0.1\%$). Therefore, we chose a $\gamma$ value close to 1 in our real data application.} At the end of step 1, we will have an estimate of $\pi_0$ for the BF with non-informative and with informative prior, that we denote respectively $\hat\pi_0$ and $\hat\pi_{0inf}$. 

For the second step of our BFDR procedure, we need to specify $\alpha$ in the prior distribution of $\pi_0$ above. For the BF with non informative prior, the hyperparameter estimate $\hat{\alpha}$ is  determined based on $\hat{\pi}_0$ obtained from Step 1 by letting the expectation of the prior distribution equal to $\hat{\pi}_0$,
\begin{equation}\label{pi0_mean}
\begin{split}
E(\pi_0) = \frac{{\alpha}}{{\alpha}+1} = \hat{\pi}_0  \;\; \Rightarrow \;\; \hat{\alpha}=\frac{\hat{\pi}_0}{1-\hat{\pi}_0}.
\end{split}
\end{equation} 

For the BF with informative prior, we use the following strategy to determine this hyperparameter that we denote $\alpha_{inf}$.  

For a given gene $i$, we can deduct from equation \ref{final_form} the relationship between the BF with and without informative prior as
\begin{equation}\label{BF_inf_noninf}
\begin{split}
BF_{i, inf} = BF_{i}/p_i,
\end{split}
\end{equation}
where $BF_{i, inf}$ is the BF statistic with informative prior, $BF_{i}$ is BF with non informative prior and $p_i$ the p-value from the KS test .\\

For the BF with informative prior, we write the probability $f(\pi_0 | Y)$ as
\begin{equation}\label{pi0_prior_inf}
\begin{split}
f(\pi_0 | Y) = (\alpha_{inf}+1)(\pi_{0inf})^{\alpha_{inf}},
\end{split}
\end{equation}

and the expression (7) of the main manuscript as

\begin{equation}\label{pi0_prior_inf}
\begin{split}
Pr(Z_i=1|Y,\pi_{0inf}) = \frac{(1-\pi_{0inf})BF_{i}/p_i}{\pi_{0inf}+(1-\pi_{0inf})BF_{i}/p_i}.
\end{split}
\end{equation}

We assume that under $H_0$, incorporating informative prior into BF should not change $v_i$ very much, in other words, $Pr(Z_i=1|Y,\pi_{0})$ based on non-informative prior and	$Pr(Z_i=1|Y,\pi_{0inf})$ based on informative prior should be similar. \\

Therefore, for gene $i$, we need to satisfy

\begin{equation}
\begin{split}
E \bigg( \frac{(1-\pi_{0inf})BF_{i}/p_i}{\pi_{0inf}+(1-\pi_{0inf})BF_{i}/p_i} \bigg) & = E\bigg( \frac{(1-\pi_{0})BF_{i}}{\pi_{0}+(1-\pi_{0})BF_{i}} \bigg).
\end{split}
\end{equation}
Assuming $\pi_0 \rightarrow 1$ and $\pi_{0inf} \rightarrow 1$, this leads to
\begin{equation}\label{pi0_non_inf}
\begin{split}
E((1-\pi_{0inf})/p_i) = E(1-\pi_0) &\Leftrightarrow \;\; E(\pi_{0inf}) = 1-p_i(1-E(\pi_0)).
\end{split}
\end{equation}

Equation (\ref{pi0_non_inf}) can then be written as 
\begin{equation*}
\begin{split}
\frac{\alpha_{inf}}{\alpha_{inf}+1} = & 1-p_i(1-\frac{\alpha}{\alpha+1})\\
= & \frac{\alpha+1-p_i}{\alpha+1}\\
= & \frac{\frac{\alpha+1}{p_i}-1}{\frac{\alpha+1}{p_i}}. \\
\end{split}
\end{equation*}
Thus, since $\alpha \gg 1$ and $\alpha/p_0 \gg 1$, we have
\begin{equation}\label{alpha_inf_i}
\begin{split}
\alpha_{inf} \approx \frac{\alpha+1}{p_i}-1 \approx \frac{\alpha}{p_i}.
\end{split}
\end{equation}
Let's define $\alpha_{0inf}\equiv \frac{\alpha}{p_i}$. As stated above, for single gene analysis, using $\alpha_{0inf}$ for BF with informative prior would lead to similar estimate of $v_i$ using BF with non-informative prior. However, in the genome-wide inference, the hyperparameter  $\alpha_{inf}$  has to be same across all the genes, so we replace $p_i$ with a fixed value $p_0$ in equation (\ref{alpha_inf_i}). Accordingly, we can determine an estimator of the hyperparameter ${\alpha_{inf}}$ as
\begin{equation}\label{alpha_inf}
\begin{split}
\hat{\alpha}_{inf} = \frac{\hat{\alpha}}{p_0}.
\end{split}
\end{equation}

\begin{figure}[h!]
\centering	
\includegraphics[scale=0.7]{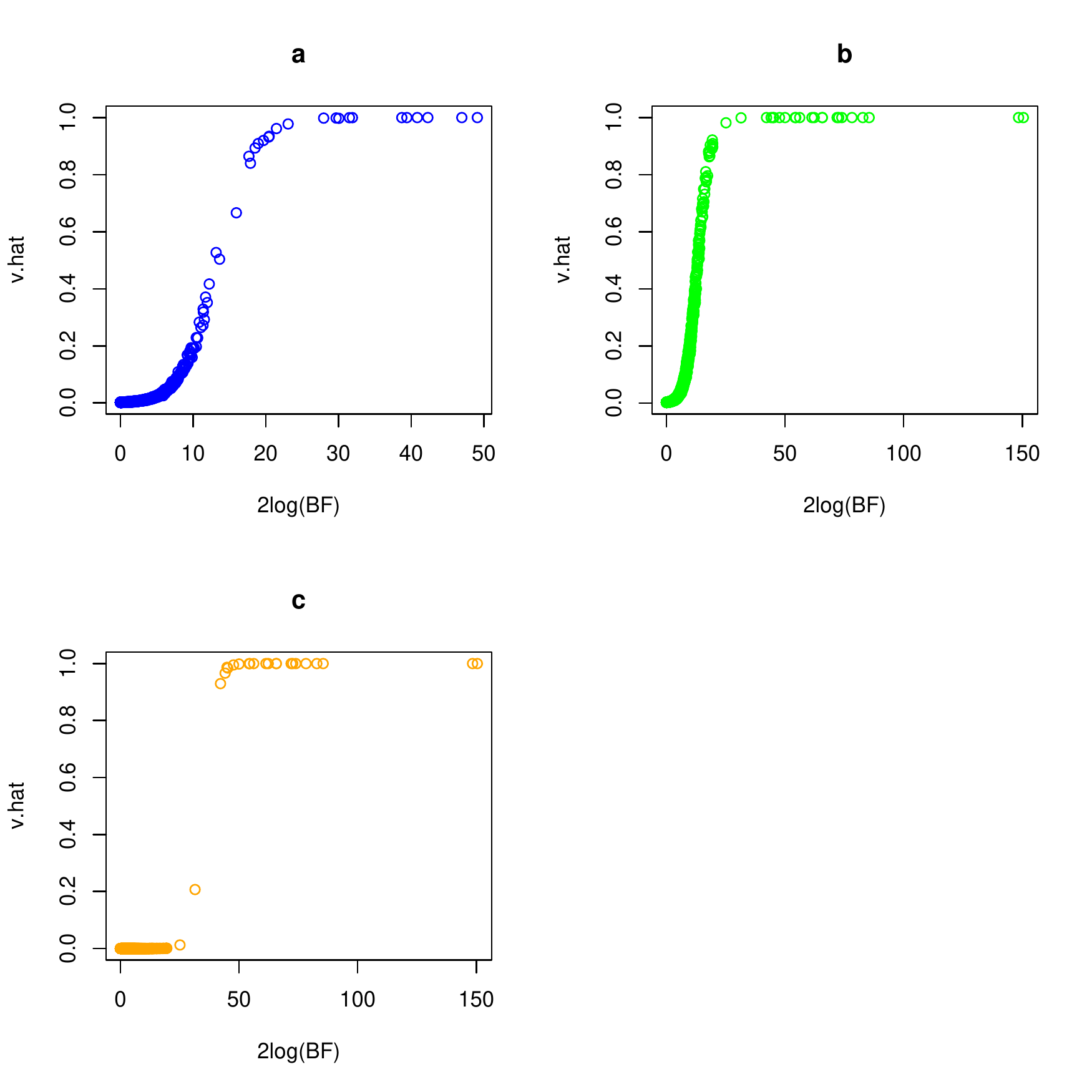}
\caption{Scatter plot of $\hat{v}$ with respect to $2\log(BF)$ based on one simulated dataset with 10,000 genes and 20 of them associated with the case-control status. Plot a) displays the scatter plot for BF with non-informative prior and $\hat{v}_i$  computed using hyperparameter $\hat{\alpha}$ (equation \ref{pi0_mean}); Plots b) display scatter plot for BF with informative prior and $\hat{v}_i$  computed using hyperparameter $\hat{\alpha}$ (equation \ref{pi0_mean}); Plot c) displays scatter plot for BF with informative prior and $\hat{v}_i$  computed using hyperparameter $\hat{\alpha}'$ (equation \ref{alpha_inf}). }
\label{v_hat}
\end{figure}

The determination of $p_0$ in equation (\ref{alpha_inf}) can be based on matching the expected number of associated genes $m (1-\hat{\pi}_{0inf})$. In particular, we propose 
\begin{equation}\label{p0}
\begin{split}
p_0 \equiv p_{(1-\hat{\pi}_{0inf}),}
\end{split}
\end{equation}
where $p_{(1-\hat{\pi}_{0inf})}$ is the $(1-\hat{\pi}_{0inf})$-quantile of the KS test $p$-value distribution.
Therefore, for gene $i$ with lowest KS $p$-values,  $p_0>p_i$ leads to having $\hat{\alpha}_{inf}<\alpha_{0inf}$, i.e. larger estimate of $v_i$ compared to using non-informative prior. Otherwise, for the genes with $p_0<p_i$, it leads to a more conservative (smaller) estimate of $v_i$ compared to using non-informative prior.
We show in Figure \ref{v_hat} that the proposed procedure adjusts well the $\hat{v}_i$ distribution in our simulated data when using the BF with informative prior. For instance, Figure \ref{v_hat}.b shows the distribution of $\hat{v}_i$ obtained using $\hat{\alpha}$ instead of $\hat{\alpha}_{inf}$ to calculate BFDR. A lot of genes have $\hat{v}_i$ values between 0.8 and 0.99 and for which the null hypothesis of no association could be rejected (i.e. genes declared associated). This problem does not happen when using $\hat{\alpha}_{inf}$ as shown in \ref{v_hat}.c, where there is a better separation between small and high values of $\hat{v}_i$.





\section{Web Appendix J: QQ plot of the BF with informative prior under $H_0$}
\begin{figure}
\centering
\includegraphics[scale=0.7]{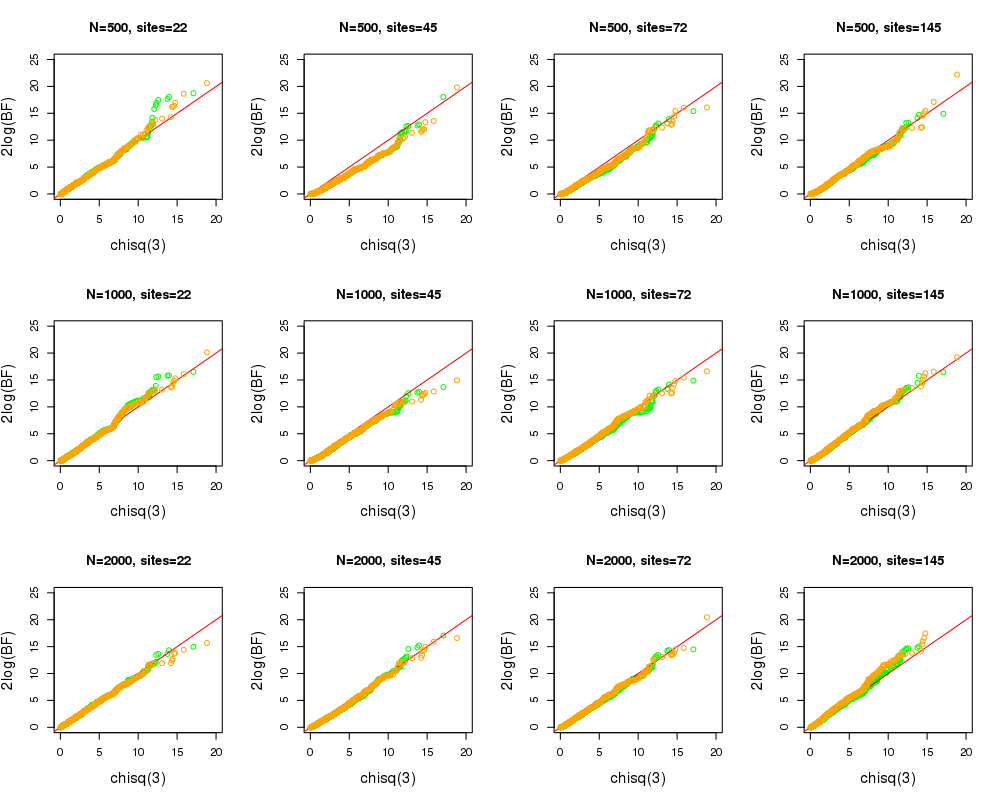}
\caption{QQ plot of the BF with informative prior under the null hypothesis assuming different sample sizes and gene sizes. The green circle symbol in each plot represents BF with beta prior and the orange circle symbol represents BF with mixture prior. }
\label{QQsim1}
\end{figure}

\section{Web Appendix K: Type I error rate and power comparison between BF and existing methods}
\begin{figure}
\centering
\includegraphics[scale=0.7]{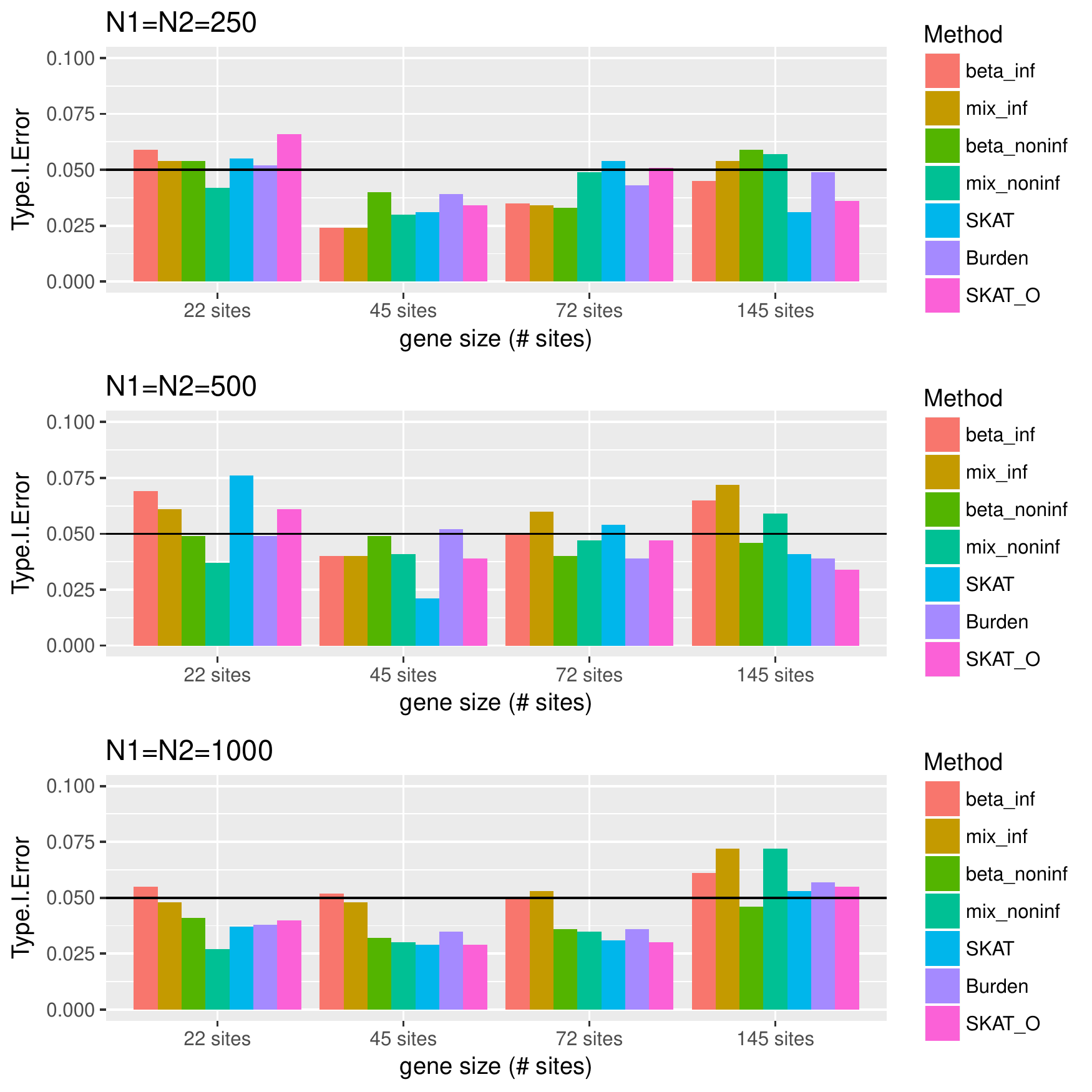}
\caption{Type I error rate for different methods ($\alpha=0.05$) including BF with beta informative prior (\texttt{beta\_inf}), BF with mixture informative prior (\texttt{mix\_inf}), BF with beta non-informative prior (\texttt{beta\_noninf}), BF with mixture non-informative prior (\texttt{mix\_noninf}), \texttt{SKAT}, \texttt{Burden} and \texttt{SKAT\_O}.}
\label{typeI_comp}
\end{figure}

\begin{figure}
\centering
\includegraphics[scale=0.7]{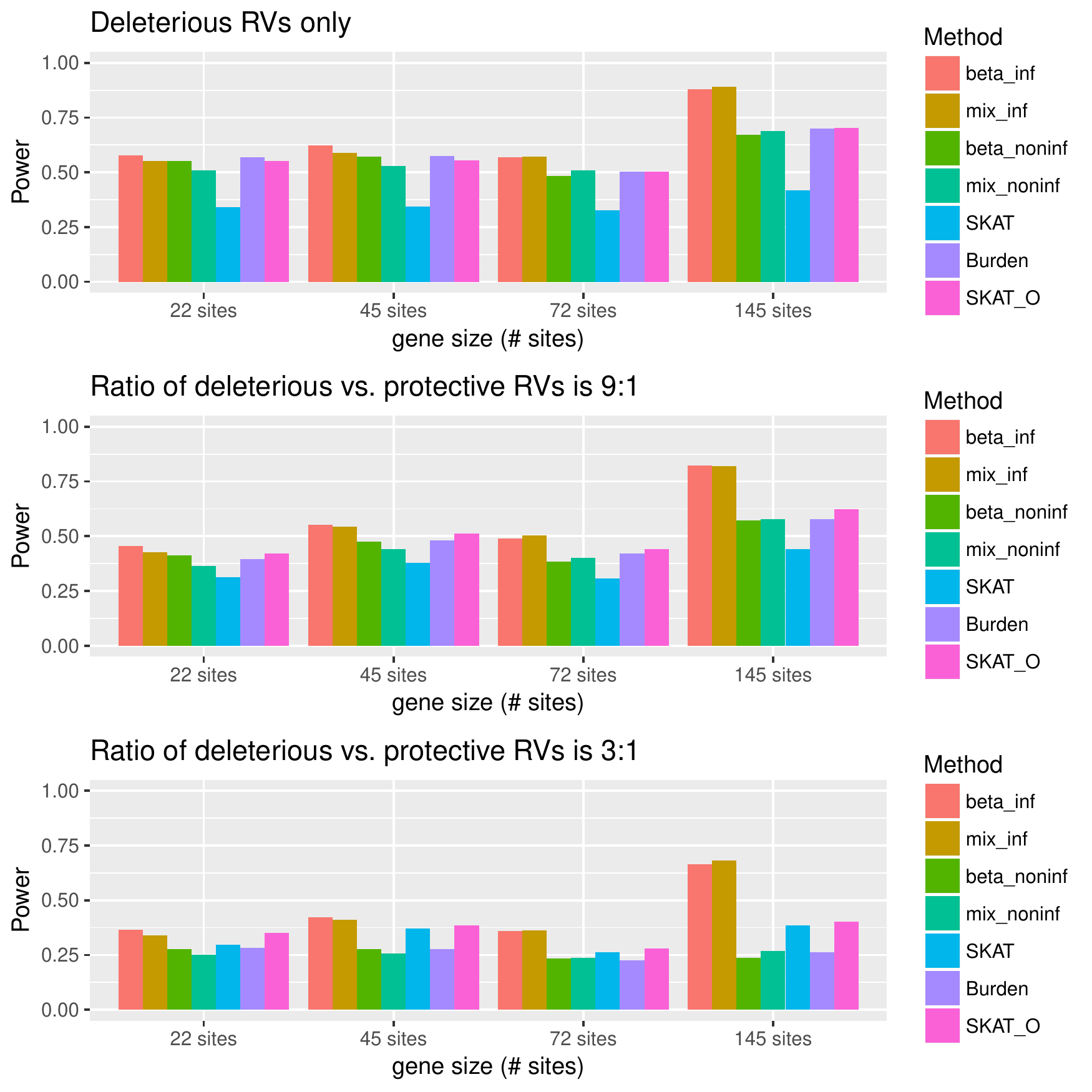}
\caption{Statistical power for different methods with sample size $N1=N2=250$ ($\alpha=0.05$) including BF with beta informative prior (\texttt{beta\_inf}), BF with mixture informative prior (\texttt{mix\_inf}), BF with beta non-informative prior (\texttt{beta\_noninf}), BF with mixture non-informative prior (\texttt{mix\_noninf}), \texttt{SKAT}, \texttt{Burden} and \texttt{SKAT\_O}.}
\label{power500_comp}
\end{figure}

\begin{figure}
\centering
\includegraphics[scale=0.7]{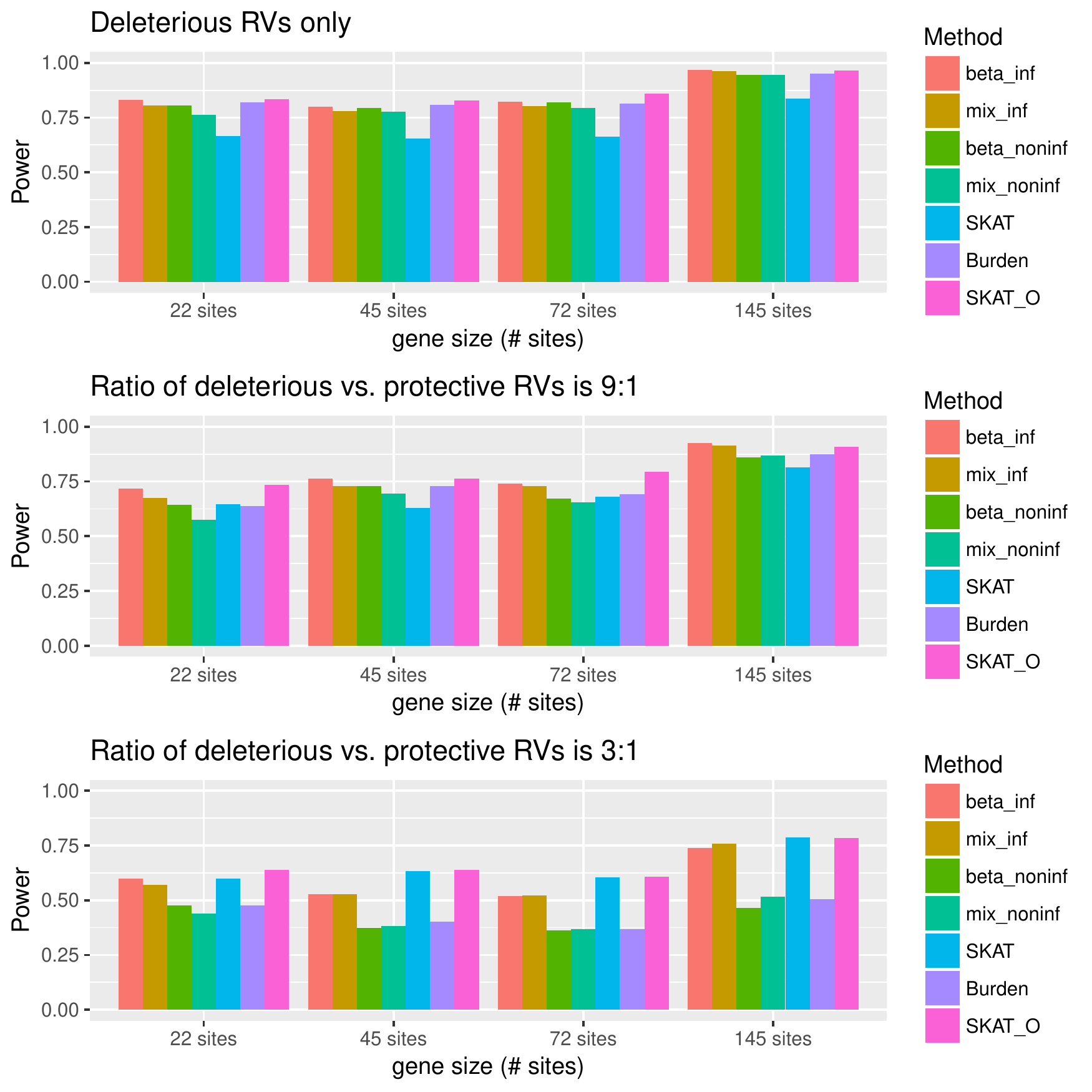}
\caption{Statistical power for different methods with sample size $N1=N2=500$ ($\alpha=0.05$) including BF with beta informative prior (\texttt{beta\_inf}), BF with mixture informative prior (\texttt{mix\_inf}), BF with beta non-informative prior (\texttt{beta\_noninf}), BF with mixture non-informative prior (\texttt{mix\_noninf}), \texttt{SKAT}, \texttt{Burden} and \texttt{SKAT\_O}. }
\label{power1000_comp}
\end{figure}

\begin{figure}
\centering
\includegraphics[scale=0.7]{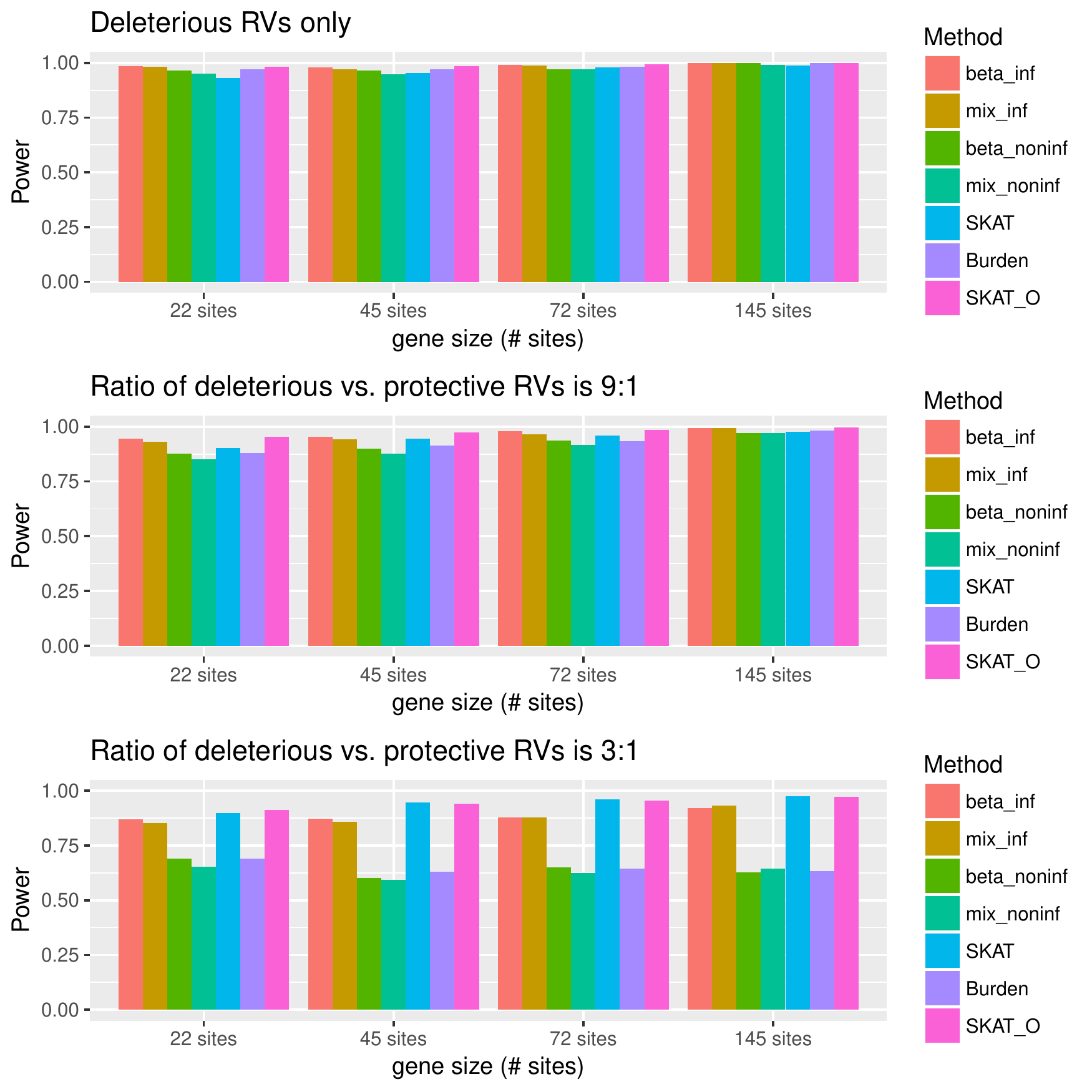}
\caption{Statistical power for different methods with sample size $N1=N2=1000$ ($\alpha=0.05$) including BF with beta informative prior (\texttt{beta\_inf}), BF with mixture informative prior (\texttt{mix\_inf}), BF with beta non-informative prior (\texttt{beta\_noninf}), BF with mixture non-informative prior (\texttt{mix\_noninf}), \texttt{SKAT}, \texttt{Burden} and \texttt{SKAT\_O}. }
\label{power2000_comp}
\end{figure}

Figure \ref{typeI_comp}, \ref{power500_comp}, \ref{power1000_comp} and \ref{power2000_comp} are graphical representation of type I error and power results for the BF and other competing methods shown in Tables 1 \& 2 of the main manuscript.

Besides, we repeated our complete simulation runs 6 times for the BF (Section 5 of main manuscript) so that we obtained a range of type I error rates and power rates under each scenario considered. Our simulation results, presented in Tables 1 and 2 of the main manuscript, fall within the range of the results obtained over the complete simulation runs. For the type I error rates, we noticed a higher variability for small sample sizes (N=500) when the number of sites $\le 45$. We also observed a slight increased in the empirical type I error when the number of sites is very large, i.e. $>145$. For the range of power rates, the variability decreases with the sample size and to some extent with the number of sites. 
\begin{figure}
\centering
\includegraphics[scale=0.7]{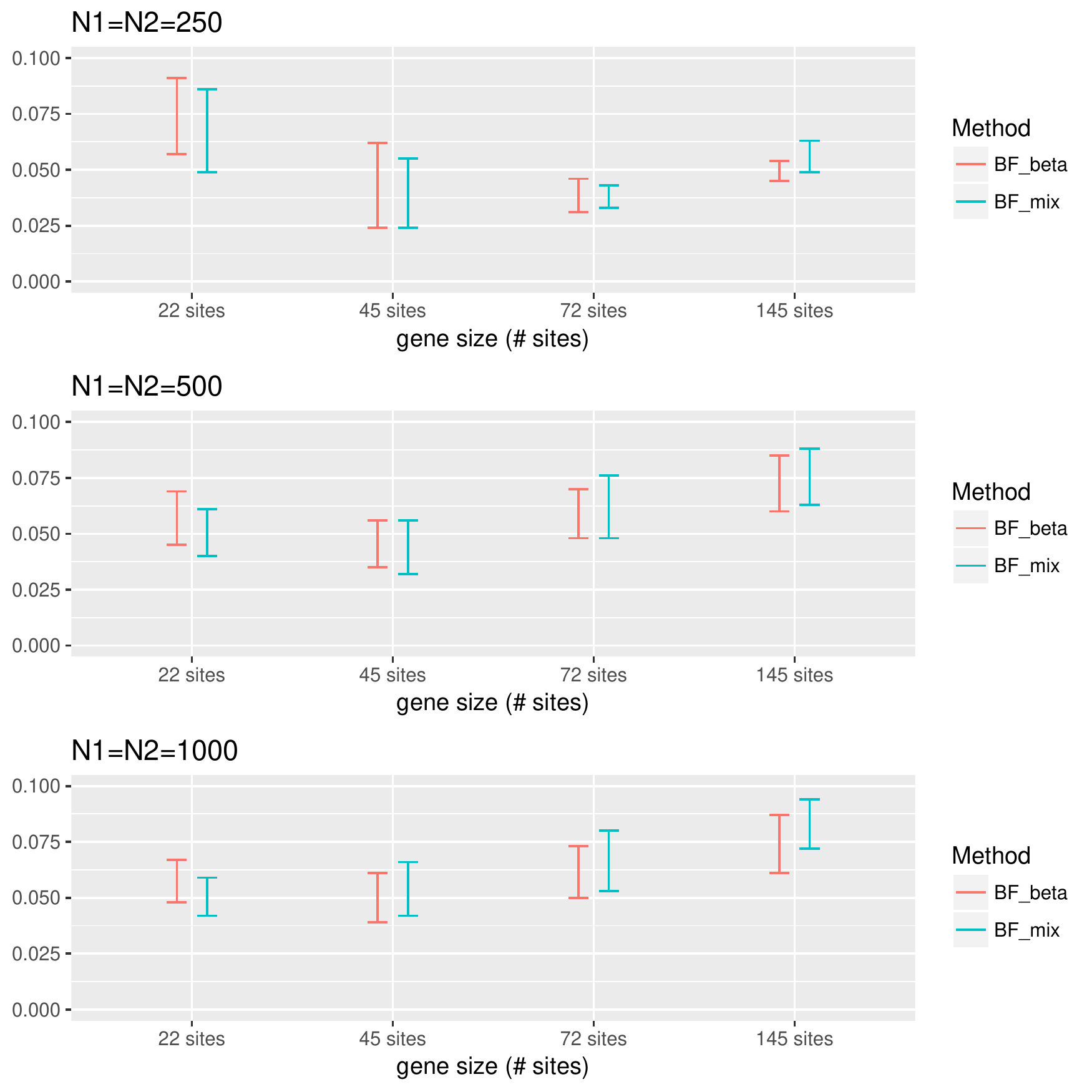}
\caption{Range of type I error rates over 6 complete simulation runs for the BF with beta (\texttt{BF\_beta}) or mixture (\texttt{BF\_mix}) informative prior under different scenarios ($\alpha=0.05$)}
\label{repeat_typeI}
\end{figure}

\begin{figure}
\centering
\includegraphics[scale=0.7]{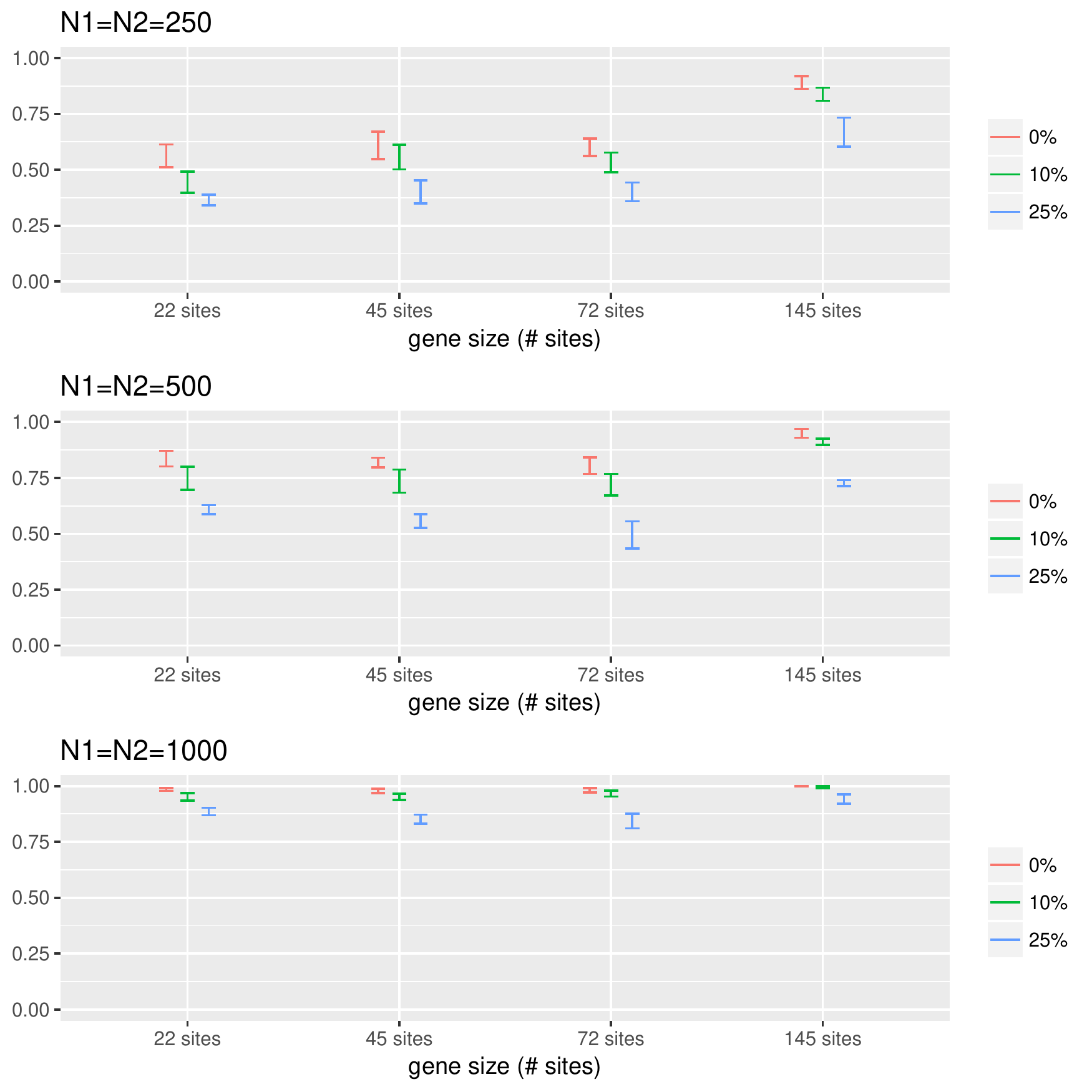}
\caption{Range of power values over 6 complete simulation runs for the BF with beta informative prior under different scenarios ($\alpha=0.05$). Red, green and blue bars indicate that $0\%$, $10\%$ and $25\%$ of the associated variants within the gene are protective.}
\label{repeat_power_beta}
\end{figure}

\begin{figure}
\centering
\includegraphics[scale=0.7]{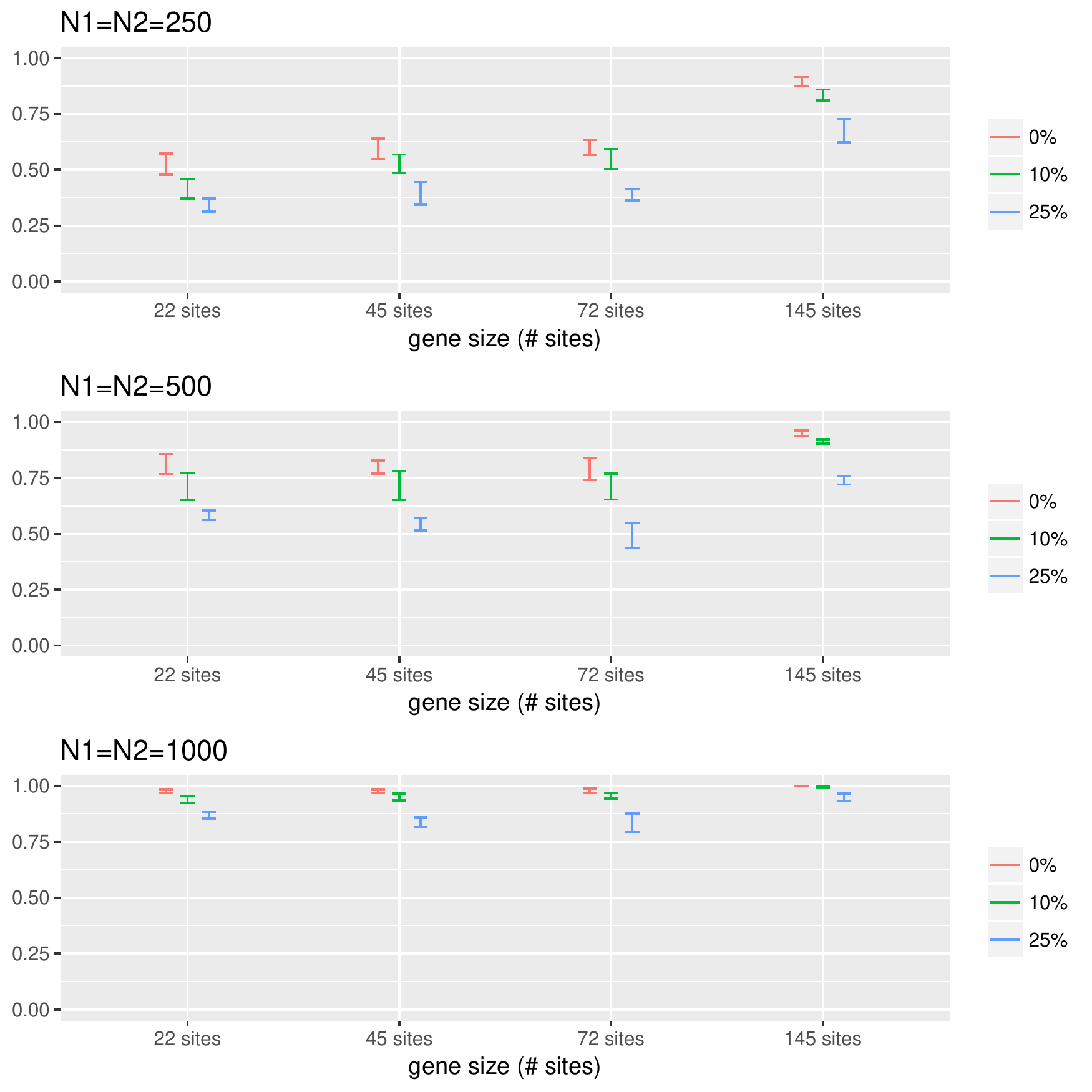}
\caption{Range of power values over 6 complete simulation runs for the BF with a mixture informative prior under different scenarios ($\alpha=0.05$). Red, green and blue bars indicate that $0\%$, $10\%$ and $25\%$ of the associated variants within the gene are protective.}
\label{repeat_power_mix}
\end{figure}
%
%

\section{Web Appendix L: Simulation results for large genes}
In the the WES real data application, less than $1\%$ of the genes (95 out of 13,738 genes) had a number of sites greater than 500. We ran some simulation studies similar to those presented in section 5 of the main manuscript to assess the type I error under $H_0$ assuming large genes. The simulated genes had 5000 sites and we assumed a total sample size of 500, 1000 and 2000 individuals with the same number of cases and controls. Each simulation run included 500 replicates. If the significance level $\alpha=0.05$, the overall type I error rate is 0.086 for $N=500$, 0.073 for $N=1000$ and 0.069 for $N=2000$ based on the BF with beta informative prior.

\section{Web Appendix M: Simulation Results with the BFDR approach}
\begin{figure}[h!]
\centering	
\includegraphics[scale=0.8]{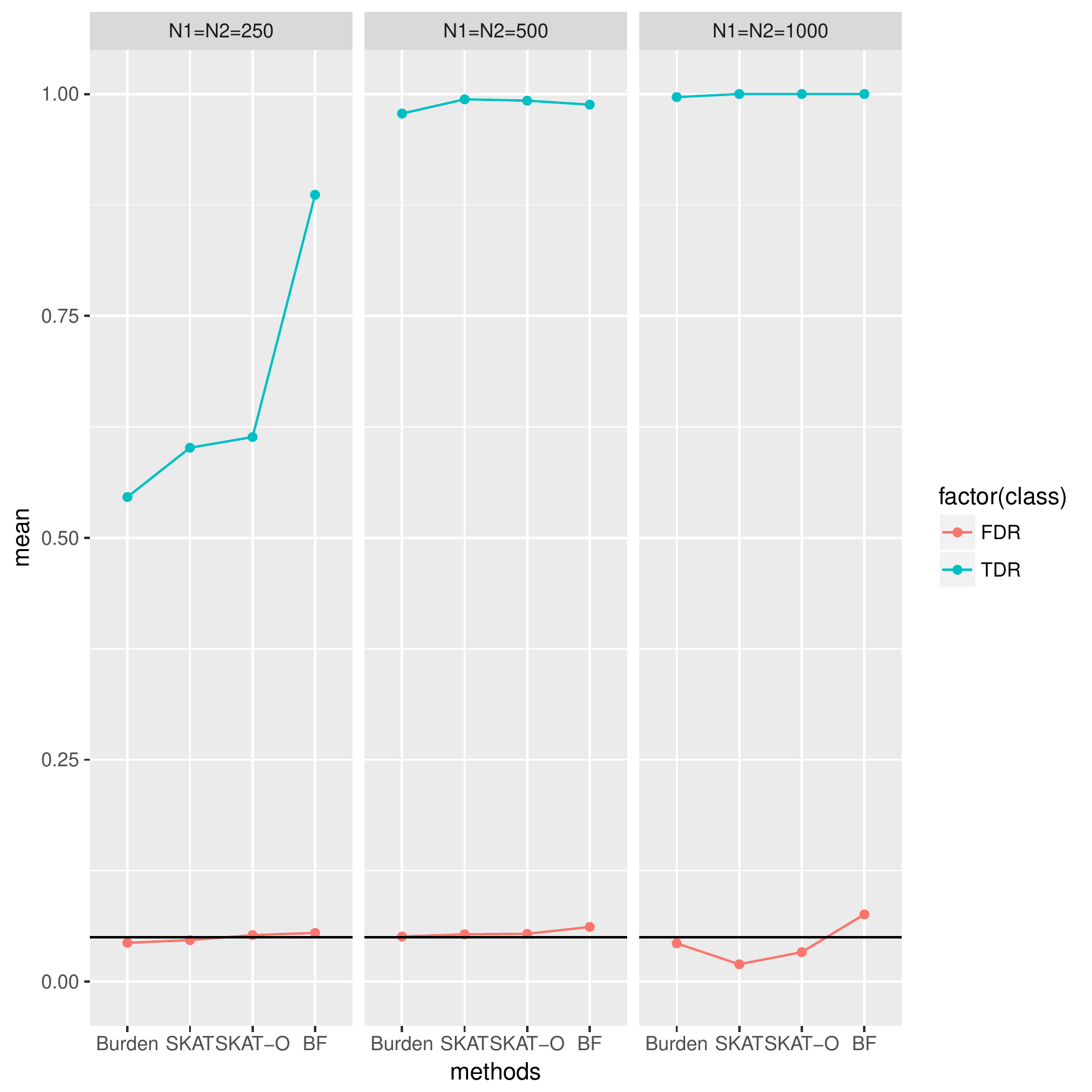}
\caption{Empirical FDR and TDR levels over 100 simulation replicates obtained with the BF, Burden test, SKAT and SKAT-O for a pre-specified FDR level of 0.05 (indicated by the horizontal line). There are 20 out of 10,000 genes associated with the case-control status in each replicate and each associated gene harbours only deleterious genetic variants (i.e., increase disease risk). BF denotes the BF with beta informative prior using BFDR procedure. All other methods are based on the Benjamini-Hochberg (BH) procedure.}
\label{FDR_TDR_result1}
\end{figure}

\begin{figure}[h!]
\centering	
\includegraphics[scale=0.8]{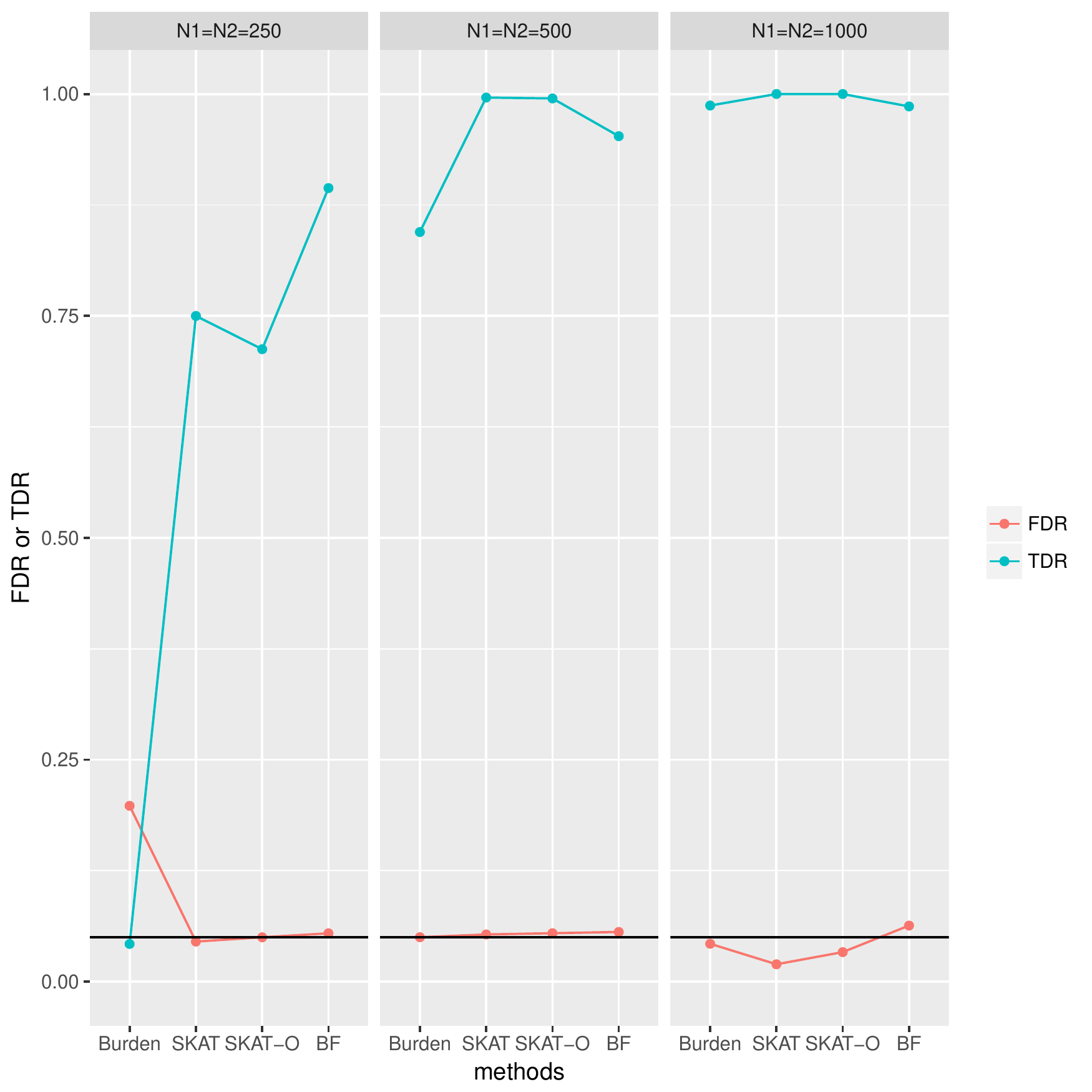}
\caption{ Empirical FDR and TDR levels over 100 simulation replicates obtained with the BF, Burden test, SKAT and SKAT-O for a pre-specified FDR level of 0.05 (indicated by the horizontal line). There are 20 out of 10,000 genes associated with the case-control status in each replicate and each associated gene has a 5:1 ratio of deleterious/protective variants. BF denotes the BF with beta informative prior using BFDR procedure. All other methods are based on the Benjamini-Hochberg (BH) procedure.}
\label{FDR_TDR_result2}
\end{figure}

We ran simulations  using the {\it sim1000G} R package \citep{Dimitromanolakis2019} to compare the BF approach based on BFDR procedure to three competing methods (SKAT, SKAT-O and the Burden test) based on frequentist alternative, { Benjamini-Hochberg (BH) procedure} \citep{Benjamini1995} in the context of genome-wide inference. The datasets were simulated assuming different sample sizes $N1=N2=250$, $N1=N2=500$, and $N1=N2=1000$. For each replicate, we simulated a total of 10,000 genes with 20 of them associated with the phenotype (case-control status). The distribution of the number of sites over the 10,000 simulated genes mimicked the distribution observed in the lung cancer study (Section 6 of the main manuscript) with $46\%$ of the genes with 20-50 sites; $36\%$ with 50-100 sites; $17\%$ with 100-500 sites; and $1\%$ with 500-1000 sites. The proportion of associated variants within the 20 associated genes were decided depending on the gene sizes: $40\%$ of associated variants for genes with 50-100 sites, $30\%$ for genes with 100-500 sites, and $20\%$ for genes with over 500 sites. In addition, we assumed that either all associated variants were deleterious (i.e., increase risk) or there was a  5:1 ratio of deleterious/protective RVs under each simulation scenario. In each scenario, we generated 100 replicated genome-wide data sets. Our goal was to compare the overall empirical FDR level to the nominal FDR level of 0.05 and also evaluate the true discovery rate (TDR), as shown in Figures \ref{FDR_TDR_result1} and \ref{FDR_TDR_result2}. The TDR is defined as the ratio of rejected true signals over the total number of true signals.

\section{Web Appendix N: Real Data Application}

{\it Quality controls}\\
First, we filtered out sites that did not pass all filters when the VCF file was created (not labeled as PASS under FILTER column) and also we only kept bi-allelic variants with Hardy-Weinberg tests $p-$value greater than $10^{-7}$, which leaves a total of 1,810,404 sites in the dataset. Second, we filtered out genotypes with low quality as assessed by GQ score (genotype quality) less than 30 or DP (read depth) less than 10. Third, we deleted variants with missing genotype rate greater than $10$ across individuals. The MAF distribution of the remaining 1,477,890 bi-allelic variants is displayed in Web Appendix Table \ref{t:second}. Fourth, we checked the gender, relatedness, heterozygosity rate and ethnicity of the individuals. One individual was removed because of relatedness and gender issues, two because they were 1st degree relative of other individuals, and one who showed extreme heterozygosity rate and was distant from European ancestry based on principal component analysis. Fifth, we removed common variants and only selected loci with MAF$<0.01$ in the dataset. Since we chose genes as the testing units, in order to conduct the testing robustly, we removed genes with less than 20 loci (sites) for the association testing. For analysis of each chromosome, we further removed individuals with missing genotype rate greater than $5$.  Finally, we excluded highly correlated RVs for the KS test, ie. Pearson correlation $>0.99$. Among the 13,738 genes with at least 20 sites in the clean dataset, the distribution of the number of sites within each gene is shown in Web Appendix Table \ref{t:three}.

\begin{table}
\caption{MAF distribution of SNPs in the lung cancer WES study}
\label{t:second}
\begin{center}
\begin{tabular}{lrrrrr}
\hline
MAF & 0 & (0,0.01) & [0.01,0.05) & [0.05,0.5) & Total \\
\hline
$\#$(SNPs) & 785995 & 488182 & 64330 & 140373 & 1477890 \\
Proportion ($\%$) & 53.2 & 33.0 & 4.4 & 9.5 & 100 \\
\hline
\end{tabular}
\end{center}
\end{table}

\begin{table}
\caption{Number of site distribution in the lung cancer WES study.}
\label{t:three}
\begin{center}
\begin{tabular}{lrrrrr}
\hline
Number of sites & [20,50) & [50,100) & [100,500) & 500+ & Total \\
\hline
$\#$(Genes) & 7201 & 4364 & 2078 & 95 & 13738 \\
Proportion ($\%$) & 52.4 & 31.8 & 15.1 & 0.7 & 100 \\
\hline
\end{tabular}
\end{center}
\end{table}

{\it Additional results}\\

\begin{sidewaystable}
\caption{Top 20 genes identified by the BF with mixture informative prior}
\label{t:four}
\centering
\begin{tabular}{rlrrrrrrr}
\hline
Chromosome & Genes & $\#$ of sites & KS test$^1$  & BF (info)$^2$ & SKAT$^3$ & Burden$^4$ & SKAT-O$^5$ & BF (noninfo) $^6$ \\ 
& & in the gene & p-value & p-value &  rank & rank & rank & rank \\	
\hline
4 & KCNIP4 & 15904 & 9.54E-11 & 5.12E-10 & 459 & 10745 & 796 & 8010 \\ 
13 & GPC5 & 17067 & 3.23E-08 & 1.55E-07 & 5955 & 8601 & 8813 & 9276 \\ 
16 & PAQR4 & 37 & 4.43E-04 & 1.12E-04 & 292 & 262 & 265 & 193 \\ 
12 & PLEKHG7 & 51 & 1.97E-02 & 1.25E-04 & 233 & 10 & 2 & 5 \\ 
17 & ATP6V0A1 & 75 & 5.07E-03 & 4.22E-04 & 1094 & 71 & 150 & 67 \\ 
3 & ERC2 & 78 & 6.60E-03 & 4.94E-04 & 852 & 82 & 144 & 60 \\ 
11 & ST14 & 111 & 5.46E-04 & 5.00E-04 & 416 & 3891 & 714 & 1037 \\ 
17 & TUBG2 & 37 & 5.40E-03 & 5.91E-04 & 652 & 195 & 356 & 90 \\ 
7 & DOCK4 & 259 & 2.80E-01 & 7.30E-04 & 3202 & 3 & 9 & 3 \\ 
16 & SDR42E1 & 26 & 4.62E-03 & 8.00E-04 & 311 & 190 & 253 & 141 \\ 
17 & EVPLL & 20 & 9.82E-03 & 8.95E-04 & 1362 & 77 & 178 & 77 \\ 
7 & KIAA1324L & 103 & 2.98E-03 & 9.49E-04 & 81 & 607 & 139 & 275 \\ 
19 & ATP1A3 & 83 & 5.88E-01 & 1.00E-03 & 73 & 7 & 13 & 1 \\ 
16 & MSRB1 & 32 & 2.70E-03 & 1.09E-03 & 68 & 513 & 105 & 372 \\ 
19 & ZNF556 & 42 & 2.58E-02 & 1.09E-03 & 853 & 38 & 68 & 33 \\ 
9 & PKN3 & 107 & 2.72E-02 & 1.30E-03 & 27 & 200 & 42 & 36 \\ 
6 & ESR1 & 54 & 3.12E-03 & 1.37E-03 & 542 & 632 & 632 & 434 \\ 
10 & SLF2 & 56 & 2.58E-02 & 1.40E-03 & 778 & 47 & 78 & 44 \\ 
5 & ERAP2 & 79 & 7.85E-01 & 1.44E-03 & 3096 & 1 & 3 & 2 \\ 
17 & TRIM47 & 33 & 3.23E-03 & 1.65E-03 & 1560 & 613 & 1151 & 531 \\ 
\hline
\multicolumn{9}{l}{ {\small 1. KS p-value used as the p random variable;}}\\
\multicolumn{9}{l}{ {\small 2. The p-value of BF with informative prior is calculated based on the null distribution of $2log(BF)$ of $\chi^2(3)$;}}\\
\multicolumn{9}{l}{{\small 3. Ranking of genes according to SKAT;}}\\
\multicolumn{9}{l}{ {\small 4. Ranking of genes according to the Burden test;}}\\
\multicolumn{9}{l}{ {\small 5. Ranking of genes according to SKAT-O;}}\\	
\multicolumn{9}{l}{ {\small 6. Ranking of genes according to BF with non-informative prior as shown in equation (15).}}\\	
\end{tabular}
\end{sidewaystable}

\begin{table}[ht]
\caption{Top 20 genes identified by SKAT, Burden and SKAT-O methods}
\label{t:five}
\centering
\begin{tabular}{lll | lll | lll}
\hline
SKAT$^1$ & SKAT & BF$^2$ & Burden$^3$ & Burden & BF$^4$ & SKAT-O$^5$ & SKAT-O & BF$^6$ \\ 
& p-value & rank & & p-value & rank & & p-value & rank \\
\hline
TERT & 2.63E-05 & 220 & ERAP2 & 2.97E-05 & 18 & TERT & 1.55E-05 & 220 \\ 
INPP4B & 4.28E-05 & 2015 & TET2 & 7.83E-05 & 50 & PLEKHG7 & 1.78E-05 & 5 \\ 
DPF2 & 1.88E-04 & 586 & DOCK4 & 1.96E-04 & 29 & ERAP2 & 9.01E-05 & 18 \\ 
CEACAM7 & 2.44E-04 & 6809 & TM4SF5 & 2.25E-04 & 36 & INPP4B & 1.32E-04 & 2015 \\ 
YAP1 & 3.50E-04 & 2880 & TLR6 & 2.28E-04 & 7 & TET2 & 2.19E-04 & 50 \\ 
NCKIPSD & 3.56E-04 & 1024 & PLA2G5 & 2.50E-04 & 992 & TM4SF5 & 2.43E-04 & 36 \\ 
PPP1R13L & 4.70E-04 & 1332 & ATP1A3 & 2.93E-04 & 99 & PLA2G5 & 2.49E-04 & 992 \\ 
MARK4 & 6.16E-04 & 10646 & AKR1A1 & 3.16E-04 & 24 & CEACAM7 & 3.65E-04 & 6809 \\ 
PPP2R2C & 6.21E-04 & 1644 & TERT & 3.49E-04 & 220 & DOCK4 & 3.79E-04 & 29 \\ 
SLC30A4 & 7.30E-04 & 381 & PLEKHG7 & 3.80E-04 & 5 & TLR6 & 4.10E-04 & 7 \\ 
STARD3 & 7.55E-04 & 7842 & PIP5K1B & 4.43E-04 & 42 & DPF2 & 4.58E-04 & 586 \\ 
INPPL1 & 7.90E-04 & 1099 & SNCA & 4.90E-04 & 35 & SCYL1 & 4.65E-04 & 188 \\ 
ELL & 8.36E-04 & 1522 & LCMT2 & 5.37E-04 & 81 & ATP1A3 & 4.99E-04 & 99 \\ 
ABCD4 & 8.94E-04 & 5730 & UNC45B & 5.48E-04 & 38 & REG4 & 5.91E-04 & 6 \\ 
REG4 & 9.25E-04 & 6 & PID1 & 5.61E-04 & 111 & ABCD4 & 6.40E-04 & 5730 \\ 
CD1C & 9.47E-04 & 336 & PEPD & 6.98E-04 & 67 & AKR1A1 & 6.58E-04 & 24 \\ 
COMT & 9.97E-04 & 3712 & SNRPN & 7.22E-04 & 94 & YAP1 & 8.55E-04 & 2880 \\ 
PIP5K1C & 1.08E-03 & 7190 & SNURF & 7.60E-04 & 94 & NCKIPSD & 8.72E-04 & 1024 \\ 
ZNF638 & 1.11E-03 & 1327 & ZC3HC1 & 8.00E-04 & 148 & NUBPL & 8.88E-04 & 219 \\ 
RASA1 & 1.13E-03 & 3547 & OVCH2 & 8.00E-04 & 223 & SNCA & 9.56E-04 & 35 \\ 
\hline
\multicolumn{9}{l}{ {\small 1. Top 20 genes identified by SKAT;}} \\
\multicolumn{9}{l}{ {\small 2. Ranking of BF with beta informative prior for the top-ranking genes identified by SKAT;}} \\
\multicolumn{9}{l}{ {\small 3. Top 20 genes identified by Burden;}} \\
\multicolumn{9}{l}{ {\small 4. Ranking of BF with beta informative prior for the top-ranking genes identified by Burden;}} \\
\multicolumn{9}{l}{ {\small 5. Top 20 genes identified by SKAT-O;}} \\
\multicolumn{9}{l}{ {\small 6. Ranking of BF with beta informative prior for the top-ranking genes identified by SKAT-O.}} \\										
\end{tabular}
\label{method_compete}
\end{table}

\begin{table}
\caption{For the top 100 genes identified by the BF and other competing methods, proportion of genes falling in each gene size category in the lung cancer WES study.}
\label{t:six}
\begin{center}
\begin{tabular}{lrrrr}
\hline
Number of sites in the gene & [20,50) & [50,100) & [100,500) & 500+ \\
\hline
BF (beta informative prior) & $53\%$ & $34\%$ & $10\%$ & $3\%$ \\
BF (mixture informative prior) & $42\%$ & $39\%$ & $16\%$ & $3\%$ \\
BF (beta non-informative prior) & $51\%$ & $39\%$ & $9\%$ & $1\%$ \\
BF (mixture non-informative prior) & $35\%$ & $39\%$ & $25\%$ & $1\%$ \\
SKAT & $40\%$ & $40\%$ & $17\%$ & $3\%$ \\
Burden & $51\%$ & $37\%$ & $11\%$ & $1\%$ \\
SKAT-O & $52\%$ & $35\%$ & $11\%$ & $2\%$ \\
\hline
\end{tabular}
\end{center}
\end{table}

\begin{figure}
\centering
\begin{minipage}[b]{0.45\textwidth}
\includegraphics[width=\textwidth]{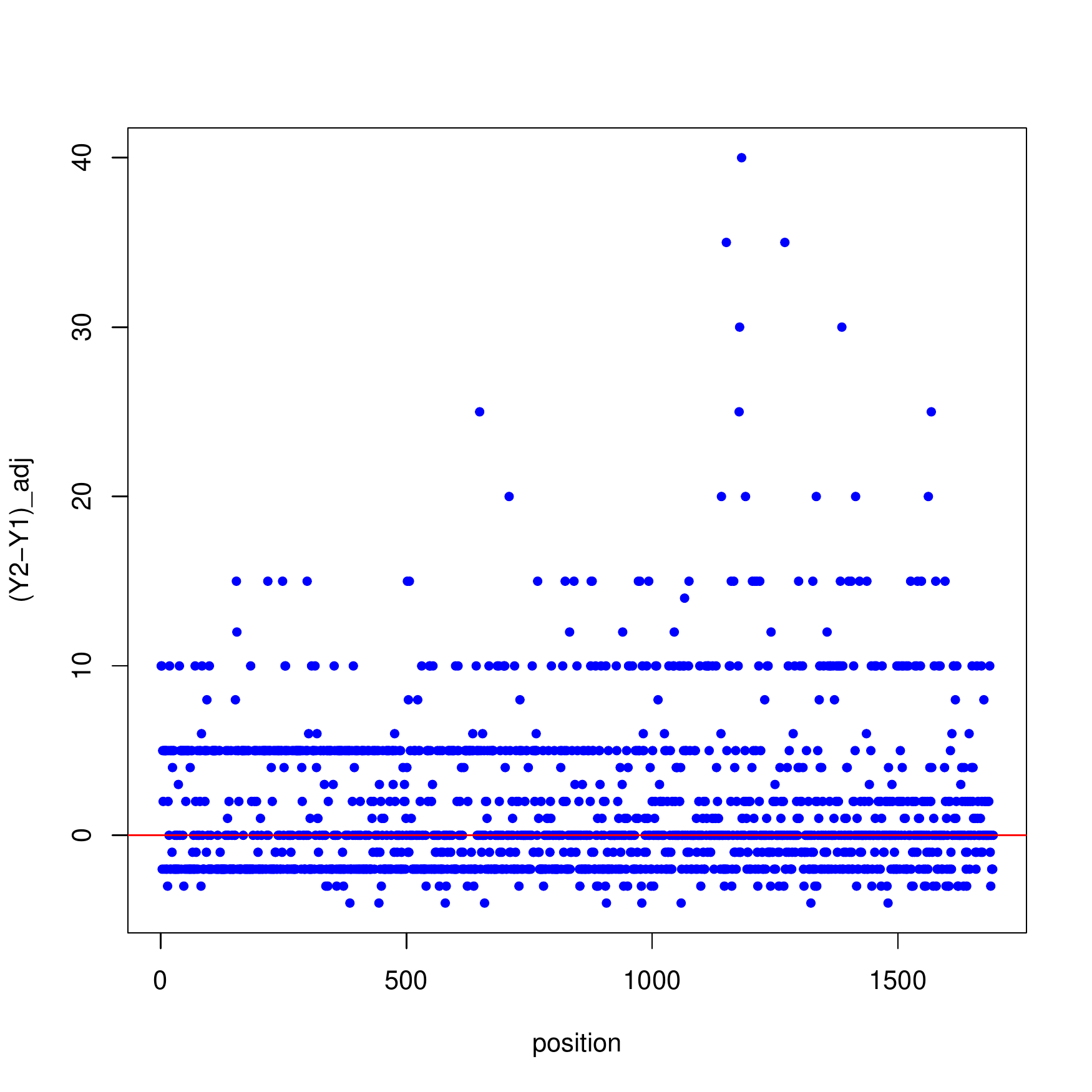}
\end{minipage}
\begin{minipage}[b]{0.45\textwidth}
\includegraphics[width=\textwidth]{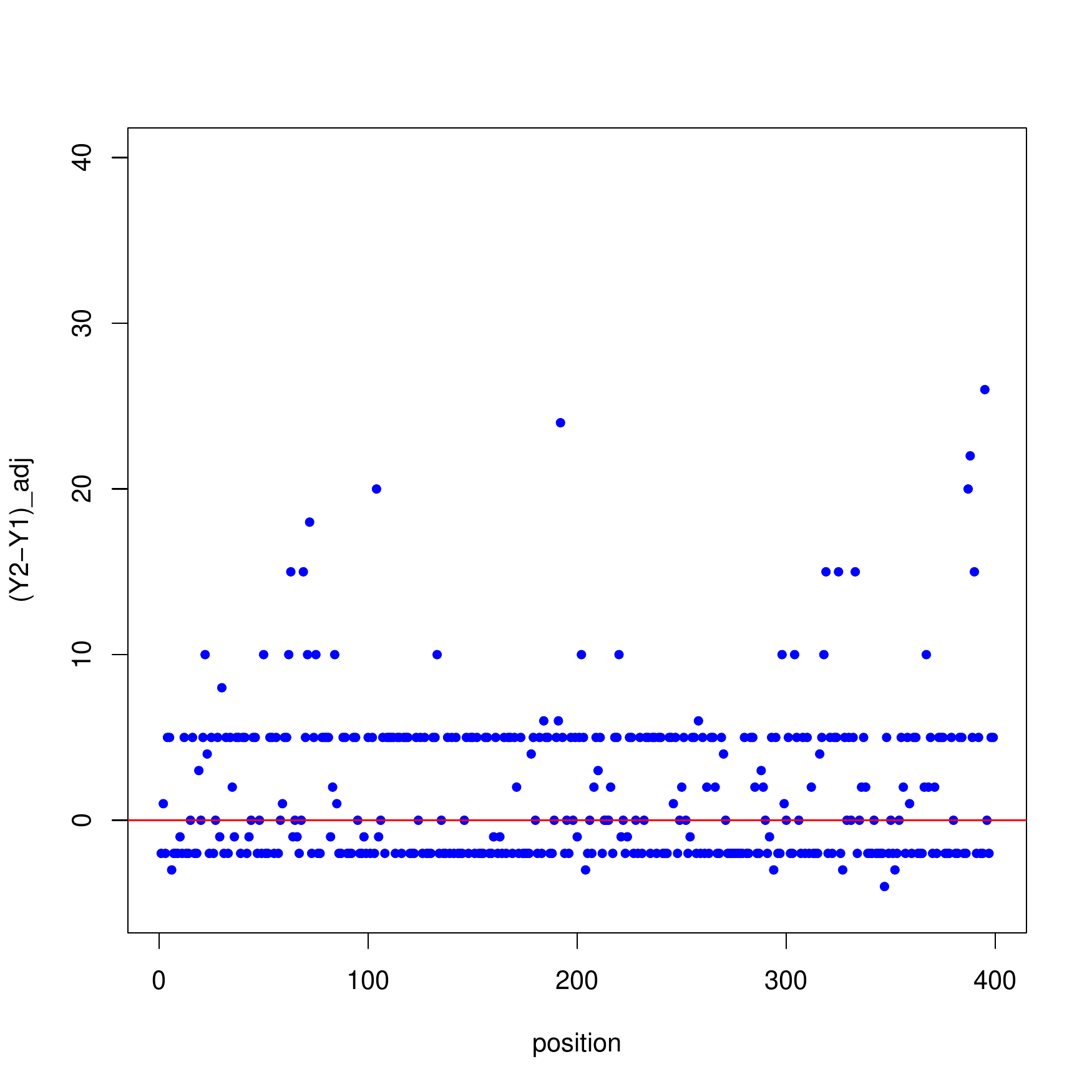}
\end{minipage}

\begin{minipage}[b]{0.45\textwidth}
\includegraphics[width=\textwidth]{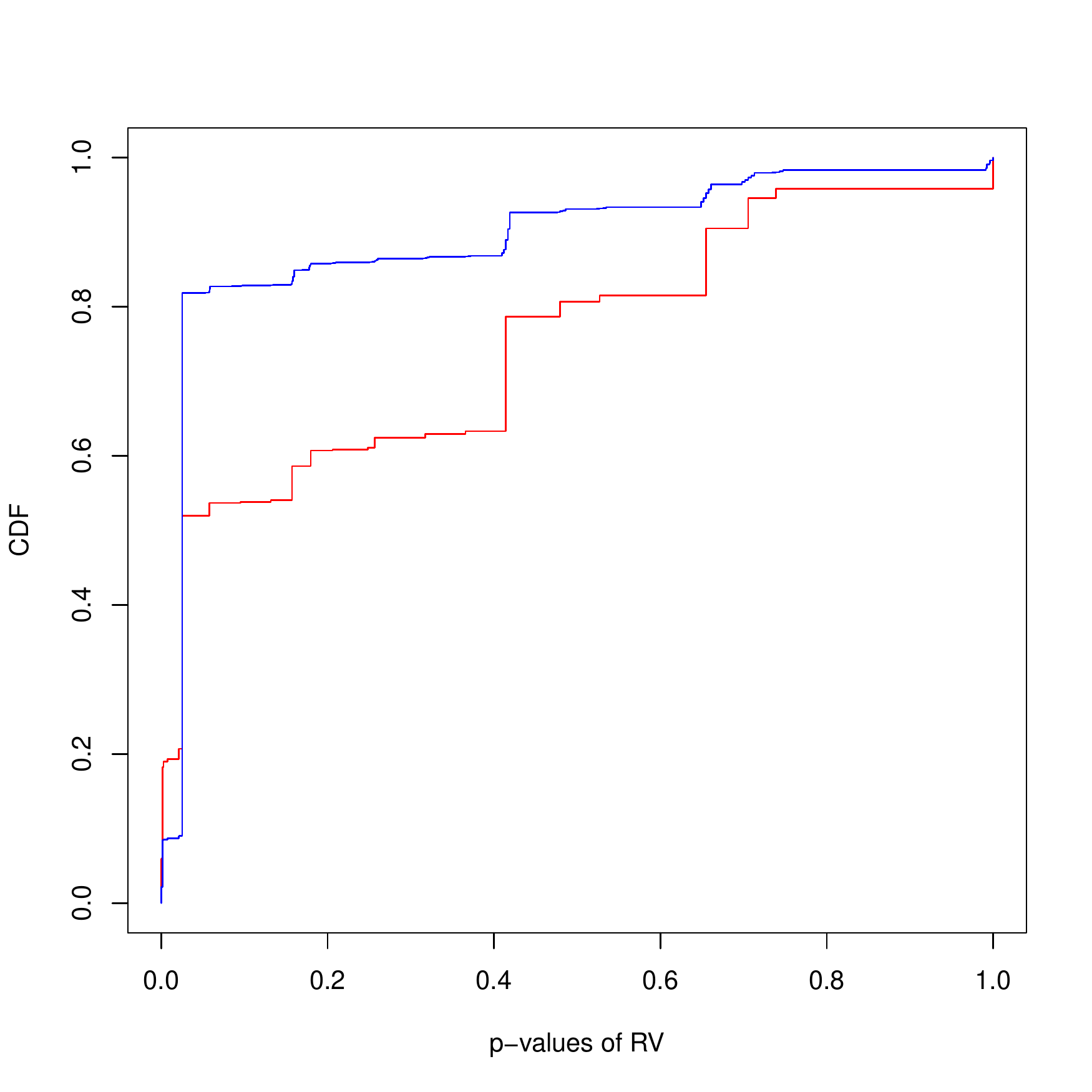}
\end{minipage}
\begin{minipage}[b]{0.45\textwidth}
\includegraphics[width=\textwidth]{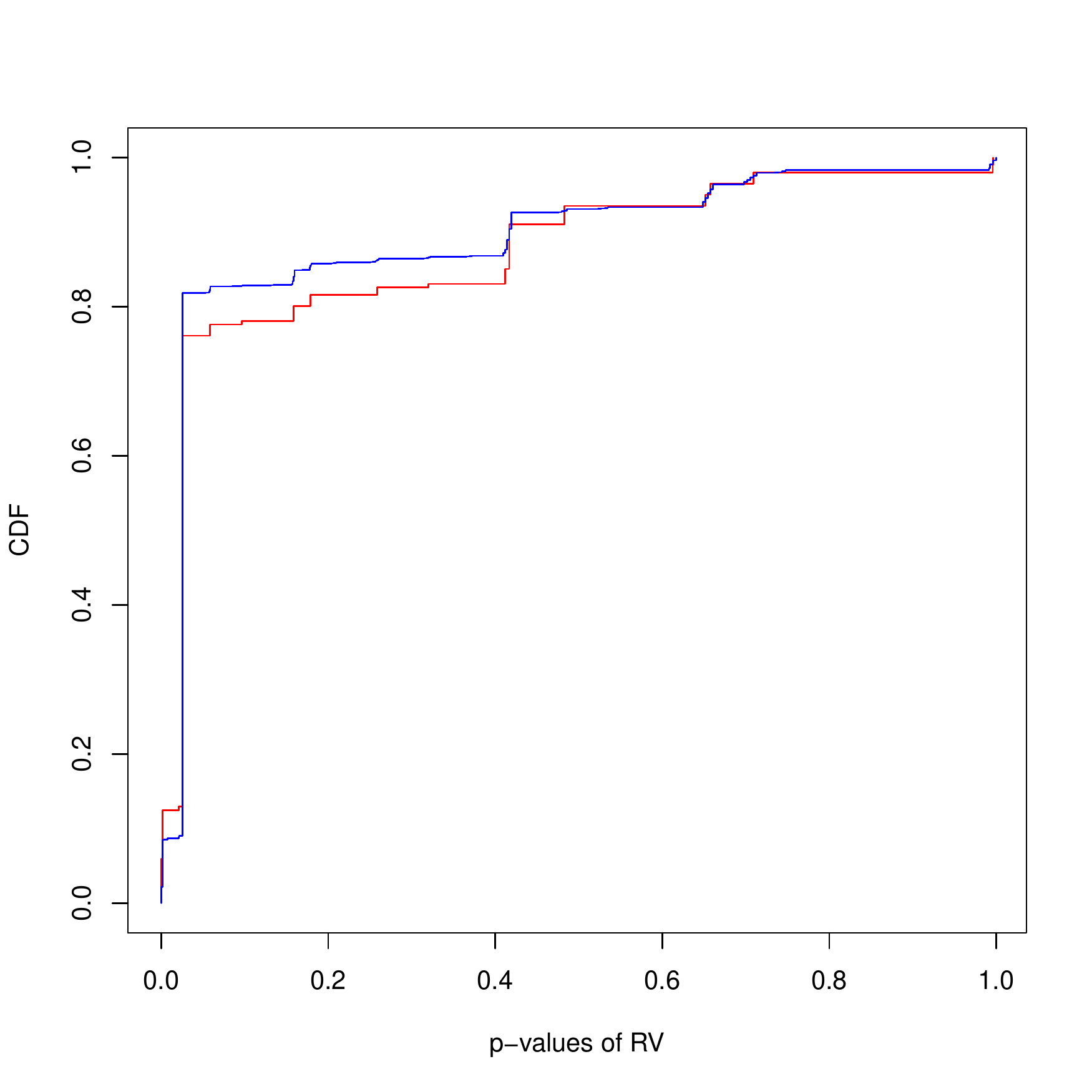}
\end{minipage}

\caption{ Association results for the top gene identified by the BF with informative prior, $KCNIP4$ (top left side) (see Section 6) and the 5th top gene identified by SKAT, $YAP1$ (top right side) (see Table \ref{method_compete}). The top two figures show the difference of standardized RV counts between cases and controls, $\tilde{Y}_{2v}-\tilde{Y}_{1v}$ (defined in section 3.5). The bottom two figures compare the CDF of $p$-values for each gene to the empirical null distribution. The blue line represents the null distribution of $p$-values based on RV tests over all genes across the genome. The red line represents the CDF of RV $p$-values from the gene displayed. The one-sided KS test is detecting an excess of small $p$-values in each gene (the red line is above the blue line for small $p$-values).}
\end{figure}

\section{Web Appendix O: Estimation of parameters $K^*$ and $K^*\eta^*$ in the hyperprior distribution}
In the Theorem 1 of the main manuscript, it is assumed that $K^*\rightarrow \infty$ and $K^*\eta^*\rightarrow \infty$. In this section, the distributions of $\log(K^*)$ and $\log(K^*\eta^*)$ for the BF with non-informative prior from the simulated data are compared across different simulation scenarios with different gene sizes and sample sizes.
\begin{figure}[h!]
\centering	
\includegraphics[scale=0.7]{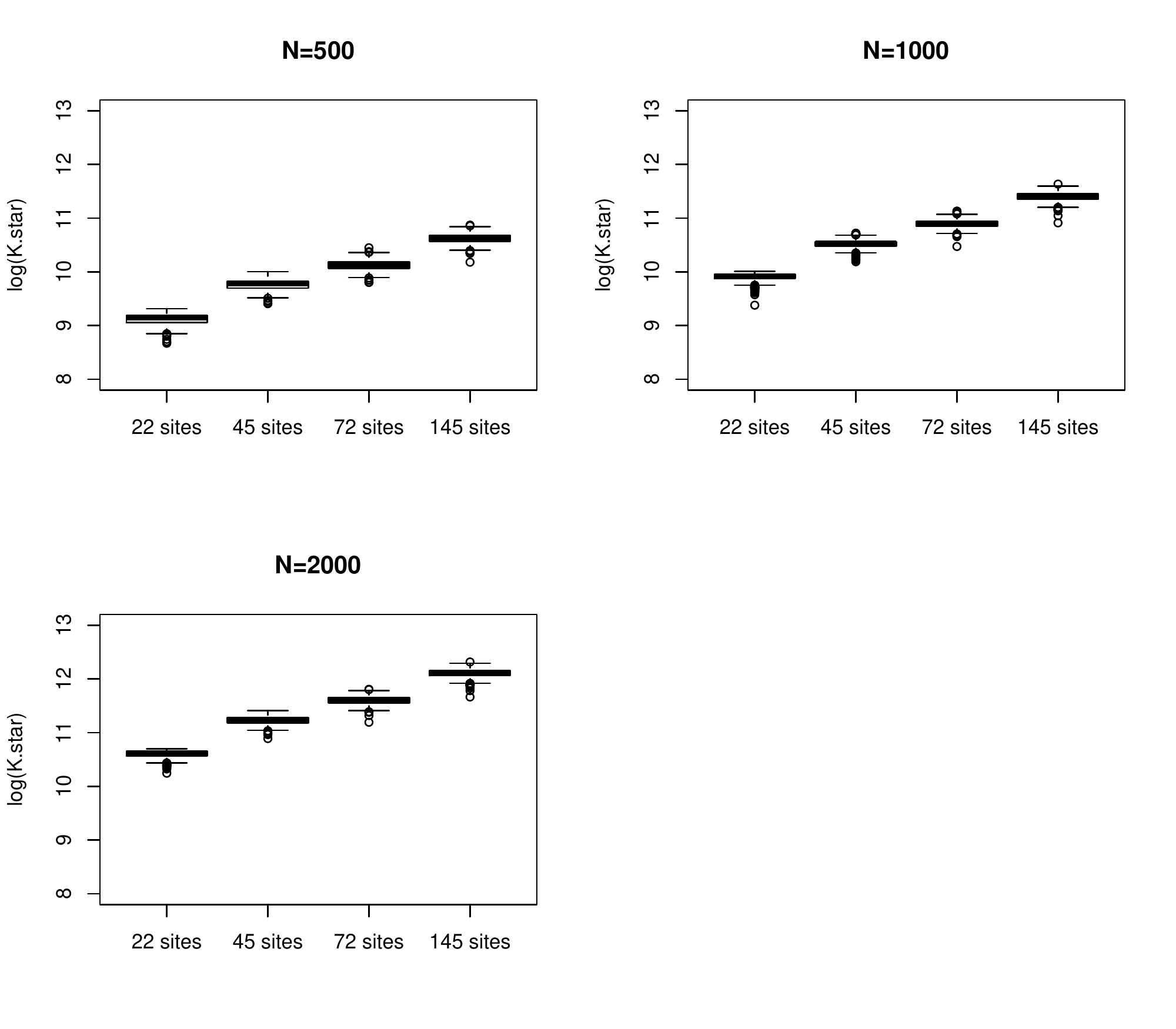}
\caption{Estimation of $K^*$ under different simulation scenarios, for BF with non-informative prior ($p=1$). The result is based on 1000 genes simulated under $H_0$. It shows that $K^*$ increases with the gene size. Based on the value of $\log(K^*)$ from the simulated data, the assumption $K^*\rightarrow \infty$ seems realistic and the asymptotic results from Theorem 1 valid.}
\label{K_star_est}
\end{figure}
\begin{figure}[h!]
\centering	
\includegraphics[scale=0.7]{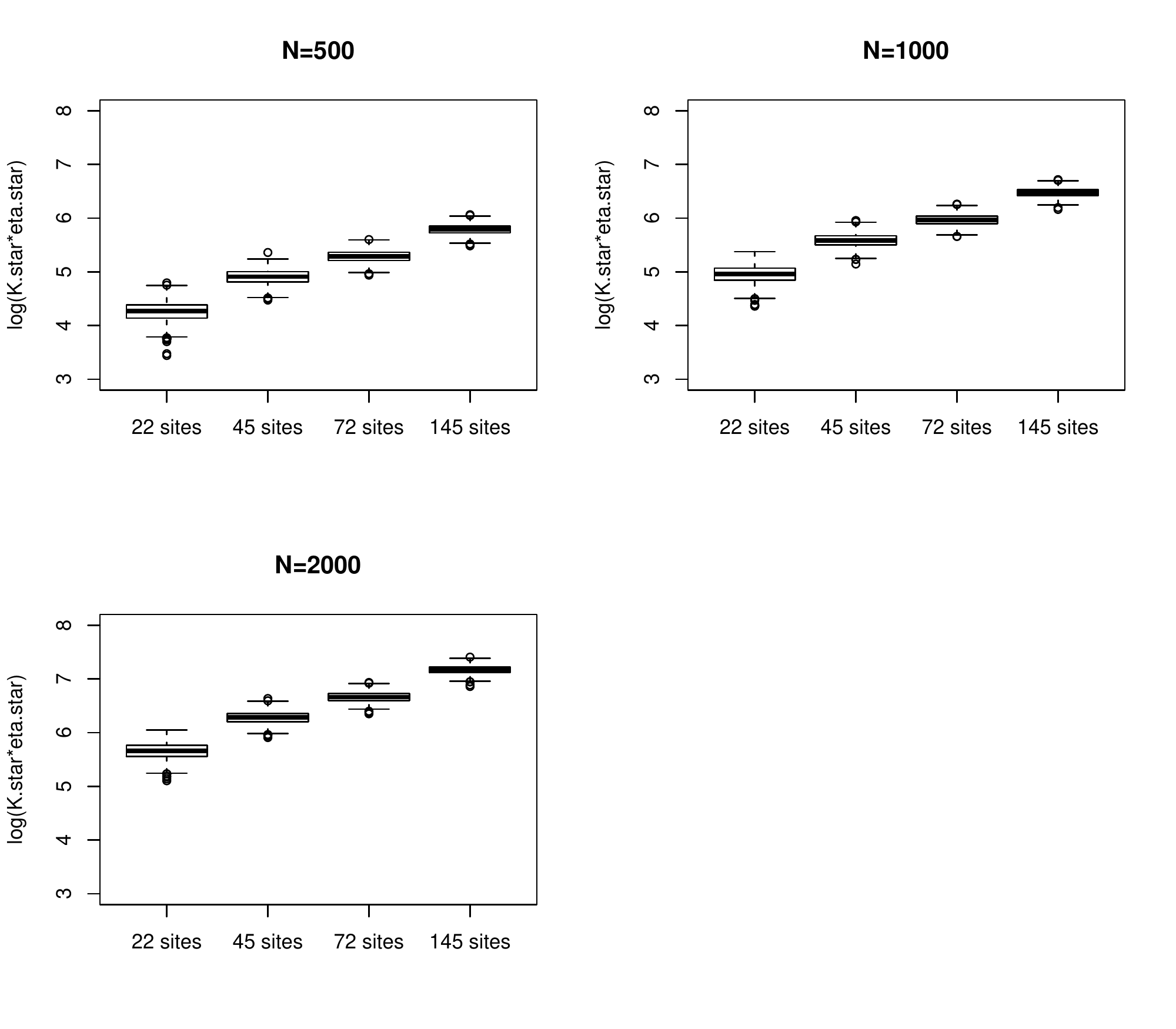}
\caption{Estimation of $K^*\eta^*$ under different simulation scenarios, for BF with non-informative prior ($p=1$). The result is based on 1000 genes simulated under $H_0$. Based on the value of $\log(K^*\eta^*)$ from the simulated data, the assumption $K^*\eta^*\rightarrow \infty$  seems realistic and the asymptotic results from Theorem 1 valid. {Caution is however necessary when both sample size and number of sites are small, as shown in the upper left panel of the figure}.}
\label{K_eta_star_est}
\end{figure}

\section{Data and code}
{Our simulation approach and BF method are implemented respectively in the $R$ package $sim1000G$ available on CRAN and $rareBF$ available on Github\\ (https://github.com/adimitromanolakis/rareBF)}. Besides, Toronto lung cancer WES data is available through dbGaP (https://www.ncbi.nlm.nih.gov/gap). 
\bibliographystyle{biom}  \bibliography{supplement}